\documentclass[a4paper,12pt]{article}

\usepackage{graphicx}
\usepackage{amsmath}
\usepackage{amssymb}
\usepackage{hyperref}
\usepackage{tabularx}
\newcolumntype{M}{>{$}c<{$}}
\usepackage[matrix,arrow,curve]{xy}
\numberwithin{equation}{section}
\numberwithin{figure}{section}
\numberwithin{table}{section}

\def\papertitlepage{\baselineskip 3.5ex\thispagestyle{empty}}
\def\Title#1{\baselineskip 1cm \vspace{1.5cm}%
  \begin{center}{\Large\bf #1}\end{center}\vspace{0.5cm}}
\def\Authors#1{\begin{center}\renewcommand{\thefootnote}{\fnsymbol{footnote}}{\it #1}\end{center}}
\def\Abstract{\vspace{1.0cm}%
  \begin{center}{\large\bf Abstract}\end{center}}

\renewenvironment{thebibliography}{\pagebreak[3]\par\vspace{0.6em}
\begin{flushleft}{\large \bf References}\end{flushleft}
\vspace{-1.0em}

\begin{enumerate}\if@twocolumn\baselineskip=0.6em\itemsep -0.2em
\else\itemsep -0.2em\fi\labelsep 0.1em}{\end{enumerate} }

\def\calA{{\cal A}} \def\calB{{\cal B}} 
  \def\calF{{\cal F}}
\def\calG{{\cal G}} \def\calH{{\cal H}} 
  
  \def\calO{{\cal O}}
  
 \def\calT{{\cal T}} 
  \def\calX{{\cal X}}
 \def\calZ{{\cal Z}}

\def\del{\partial}

\def\mathe{\mathrm{e}}
\def\mathi{i}

\def\Tr{\mathrm{Tr}}

\newcommand{\fslash}[1]{\ooalign{\hfil\ensuremath{\slash}\hfil\crcr\ensuremath{#1}}}
\newcommand{\fbslash}[1]{\ooalign{\hfil\ensuremath{\backslash}\hfil\crcr\ensuremath{#1}}}

\def\delx{\del x}

\def\Poisson[#1,#2]{\{#1,\,#2\}_{\text{P}}}
\DeclareMathDelimiter{\lcolon}{\mathopen}{operators}{"3A}{largesymbols}{"3A}
\DeclareMathDelimiter{\rcolon}{\mathclose}{operators}{"3A}{largesymbols}{"3A}

\def\+{\!\!+\!\!}
\def\delcech{\Check{\delta}}
\def\deldol{\overline{\del}_{X}}
\def\targetd{\mathrm{d}}
\def\targetdbar{\mathrm{d}}

\def\lpar{(\!(}
\def\rpar{)\!)}
\def\lbra{[\![}
\def\rbra{]\!]}

\def\dynkin(#1){(#1)}
\def\Sym{\mathbf{S}}

\def\lb{\overline\lambda}
\def\a{\alpha}
\def\t{\theta}
\def\b{\beta}
\def\g{\gamma}

\def\w{\omega}
\def\bra<#1|{\langle#1|}
\def\ket|#1>{|#1\rangle}
\def\braket<#1|#2>{\langle#1|#2\rangle}
\def\llangle{\langle\!\langle}
\def\rrangle{\rangle\!\rangle}
\def\bbra<#1|{\llangle#1|}
\def\kket|#1>{|#1\rrangle}
\def\bbraket<#1|#2>{\llangle#1|#2\rrangle}

\begin{document}
%%% * Title, author etc
{\papertitlepage
\hbox{ }\vspace*{0cm}
% Preprint numbers
{%{\begin{minipage}{3cm}\underline{{\scshape First Draft (v.1.6)}}\end{minipage}}
\hfill
\begin{minipage}{4.2cm}
IFT-P.008/2008\par\noindent
IHES-P/08/31/2008\par\noindent
ITEP-TH-26/08\par\noindent
June, 2008
\end{minipage}}
%%%%%%%%%%%%%%%%%%%%%%%%%%%%
%\vspace{1cm}
\Title{Pure Spinor Partition Function and the Massive Superstring Spectrum}
%\vspace{1cm}
\Authors{{\sc Yuri Aisaka${}^{1}$\footnote{\tt yuri@ift.unesp.br}},
{\sc E.~Aldo~Arroyo${}^{1}$\footnote{\tt aldohep@ift.unesp.br}},
{\sc Nathan~Berkovits${}^{1}$\footnote{\tt nberkovi@ift.unesp.br}}, \\
and {\sc Nikita Nekrasov${}^2$\footnote{{\tt nikitastring@gmail.com}, {\tiny on leave of absence from ITEP, Moscow, Russia}}}
\\
${}^1$Instituto de F\'{i}sica Te\'{o}rica, 
S\~{a}o Paulo State University, \\[-2ex]
Rua Pamplona 145, S\~{a}o Paulo, SP 01405-900, Brasil
\\
${}^2$Institut des Hautes Etudes Scientifiques, Le Bois-Marie, \\[-2ex]
Bures-sur-Yvette, F-91440, France
}
} % end \papertitlepage
\setcounter{footnote}{0}
\vskip-\baselineskip
%%%%%%%%%%%%%%%%%%%%%%%%%%%%%%%%%%%%%%%%%%%%%%%%%%%%%%%%%%%%%%%%
%%% * Abstract
{\baselineskip .5cm
\Abstract
We explicitly compute up to the fifth mass-level
the partition function of ten-dimensional
pure spinor worldsheet variables including the spin dependence.
After adding the
contribution from the $(x^{\mu}, \theta^{\alpha}, p_{\alpha})$ matter variables,
we reproduce the massive superstring spectrum.

Even though pure spinor variables are bosonic, the pure spinor
partition function contains fermionic states which first appear at
the second mass-level. These fermionic states come from functions
which are not globally defined in pure spinor space, and are
related to the $b$ ghost in the pure spinor formalism.
This result clarifies the proper definition of the Hilbert space for pure
spinor variables.
}
\newpage
%%%%%%%%%%%%%%%%%%%%%%%%%%%%%%%%%%%%%%%%%%%%%%%%%%%%%%%%%%%%%%%%
\tableofcontents

\section{Introduction}

Over the last seven years, the pure spinor formalism for the
superstring has been successfully used to compute multiloop
scattering amplitudes and describe Ramond-Ramond backgrounds in
a super-Poincar\'{e} covariant manner~\cite{Berkovits:2000fe}.
Nevertheless, there are some fundamental features of the new
formalism which are not yet well-understood.
Two such features are the composite $b$ ghost and the Hilbert
space for the pure spinor variables.

As in $N=2$ topological strings, the $b$ ghost in the pure spinor formalism
is not a fundamental worldsheet variable but is a composite operator
defined to satisfy $\{Q, b\} = T$ where $Q$ is the BRST operator and
$T$ is the stress tensor. However, in the pure spinor formalism,
the composite $b$ ghost involves inverse powers of $\lambda^\a$
where $\lambda^\a$ is the pure spinor variable constrained to satisfy
$\lambda\g^\mu\lambda=0$. Since this operator diverges when certain
components of $\lambda^\a$ are zero, the
$b$ ghost is not a globally defined function on the space of pure
spinors;
rather, it must be described using a certain extension of
the higher cohomologies of this space.

Although it was shown in~\cite{Berkovits:2005bt}\cite{Berkovits:2006vi}
how to functionally integrate over $\lambda^\a$
in the presence of such operators, it was unclear how to properly
define the Hilbert space of allowable functions of pure spinor variables.
In this paper, this Hilbert space question will be answered by explicitly
computing the partition function for the pure spinor variables and
studying properties of the states in the Hilbert space.
It will be shown that only states which diverge slower than $(\lambda)^{-4}$
when $\lambda\to 0$ contribute to the pure spinor partition function. Since
the functional integral $\int\targetd^{11}\lambda ~f(\lambda)$ is well-defined as long
as $f(\lambda)$ diverges slower\footnote{Since ${\lambda}=0$ is a point
in the complex variety,  the expression ``{\it $f({\lambda})$ diverges
as ${\lambda}^{-k}$}\,'' has to be taken in the algebraic geometry sense, i.e. as
follows: $f({\lambda})$ is singular on some subvariety containing ${\lambda}=0$, 
but ${\lambda}_{1}{\lambda}_{2}\ldots {\lambda}_{k}\, f({\lambda})$ is regular at ${\lambda}=0$
where ${\lambda}_{i}$, $i=1, \ldots , k$ are some, not necessarily
independent, linear functions in the ambient vector space
${\mathbb{C}}^{16}$.}
than $(\lambda)^{-11}$ when $\lambda\to 0$
(or slower than ${(\lambda)
}^{-8}$ if we use the globally defined holomorphic
top form on the pure spinor space instead of ${\rm d}^{11}{\lambda}$), this result
implies that functional integration over $\lambda^\a$ can be consistently
defined.

For states depending only on the zero modes of $\lambda^\a$,
the Hilbert
space of states in the pure spinor formalism is easily understood
and is given by arbitrary polynomials in $\lambda^\a$. This follows
from the fact that ${\lambda}=0$ is the point of high codimension, so 
that any holomorphic function extends to ${\lambda}=0$. 
Since $\lambda^\a$ is constrained
to satisfy $\lambda\g^\mu\lambda=0$, these
polynomials are
parameterized by constants $f_{\lpar \a_1 ...\a_n \rpar}$ for $n=0$ to $\infty$
which are symmetric in their spinor indices and which satisfy
$\g_\mu^{\a_1 \a_2} f_{\lpar \a_1 ... \a_n \rpar}=0$.

As shown in~\cite{Berkovits:2005hy}, this Hilbert
space for the zero modes is described by the partition function
$$Z_0(t) = (1-t)^{-16}(1 -10 t^2+ 16 t^3 - 16 t^5 +10 t^6 -t^8)$$
where $\lambda^\a$ carries $+1$ $t$-charge. Expanding $Z_0(t)$ in powers of
$t$, one reproduces the independent number of $f_{\lpar \a_1 ...\a_n \rpar}$'s at
order $t^n$. After multiplying by $(1-t)^{16}$ which comes from the
partition function for the 16 $\t^\a$ zero modes, $(1-t)^{16}Z_0(t)$
describes the $x$-independent degrees of freedom for the massless sector
of the open superstring. For example, 1 describes the Maxwell ghost,
$-10 t^2$ describes the photon, $+16 t^3$ describes the photino, and
the remaining terms describe the antifields for these states.
Note that $Z_0(t)$ satisfies the identity $Z_0(1/t) = - t^8 Z_0(t)$
which implies a symmetry between the fields and antifields.

In this paper, we shall perform a similar analysis for the non-zero modes
of $\lambda^\a$, as well as the modes of its conjugate momentum $\w_\a$.
The partition function for the lowest non-zero mode was already computed
in~\cite{Grassi:2005jz}, and we shall extend this computation up to the first five non-zero
modes. The computation will be performed in two ways, firstly using
the ghost-for-ghost method and secondly using the fixed-point method.
After including the contribution from the matter variables $(x^{\mu},\t^\a,p_\a)$,
we will show that the complete partition function correctly describes the
first five massive levels of the open superstring spectrum.

In computing the partition function for the non-zero modes of $\lambda^\a$
and $\w_\a$, we will discover a surprise. Because the constraint
$\lambda\g^\mu\lambda=0$ generates the gauge transformation
$\delta_{\Lambda} \w_\a = \Lambda^\mu (\g_\mu\lambda)_\a$
for the conjugate momentum,
one naively expects that the Hilbert space is described by polynomials
of $\lambda^\a$ and $\w_\a$ (and their worldsheet
derivatives) which are invariant under
this gauge transformation. However, in addition to these ordinary gauge invariant
states, we will discover that field-antifield symmetry implies that
there are additional states starting at the second mass level which
contribute to the partition function with a minus sign. These additional
states should therefore be interpreted as fermions, which is surprising
since $\lambda^\a$ and $\w_\a$ are bosonic variables.

We will argue that these extra fermionic states are related to the
$b$ ghost in the pure spinor formalism, and come from functions which
are not globally defined on the space of the pure spinors.
As discussed in~\cite{Nekrasov:2005wg}, the constrained pure spinor
ghosts can be treated as a
$\beta\gamma$ system where one solves the pure spinor constraint locally in
terms of unconstrained worldsheet variables $(\b_i,\g^i)$ for
$i=1$ to 11. This solution in terms of unconstrained variables is
well-defined only when a certain component of $\lambda^\a$ is non-vanishing.
One can therefore patch together different solutions where the different
patches correspond to regions in the space of pure spinors
where different components of $\lambda^\a$
are assumed to be non-zero.

The gauge invariant polynomials are globally defined on all patches,
however, one can also consider functions which are only well-defined
on the overlap of two patches, on the overlap of three patches, etc.
When the function is defined on the overlap of $N$ patches, it is natural
to identify the state with a fermion/boson if $N$ is even/odd.
This can be understood if one converts from the patching
language of \v{C}ech cohomology to the differential form language of
Dolbeault cohomology. Using Dolbeault language, functions defined on
the overlap of $N$ patches are associated with $(N-1)$-forms which
have the standard fermionic/bosonic statistics for differential forms
when $N-1$ is odd/even.

The extra fermionic states which start to appear at the second mass level
will all be identified in Dolbeault language
with differential three-forms, and are therefore
fermionic. Furthermore, it will be argued that all these states are
related to a certain term in the composite operator for the $b$ ghost.

In the pure spinor formalism, the $b$ ghost satisfying $\{Q,b\}=T$
is a composite operator constructed from both the matter variables
$(x^\mu,\t^\a,p_\a)$ and ghost variables
$(\lambda^\a,\w_\a)$. This composite operator cannot be globally defined on
all patches, and in Dolbeault language, is described by the sum of
a zero-form, one-form, two-form and three-form. The three-form in the
$b$ ghost is independent of the matter variables $(x^{\mu},\t^\a,p_\a)$,
and will be identified with a fermionic scalar in the pure spinor
partition function at the
second mass level. At higher mass levels, the extra fermionic states
in the pure spinor partition function
can be similarly identified with products of this fermionic three-form with
polynomials of $\lambda^\a$ and $\w_\a$ (and their worldsheet derivatives).

In hindsight, the appearance of the $b$ ghost in the $(\lambda^\a,\w_\a)$
partition function is not surprising since any covariant description
of massive states is expected to include auxiliary spacetime
fields whose vertex operator involves the $b$ ghost. Nevertheless,
the manner in which the $b$ ghost appears in a partition function for
bosonic worldsheet variables is quite remarkable and suggests that
many important features of the $b$ ghost can be learned by studying
the pure spinor partition function.

\bigskip
The plan of this paper is as follows:
We begin in section~\ref{sec:revps} by reviewing the basics of the pure spinor formalism.
Due to the non-linear nature of the pure spinor constraint,
there is a subtlety in defining the pure spinor Hilbert space.
We shall recall two appropriate languages---the
\v{C}ech description and
its Dolbeault (or non-minimal) cousin---that can be used to address this subtlety,
and also introduce Chesterman's BRST method with an
infinite tower of ghosts-for-ghosts~\cite{Chesterman:2004xt}.
Following our discussion of the toy models in~\cite{Toymodels},
we then indicate how these descriptions are related.
This will serve as an introduction to the picture we are going to establish.

In section~\ref{sec:partPS}, the partition function of gauge invariant polynomials are computed
by explicitly constructing them at lower levels.
We point out that the space of gauge invariants is insufficient if one
requires field-antifield symmetry;
in particular, a fermionic state is found to be missing at level $2$,
which later will be identified as a term in the composite $b$ ghost.

Section~\ref{sec:PartPS} is devoted to the computation of the partition function
including the missing states found in section~\ref{sec:partPS}.
(The results are listed in appendix~\ref{app:characters}.)
We use two methods for the computation,
each with its advantages and disadvantages.
The first method utilizes Chesterman's BRST description of the pure spinor system~\cite{Chesterman:2004xt} involving ghosts-for-ghosts.
A nice feature of this method is that two important symmetries---%
field-antifield and ``$\ast$-conjugation'' symmetries---%
are (formally) manifest.
However, since this description
requires an infinite tower of ghosts-for-ghosts,
the expression for the partition function is not rigorously defined.
Nevertheless, we show that there is an unambiguous way to compute
the partition function level by level respecting the two symmetries.
The second method uses a fixed point formula
which generalizes the zero mode result of~\cite{Berkovits:2005hy}.
The formula includes the spin dependence of the states,
and the computation is fairly straightforward.
However, it misses some finite number of states that must be recovered
by imposing the two symmetries.

We then explain in section~\ref{sec:SymPart}
how the field-antifield and $\ast$-conjugation symmetries
can be understood from
the structure of pure spinor cohomologies.

In section~\ref{sec:lightcone} we relate the partition function
and the superstring spectrum.
After including the contribution from the matter variables,
we show that a simple twisting of the charges gives rise to the partition function
of lightcone fields and their antifields. Furthermore, we show
up to the fifth massive level
that the partition function thus obtained reproduces
the usual lightcone superstring spectrum (without the on-shell condition).

We conclude in section~\ref{sec:summary} and indicate some possible
applications of our findings.
Several appendices are included for convenience.
Some group theoretical formulas are collected in appendix~\ref{app:conventions},
and a list of partition functions can be found in appendix~\ref{app:characters}
Finally in appendix~\ref{app:reducibility} we present some details
of the reducibility analysis of the pure spinor constraint.

%%%%%%%%%%%%%%%%%%%%%%%%%%%%%%%%%%%%%%%%%%%%%%%%%%%%%%%%%%%%%%%%
\section{A brief review of the pure spinor formalism}
\label{sec:revps}

Let us begin by reviewing certain aspects of the pure spinor formalism
and indicating the results we are going to establish in the present paper.
This is not intended to be a complete overview of the formalism,
as we only cover issues which are relevant to the partition function
computation.
On the other hand, we will also include a summary of our results
obtained from the analysis of simple toy
models with quadratic constraints~\cite{Toymodels}.
The essential features of these simpler toy models
are very similar to those of the more complicated pure spinor model.

\subsection{Basics and a subtlety}
\label{sec:buzz}

The worldsheet variables of the pure spinor formalism
consist of the following three sectors:
\begin{align}
  x^{\mu},\quad (p_{\alpha},\theta^{\alpha}),\quad (\omega_{\alpha},\lambda^{\alpha}),\quad (\mu=0, \ldots ,  9,\;\alpha=1, \ldots , 16)\,.
\end{align}
(We restrict ourselves to the left-moving sector of closed strings,
or the open string.)
The first two sectors are the Green-Schwarz-Siegel variables
describing the string propagation in ten-dimensional superspace,
and they satisfy the usual free field operator product expansions~\cite{Siegel:1985xj}:
\begin{align}
 \label{eq:xpthetaOPE}
  x^{\mu}(z)x^{\nu}(w) &= -\eta^{\mu\nu}\log(z-w)\,,\quad p_{\alpha}(z)\theta^{\beta}(w) = {\delta_{\alpha}{}^{\beta} \over z-w} \,.
\end{align}
In addition to these ``matter'' sectors, there is a bosonic ``ghost'' sector
consisting of the pure spinor variable $\lambda^\a$
subject to quadratic constraints~\cite{Cartan}
\begin{align}
  \label{eq:PSconstraint1}
  \lambda^{\alpha}\gamma^{\mu}_{\alpha\beta}\lambda^{\beta} = 0 \quad (\mu=0, \ldots ,\, 9 )\,,
\end{align}
and its conjugate $\omega_{\alpha}$.

Physical states are defined by the cohomology of the ``physical BRST operator''\footnote{%
There are some arguments how this BRST operator arises
from gauge fixing a fermionic local symmetry of a Green-Schwarz-like classical action,
but in this paper we will not worry about the ``origin'' of pure spinors.
Readers interested in this issue are referred to~\cite{Berkovits:2004tw}\cite{Aisaka:2005vn}.}
\begin{align}
\begin{split}
\label{physb}
  Q &= \int\lambda^{\alpha}d_{\alpha}\,, \\ {\rm where ~~~}
  d_{\alpha} &= p_{\alpha} + (\gamma^{\mu}\theta)_{\alpha}\delx^{\mu} - {1\over2}(\gamma^{\mu}\theta)_{\alpha}(\theta\gamma_{\mu}\del\theta)\,.
\end{split}
\end{align}
$Q$ can be checked to be nilpotent using the free field operator products~(\ref{eq:xpthetaOPE})
and the pure spinor constraint~(\ref{eq:PSconstraint1}).
The massless vertex operator, for example, can be written in a manifestly
super-Poincar\'{e} invariant manner by coupling a spinor superfield to a pure spinor as
\begin{align}
  U &= \lambda^{\alpha}A_{\alpha}(x,\theta) \,.
\end{align}
Expanding in powers of $\theta$, one finds the (zero-momentum) vertex operators for the photon and photino
to be $(\lambda\gamma^{\mu}\theta)$ and $(\lambda\gamma^{\mu}\theta)(\gamma_{\mu}\theta)_{\alpha}$. A
similar construction has been
done for states at the first massive level~\cite{Berkovits:2002qx},
and by now there are various arguments that the cohomology of $Q$ reproduces the full superstring spectrum
in a covariant manner.

However, there is an important subtlety that has to be explained.
Namely, we have not yet specified the Hilbert space in which the cohomology of
$Q$ is computed.
Classically, because of the constraint~(\ref{eq:PSconstraint1}),
the conjugate $\omega_{\alpha}$ must appear in combinations
invariant under the ``gauge transformation'' generated by the constraint $\lambda\gamma^{\mu}\lambda=0$:
\begin{align}
  \delta_{\Lambda}\omega_{\alpha} = \Lambda^{\mu}(\gamma_{\mu}\lambda)_{\alpha}\,.
\end{align}
Examples of
such gauge invariants are the $\lambda$-charge\footnote{%
Although the charge measured by $J_{0}$ is often called the ``ghost number'',
we shall call it the ``$\lambda$-charge''
to avoid confusion with another ghost number which will be introduced later.}
current $J$,
the Lorentz current $N^{\mu\nu}$,
and the energy-momentum tensor $T$.
Their classical expressions are given by
\begin{align}
 \label{eq:clcurrents}
  J &= -\omega\lambda\,,\quad N^{\mu\nu}=-{1\over2}\omega\gamma^{\mu\nu}\lambda\,,\quad T = -\omega\del\lambda\,.
\end{align}
Quantum mechanically, however, since $\omega$ and $\lambda$ are {\em not} free fields,
it is not obvious how to define these composite operators.
One way is to parameterize $\lambda$ (and $\omega$) by genuine free fields.
Using the decomposition $U(5)\subset SO(10)$,
the pure spinor constraint~(\ref{eq:PSconstraint1}) implies the
$5$ conditions
\begin{align}
\label{defla}
  \lambda^{a} = {1\over8}(\lambda_+)^{-1}\epsilon^{abcde}\lambda_{bc}\lambda_{de}
\end{align}
where $\lambda^\a$ decomposes under $U(5)$ as
\begin{align}
  \lambda &= (\lambda_{+},\lambda_{ab},\lambda^{a}) =(\mathbf{1},\mathbf{10},\mathbf{\overline{5}}) \,.
\end{align}
So $(\omega,\lambda)$ can be parameterized by $11$ free $\beta\gamma$ pairs
which describe $(\lambda_+,\lambda_{ab})$ and their conjugate momenta.

However, one now runs into a subtlety concerning inverse powers
of $\lambda_+$ in the definition of  $\lambda^a$ in~(\ref{defla}). Recall that
inverse powers of
$\lambda$
are also required
to construct the composite ``reparameterization $b$-ghost''
that satisfies~\cite{Berkovits:2004px,Berkovits:2005bt}
\begin{align}
  \{Q,\, b\} = T \,.
\end{align}
Once inverse powers of $\lambda$ are allowed,
it naively appears that the cohomology of $Q$ becomes trivial due to the relation
\begin{align}
  \{Q,\, \lambda_{+}^{-1}\theta_{+} \} = 1 \,.
\end{align}
Of course, the expression ${\lambda}_{+}^{-1}{\theta}_{+}$ is not globally
well-defined on $X_{10}$, but so is the composite $b$-ghost.
Thus, one has to clarify which 
type of poles in $\lambda$ are allowed and which are not, what
global properties the allowed expressions should have and so on.
One of our aims in the present paper is to clarify this issue
by applying the general framework of curved $\beta\gamma$ systems~\cite{GCDO,Witten:2005px,Nekrasov:2005wg}
to pure spinors.
(For the mathematically better developed theory of $\beta\gamma$-systems on
superspaces of the form ${\Pi}TX$ or ${\Pi}T^{*}X$, see
e.g.~\cite{Malikov:1998dw, susybg};
for the treatment of instanton effects, see~\cite{instantons}.)

\subsection{Pure spinor sector as a curved $\beta\gamma$ system}

A standard way to construct a general curved $\beta\gamma$ system on a complex manifold $X$
is to employ a \v{C}ech description of $X$.
Namely, one starts with a set of free conformal field theories
taking values in the coordinate patches $\{U_A\}$ of $X$,
and tries to glue them together%
~\cite{GCDO, Witten:2005px, Nekrasov:2005wg}.
The field content of each conformal field theory is described by
the (holomorphic) coordinates of a patch $u^{a}$
and its conjugate $v_{a}$ satisfying the
free field operator product expansion
\begin{align}
  \label{eq:localfreeOPE}
  u^{a}(z)v_{b}(w) &= {{\delta^{a}}_b \over z-w} \,.
\end{align}
Unlike conventional sigma models on complex manifolds,
one need not introduce antiholomorphic coordinates.

Not all manifolds $X$, however, lead to a consistent worldsheet theory.
A basic requirement is that one must be able to consistently glue
the operator products~(\ref{eq:localfreeOPE}) on overlaps.
Gluing on double overlaps $U_{A}\cap U_{B}$ can always be
done (though they are not quite unique),
but the gluing on $U_{A}\cap U_{B}$, $U_{A}\cap U_{C}$ and $U_{B}\cap U_{C}$
must be consistent on the triple overlap $U_{A}\cap U_{B}\cap U_{C}$ (cocycle condition).
In order that there is no topological obstruction for this,
the first Pontryagin class $p_{1}(X)$ must be vanishing.
Analogous obstructions can be present for the global existence of worldsheet currents
that generate the symmetries of $X$ (``equivariant version'' of $p_{1}(X)$).
Also, to be able to define the energy-momentum tensor $T$ globally
(i.e. to have a {\em conformal}\/ field theory),
$X$ must possess a nowhere vanishing holomorphic top-form
and hence the first Chern class $c_{1}(X)$ must also be vanishing.

In the case of pure spinors all these obstructions turn out to be absent~\cite{Nekrasov:2005wg}.
The target space is basically the space of $SO(10)$ pure spinors, with the origin removed:
\begin{align}
  X_{10} &= \{ \lambda^{\alpha} \;|\: 
  \lambda^{\alpha}\gamma^{\mu}_{\alpha\beta}\lambda^{\beta} = 0 \, , {\lambda} \neq 0 \} \,,
\end{align}
which is a complex cone over a compact projective space $\calX_{10}$.
It is well known that $\calX_{10}$ is the homogeneous space
\begin{align}
  \calX_{10} &= SO(10)/U(5) \,,
\end{align}
and has ten (complex) dimensions. The origin ${\lambda}=0$ is removed from the space
of all solutions to the equations ${\lambda}{\gamma}^{\mu}{\lambda}=0$ in order to 
meet the general criteria above, $p_1=c_1=0$ etc.
That is, $X_{10}$ is regarded as a $\mathbb{C}^{\ast}$-bundle over the base $\calX_{10}$ (thus we
are dealing with the $\beta\gamma$-system which is not covered by the general analysis of~\cite{CDO}).
With this removal of the origin understood, $X_{10}$ can be covered by 16 patches $\{U_{A}\}$ (${A=1, \ldots , 16}$)
where in each patch at least one component of $\lambda$ (which we denote $\lambda^{A}$) is non-vanishing.
Very explicit formulas for the gluing of operator products, symmetry currents $J$ and $N^{\mu\nu}$,
and the energy-momentum tensor $T$ can be found in~\cite{Nekrasov:2005wg}.

\bigskip
Given a space $X$ on which the curved $\beta\gamma$ system can be consistently defined,
the space of observables, or simply the Hilbert space of the model, is defined as
the  cohomology of the difference operator $\delcech$, also known as \v{C}ech operator.
Let us recall that a \v{C}ech $n$-cochain $\psi=(\psi^{A_{0}A_{1}\cdots A_{n}})$
refers to the data assigned to every $n$th overlaps, $U_{A_{0}A_{1}\cdots A_{n}}=U_{A_{0}}\cap \cdots \cap U_{A_{n}}$,
and $\delcech$ sends an $n$-cochain to an $(n+1)$-cochain:
\begin{align}
\label{defcech}
\begin{split}
(\delcech \psi)^{A_{0}\cdots A_{n+1}} &=
 \sum_{i=0}^{n}(-1)^{i} \psi^{A_{0}\cdots \Check{A}_{i}\cdots A_{n+1}}\,.
\end{split}
\end{align}
The $n$th \v{C}ech cohomology $H^{n}(\delcech)$ is defined
as the space of $\delcech$-closed $n$-cochains ($n$-cocycles)
modulo $\delcech$-exact elements ($n$-coboundaries).
In particular, the zeroth cohomology $H^{0}(\delcech)$ is simply
the space of ``gauge invariant'' operators defined globally on $X$.
But experience with models with a simple quadratic constraint of the form $\lambda^{i}\lambda^{i}=0$ suggests
that higher cohomologies are important as well~\cite{Toymodels}.

With respect to the higher cohomologies,
there are several lessons to be learned from the analysis in~\cite{Toymodels}.
First, for the models considered in~\cite{Toymodels},
only the zeroth cohomology $H^0(\delcech)$
and the first cohomology $H^{1}(\delcech)$ were non-empty.
Second, there was a one-to-one mapping between $H^{0}(\delcech)$ and $H^{1}(\delcech)$.
Finally $H^{1}(\delcech)$ was essential for
having ``field-antifield symmetry'' after coupling the system to the fermions $(p_{i},\theta^{i})$.
(In the pure spinor formalism, ``field-antifield symmetry'' literally refers to the
symmetry between spacetime fields and antifields,
and is essential for the consistent definition of scattering amplitudes.)

Somewhat surprisingly, the situation is almost identical for the pure spinor system,
except that $H^1(\delcech)$ is replaced by the third cohomology $H^{3}(\delcech)$.
More precisely, only the zeroth cohomology and the third cohomology will
contribute to the partition function, and there will be a conjectured
one-to-one mapping between states in $H^0(\delcech)$ and in $H^3(\delcech)$.
In the pure spinor formalism, an additional reason why $H^{3}(\delcech)$
is important is that a nontrivial element in
$H^{3}(\delcech)$ is essential for the construction
of the composite reparameterization $b$-ghost.

\subsection{Dolbeault or non-minimal description}

The description of the curved $\beta\gamma$ system in the previous subsection
was done using the \v{C}ech language
by patching together a collection of free conformal field theories~\cite{Malikov:1998dw}.
There is a closely related formulation which uses the Dolbeault language.
The two are related in the same manner
as the standard \v{C}ech and Dolbeault cohomologies
of a complex manifold are related.
In the
\v{C}ech description, only the holomorphic local coordinates $u_{a}$ of $X$ were used,
but the Dolbeault description utilizes the antiholomorphic variable $\overline{u}^{\overline{a}}$ as well.
This allows the construction
of a partition of unity on $X$ and,
by considering the cohomology of an extension of the Dolbeault operator $\deldol$,
one can deal exclusively with globally defined objects~\cite{susybg,Witten:2005px,Nekrasov:2005wg}.

In the pure spinor formalism, the so-called non-minimal formulation
corresponds to this Dolbeault formulation~\cite{Berkovits:2005bt}\cite{Berkovits:2006vi}.
There, one introduces another set of pure spinor variables
$\overline{\lambda}_{\alpha}$
and its (target space) differential $r_{\alpha}=\mathrm{d}\overline{\lambda}_{\alpha}$ which are constrained to satisfy
\begin{align}
  \overline{\lambda}_{\alpha}\gamma^{\mu\alpha\beta}\overline{\lambda}_{\beta} &= 0\,,\quad
  \overline{\lambda}_{\alpha}\gamma^{\mu\alpha\beta}{r}_{\beta} = 0\,.
\end{align}
The conjugate momenta for the non-minimal fields are denoted by
$\overline{\omega}^{\alpha}$ and $s^{\alpha}$, and they must appear in combinations which are
invariant under the non-minimal gauge transformations
\begin{align}
\label{eq:nmgaugetransf}
\begin{split}
  \delta_{\overline{\Lambda}}\overline{\omega}^{\alpha} &= \overline{\Lambda}^{\mu}(\gamma_{\mu}\overline{\lambda})^{\alpha}\,,\quad
  \delta_{\Psi}\overline{\omega}^{\alpha} = \Psi^{\mu}(\gamma_{\mu}r)^{\alpha}\,, \\
  \delta_{\Psi}s^{\alpha} &= \Psi^{\mu}(\gamma_{\mu}\overline{\lambda})^{\alpha}\,,
\end{split}
\end{align}
with $\overline{\Lambda}_{\mu}$ and $\Psi_{\mu}$ being bosonic and fermionic gauge parameters.

The Dolbeault operator $\deldol$ can be defined
as a natural extension of the Dolbeault differential in complex geometry:
\begin{align}
\label{defdol}
  \deldol = -r_{\alpha}\overline{\omega}^{\alpha} \sim \mathrm{d}\overline{\lambda}_{\alpha}{\del \over \del \overline{\lambda}_{\alpha}}\,.
\end{align}
Note that $\deldol$ is gauge invariant under~(\ref{eq:nmgaugetransf}).
If one wishes to be more rigorous,
the expression for $\deldol$ should be understood in terms of its local expressions
that are consistently glued.
Also, note that only the zero-modes for the non-minimal sector
are relevant for the $\deldol$-cohomology due to the relation
\begin{align}
  \deldol(s\del\overline{\lambda}) = \overline{\omega}\del\overline{\lambda} + s\del r = -T_{\text{non-min}} \,.
\end{align}
Whenever there is a $\deldol$-closed operator $F$
with positive weight $h$ carried by the non-minimal sector,
it can be written as $\deldol$ of itself
multiplied by the zero-mode of $s\del\overline{\lambda}$:
\begin{align}
  -{1\over h}\deldol\bigl( (s\del\overline{\lambda})_{0}F\bigr) = F \,.
\end{align}

\bigskip
The minimal (\v{C}ech) and non-minimal (Dolbeault) formulations
can be related by imitating the argument that establishes the usual \v{C}ech-Dolbeault isomorphism.
That is, the cohomologies of $\delcech$ and $\deldol$ are related
using the partition of unity $\{\rho_{A}\}$ ``subordinate to'' the coordinate patches $\{U_{A}\}$:
\begin{align}
\begin{split}
  \rho_{A} &= {\overline{\lambda}_{A}\lambda^{A} \over \lambda\overline{\lambda}}\quad\to\quad
  \text{$\sum_{A}\rho_{A} = 1$ and $\rho_{A}=0$ outside $U_{A}$} \,, \\
  \targetdbar \rho_{A} &=
  \deldol(\rho_{A}) = {(\lambda\overline{\lambda})r_{A}\lambda^{A} - (\lambda r)\overline{\lambda}_{A}\lambda^{A}\over (\lambda\overline{\lambda})^{2}} \,.
\end{split}
\end{align}
(Here and hereafter, Einstein summation convention does not apply for the index $A$;
when needed, we will always write the summation over $A$ explicitly.)
A \v{C}ech $n$-cochain 
$\Check{\psi}=(\psi^{A_{0}\cdots A_{n}})$ is described in the 
Dolbeault language
by an $n$-form
\begin{align}
  \Bar{\psi} = {1\over (n+1)!}\sum_{A_{0},\cdots, \, A_{n}}   \psi^{A_{0}\cdots A_{n}}
  \rho_{A_{0}}\targetdbar \rho_{A_{1}}\wedge\cdots\wedge \targetdbar \rho_{A_{n}} \,.
\end{align}
Since $\Check{\psi}$ is holomorphic (i.e. $\deldol\psi^{A_{0}\cdots{A_{n}}}=0$),
the usual argument relating the \v{C}ech and Dolbeault cohomologies
can be applied
(provided one uses a good cover so that $\deldol$-cohomology is locally trivial).

\subsection{Cohomology of the pure spinor superstring}

In order to include the contribution of states that are not globally
defined on the space of pure spinors, the physical BRST operator $Q=\int
\lambda^\a d_\a$ of (\ref{physb}) should be modified either to
\begin{align}
  \Check{Q} = Q + \delcech\quad \text{or}\quad  \overline{Q} = Q + \deldol\,
\end{align}
where $\delcech$ and $\deldol$ are defined in (\ref{defcech}) and (\ref{defdol}).

The space on which $Q+\delta$ (where $\delta$ is either $\delcech$ or $\deldol$)
acts naturally has two gradings,
one for $\delta$ (which will be called ghost number)
and another for $Q$ (which will be called $\lambda$-charge).
The cohomology is thus graded by the sum of these two charges,
\begin{align*}
\xymatrix{
 & \ar@{->}[d] & \ar@{->}[d]  & \ar@{.}[ld]|<{k=m+n+1} \\
{}\ar@{->}[r] & \calF^{m,n} \ar@{->}[r]^{\delta}\ar@{->}[d]^{Q} & \calF^{m,n+1} \ar@{->}[r]\ar@{->}[d]^{Q} {\ar@{.}[ld]} & \\
{}\ar@{->}[r] & \calF^{m,n+1} \ar@{->}[r]^{\delta}\ar@{->}[d] {\ar@{.}[ld]}& \calF^{m+1,n+1}\ar@{->}[r]\ar@{->}[d] &\\
 & &  &}
\end{align*}
where a cohomology element with degree $k$ takes the form
\begin{align}
\begin{split}
  \psi &= \sum_{m+n=k}\psi_{m,n} \,,\\
  &\psi_{m,n}\in \calF^{m,n}=\calH_{\text{ps}}^{m,n}\otimes \calH_{x,p,\theta}\;\;(\text{ghost number $m$, $\lambda$-charge $n$}) \,.
\end{split}
\end{align}
The ghost number corresponds to the chain degree in \v{C}ech language
and to the form degree (measured by $J_{rs}=-rs$) in Dolbeault language.
For both cases, the $\lambda$-charge is measured by $J_{\omega\lambda}=-\omega\lambda$.
Hence, the summand $\psi_{m,n}$ in each descriptions are schematically,
\begin{align}
\text{\v{C}ech}\colon&\quad  \psi_{m,n} =  (\psi^{A_{0}\cdots A_{m}}_{n}) \,,\quad
\text{Dolbeault}\colon\quad  \psi_{m,n} = (r)^{m}\psi_{n} \,.
\end{align}
An important point is that a cohomology element
in general consists of several pieces with {\em different} $\delta$-degrees.
Nevertheless, as we shall argue momentarily, the cohomology of $\delta$
(i.e. \v{C}ech or Dolbeault cohomologies)
plays a central role in studying $(Q+\delta)$-cohomology,
and we shall spend considerable time computing those cohomologies
in the forthcoming sections.

\bigskip
The conditions for an operator $\psi$ to be
in the $(Q+\delta)$-cohomology is as follows.
For $\psi$ to be $(Q+\delta)$-closed, it must satisfy the master equation
\begin{align}
(Q+\delta)\psi=0\quad\Leftrightarrow\quad
\left\{
\begin{array}{rcl}
  Q\psi_{p,k-p}&=&0\,,  \\
  Q\psi_{p+1,k-1} + \delta\psi_{p,k-p}&=&0 \,,\\
    & \vdots& \\
  Q\psi_{q,k-q} + \delta\psi_{q-1,k-1}&=&0 \,,\\
 \delta\psi_{q,k-q}&=&0 \,,
\end{array}
\right.
\end{align}
for some $(p,q)$, or, more pictorially,
\begin{align*}
 \xymatrix@!=3ex{
&  \psi_{p,k-p} \ar@{->}[ld]|{Q} \ar@{->}[rd]|{\delta} & { + }
&  \psi_{p+1,k-1} \ar@{->}[ld]|{Q} \ar@{->}[rd]|{\delta} & { +\;\;\cdots } & { + }
&  \psi_{q,k-q} \ar@{->}[ld]|{Q} \ar@{->}[rd]|{\delta}  \\
  0 &  & 0 &  & 0 & 0 &  & 0}
\end{align*}
In particular, the ``head'' element $\psi_{p,k-p}$ is $Q$-closed
and the ``tail'' element $\psi_{q,k-q}$ is $\delta$-closed.
For $\psi$ to represent a non-trivial cohomology
it must not be $(Q+\delta)$-exact.
Then, since $Q$ and $\delta$ commute,
one can without loss of generality
assume that the head $\psi_{p,k-p}$ is $Q$-non-exact
and the tail $\psi_{q,k-q}$ is $\delta$-non-exact.
Since $\delta$ does not act on the physical sector $(x,p,\theta)$,
the latter implies that the tail is an element of the $\delta$-cohomology
(multiplied by some function of $(x,p,\theta)$).

So when studying the $(Q+\delta)$-cohomology,
one can simply restrict the tail element
to be in the $\delta$-cohomology.
More specifically, when analyzing an exactness relation $\psi=(Q+\delta)\phi$,
it can be assumed that $\phi$ has a $\delta$-closed tail which is  ``longer'' than $\psi$:
\begin{align*}
\xymatrix{
&\phi_{p,k-p-1}\ar@{->}[ld]|{Q}\ar@{}[r]|{\cdots}
  &\phi_{q-1,k-q}\ar@{->}[rd]|{\delta} && {\phi_{q,k-q-1}}\ar@{->}[ld]|{Q}
   & \ar@{}[l]|{\cdots}{\phi_{q',k-q'-1}}\ar@{->}[r]^-{\delta} & 0\\
*[r]{\psi_{p,k-q}}&\ar@{}[r]|{\cdots}& & {\psi_{q,k-q}}\ar@{->}[r]_{\delta} & 0 & &
}
\end{align*}

\bigskip
When $\psi$ is in $(Q+\delta)$-cohomology, it can happen that
both the head and tail of $\psi$ carry ghost number $0$.
In this case, $\psi$ is in the cohomology of $Q$ computed in the
space of globally defined operators, or simply:
\begin{align}
  \delta\psi=Q\psi=0 \,.
\end{align}
For example, the super-Maxwell vertex operator $\lambda^{\alpha}A_{\alpha}(x,\theta)$ is of this type.
However, the $(Q+\delta)$-cohomology can also be affected by
the higher cohomologies of $\delta$.
An important example of this phenomenon
is the relation between the energy momentum tensor $T$
and the $b$-ghost in the pure spinor formalism.
In this case, $b$ is an object with its tail in the third cohomology $H^{3}(\delta)$
and satisfies $(Q+\delta)b=T$:
\begin{align*}
\xymatrix@=3ex{
    & b_{0}\ar@{->}[ld]|{Q}\ar@{->}[rd]|{\delta}
    && b_{1}\ar@{->}[ld]|{Q}\ar@{->}[rd]|{\delta}
    && b_{2}\ar@{->}[ld]|{Q}\ar@{->}[rd]|{\delta}
    && b_{3}\ar@{->}[ld]|{Q}\ar@{->}[rd]|{\delta} \\
  T &  &  0 && 0 && 0 && 0
}
\end{align*}

\paragraph{Non-triviality of the cohomology}

In the above example involving the $b$ ghost, it is crucial
that $b_3$ is a nontrivial element in the $\delta$-cohomology.
In fact,
we shall argue later that $\delta$-cohomology is non-empty
only at ghost numbers $0$ and~$3$,
i.e. $H^{n}(\delta)=0$ when $n\ne0,3$.

As described earlier,
we do not wish to have the ``inverse'' of $(Q+\delta)$ in the Hilbert space
since such an operator would trivialize the $(Q+\delta)$ cohomology.
The troublesome operator satisfying $\{Q+\delta,\xi\}=1$ is
\begin{align}
\xi &= \Bigl({ \theta^{A} \over \lambda^{A} }\Bigr)
+ \Bigl({ \theta^{A_{1}}\theta^{A_{2}} \over \lambda^{A_{1}}\lambda^{A_{2}} }\Bigr)
+ \cdots + \Bigl({ \theta^{A_{1}}\cdots\theta^{A_{16}}  \over \lambda^{A_{1}}\cdots \lambda^{A_{16}}  }\Bigr)
\sim {\theta_{+}\over \lambda_{+}}\,,
\end{align}
in the minimal formalism, and is
\begin{align}
\xi &= {\overline{\lambda}\theta \over \lambda\overline{\lambda} + r\theta}\,,
\end{align}
in the non-minimal formalism.
As described in~\cite{Berkovits:2005bt}\cite{Berkovits:2006vi}, this operator
can be excluded by restricting the order of divergence in $(\lambda\overline{\lambda})^{-n}$
(or more precisely the ghost number $n$) to be less than $n=11$,
which is also needed for defining the path integral over $\lambda$ and $\overline{\lambda}$ zero-modes
as $(\lambda\overline{\lambda})\to0$.
One possible problem with this restriction is that, since the
$b$-ghost diverges as fast as $(\lambda\overline{\lambda})^{-3}$,
one needs to introduce a regularization
in computing higher loop amplitudes that require more than $3$ $b$'s.
A regularization procedure was explained in~\cite{Berkovits:2006vi},
but it was complicated to use in explicit calculations.

As mentioned above, the $\delta$-cohomology will be argued to be
empty
except for ghost numbers $0$ and $3$.
This implies that the worrisome divergence coming from fusing multiple $b$'s
are in fact BRST trivial and can be simply discarded,
provided there is no divergence arising at the boundary of the moduli space.
In other words, the trivial cohomology of $H^n(\delta)$ for $n>3$ allows
one to consistently remove operators which diverge faster than
$(\lambda\overline{\lambda})^{-3}$.

We will begin our analysis of the cohomology of $\delta$ in section~\ref{sec:partPS}.
But before entering into the details,
let us explain another method for computing the $\delta$-cohomology
and its relation with the \v{C}ech/Dolbeault cohomologies described earlier.
This alternative method utilizes Chesterman's ghosts-for-ghosts
introduced in his BRST approach to
the pure spinor constraint~\cite{Chesterman:2004xt}.

\subsection{Ghost-for-ghost versus \v{C}ech/Dolbeault descriptions}
\label{subsec:BRSTCechDolbeault1}

For a curved $\beta\gamma$ system defined by homogeneous constraints
$G^{I}(\lambda)=0$ ($I=1, \ldots ,  N$),
an alternative to the \v{C}ech/Dolbeault formulation
would be to apply the BRST formalism
to describe the constraint.
Analysis of simple models~\cite{Toymodels}
suggests that classically both descriptions lead to
the same Hilbert space (phase space)
including the operators in higher cohomologies.
(See~\cite{Grassi:2006wh,Grassi:2007va} for a comparison of
ordinary gauge invariant operators.)
Although the two descriptions differ in general
quantum mechanically, our partition function
$\Tr[(-1)^{F}\cdots]$ is insensitive to the discrepancy.

In the BRST framework,
ghost pairs $(b_{I},c^{I})$
(and ghosts-for-ghosts if necessary)
are introduced to impose the constraint indirectly,
and the Hilbert space is defined as the
cohomology of the BRST operator
\begin{align}
  \label{eq:naiveBRSTop}
  D &= \int b_{I}G^{I} \,.
\end{align}
The ghost numbers are assigned $g(b_I,c^I;\omega,\lambda)=(1,-1;0,0)$ so that $g(D)=1$
and the cohomology $H^{n}(D)$ is graded accordingly.

A very nice feature of this ghost description of the constraints
is that one can describe the system entirely in terms of free fields.
However, a difficulty arises when applying it to the pure spinor system
since the constraints are infinitely reducible.

A set of constraints is called reducible
if not all the constraints are independent,
i.e. if there exist non-trivial relations among them.
Depending on how one chooses to represent the reducibility relations,
there can be relations-for-relations.
(This often happens if one wishes to keep the symmetries of the system manifest.)
For the pure spinor system, the constraint is
infinitely reducible meaning there is an infinite chain of
relations-for-relations.
Thus, infinite generations of ghosts have to be introduced
and the BRST operator $D$~(\ref{eq:naiveBRSTop}) will have an
infinite number of terms.
It can be cumbersome in practice but
a systematic procedure for handling reducible constraints is known, and in fact,
the ghost-for-ghost
method is quite useful
for computing the full partition function of pure spinors.

Note that, when applied to the pure spinor case,
the operator $D$ is used to implement 
the pure spinor constraint via its cohomology,
and is unrelated to the physical BRST operator $Q=\int\lambda^{\alpha}d_{\alpha}$.
However, $D$ can be combined with $Q$ to define a single nilpotent operator
of the form $\Hat{Q}=D+Q+\cdots$,
where the ellipses  can be fixed by requiring nilpotency
and are essentially unique.
Then, the so-called ``cohomological perturbation theory''
(formally) assures that the constrained cohomology of $Q$ ($D=0$) is
equivalent to the unconstrained cohomology of $\hat{Q}$~\cite{Henneaux:1992ig}.
We will call this auxiliary BRST operator $D$ in the ghost-for-ghost method
as ``mini-BRST operator'',
and sometimes refer to its cohomology as ``mini-BRST cohomology''.

The mini-BRST operator $D$ was first introduced by Chesterman
in an attempt to construct
the ghost-extended physical BRST operator $\Hat{Q}=D+Q+\cdots$.%
\footnote{The idea of working with unconstrained $\lambda^{\alpha}$ variables covariantly was
originally developed by Grassi, Policastro, Porrati and van Nieuwenhuizen
in~\cite{Grassi:2001ug,Grassi:2002aa}
but with a truncation on the mini-BRST operator $D$.
Unfortunately, due to the truncation,
it appeared difficult to assure that $\Hat{Q}$ reproduces
the superstring spectrum.
References~\cite{Berkovits:2000nn} and~\cite{Aisaka:2002sd}
discusses the use of $SO(8)$ and $U(5)$ version of the mini-BRST operator,
respectively.}
The idea of having a single physical BRST operator
acting on a totally unconstrained space is attractive.
But as $D$ already contains infinite number of terms, actual construction
of $\Hat{Q}$ is not feasible.
Thus, although we study the cohomology of the mini-BRST operator $D$,
we will not attempt to study the cohomology
of the ghost-extended physical BRST operator $\Hat{Q}$ directly
(except in the last section~\ref{sec:lightcone}
where we make use of an $SO(8)$ version of $\Hat{Q}$ to derive
the lightcone spectrum).

\bigskip
One of our main goals is to establish a classical equivalence
between the ghost-for-ghost and \v{C}ech-Dolbeault descriptions.
Although some portions are left conjectural,
we claim that the equivalence can be established
using exactly the same arguments that
were presented for the simpler toy models~\cite{Toymodels}.
That is, the cohomologies in question can be related
by defining a ``non-minimal'' version of Chesterman's mini-BRST operator
\begin{align}
\overline D = D + \deldol
\end{align}
where $D$ is the usual mini-BRST operator of the ghost-for-ghost method,
and $\deldol=-r_\a \overline{\w}^\a$
is the Dolbeault operator constructed from non-minimal variables.

At first sight, it seems that the non-minimal variables added here
should be {\em unconstrained} to ensure the cohomology to be kept intact.
We however note that whether
$\overline{\lambda}_{\alpha}$ satisfies $\overline{\lambda}\gamma^{\mu}\overline{\lambda}=0$
or not is irrelevant as long as the cohomology is concerned.
In both cases, the non-minimal momenta $\overline{\omega}^{\alpha}$
and $s_{\alpha}$ cannot contribute to the cohomology,
so one can switch between the two viewpoints by simply
forgetting/imposing the non-minimal constraint.

In section~\ref{sec:SymPart}, it will be argued
that the following four cohomologies are classically equivalent:
\begin{enumerate}
\item Minimal mini-BRST (ghost-for-ghost): cohomology of $D$
\item Non-minimal mini-BRST (ghost-for-ghost): cohomology of $D+\deldol$
\item Dolbeault cohomology $\deldol$ (of gauge invariant operators)
\item \v{C}ech cohomology of $\delcech$ (of gauge invariant operators).
\end{enumerate}
Let us clarify the meaning of ``gauge invariance'' in this picture.
Suppose we momentarily forget the non-zero modes
and think about a point particle moving in the space of pure spinors $X_{10}$.
When one speaks of the gauge transformation $\delta_{\Lambda}\omega_{\alpha}=(\fslash{\Lambda}\lambda)_{\alpha}$,
it is implicitly assumed that the phase space $T^{\ast}X_{10}$
is embedded in a Euclidean space $T^{\ast}\mathbb{C}^{16}$.
Then, the gauge transformation generates a motion vertical to $T^{\ast}X_{10}$,
and the gauge invariance of an object simply means that it is living inside $T^{\ast}X_{10}$.
Now, in the curved $\beta\gamma$ language of the
\v{C}ech/Dolbeault description,
$T^{\ast}X_{10}$ is treated intrinsically and everything is manifestly gauge invariant.
So there is really no way to construct a ``gauge {\em non}-invariant object''
by using the local coordinates on the cotangent space.

However, in the ghost-for-ghost description,
$(\omega,\lambda)$ are promoted to genuine unconstrained
free fields so that
$X_{10}$ is naturally embedded in the flat space $\mathbb{C}^{16}$.
It is then sometimes convenient to use $(\omega,\lambda)$ instead of
their local parameterization to denote the operators.
But since not all expressions that can be written with $(\omega,\lambda)$ are
in $T^{\ast}X_{10}$, one needs to know when this notation makes sense.
The notion of gauge invariance does just this.
Note that, at least classically, gauge invariant operators such as
$J=-\omega\lambda$ and $N_{\mu\nu}=-(1/2)(\omega\gamma_{\mu\nu}\lambda)$
can be always translated to the intrinsic curved $\beta\gamma$ language.

To relate the four cohomologies listed above, one can follow the steps
$(a)-(d)$ in the diagram:
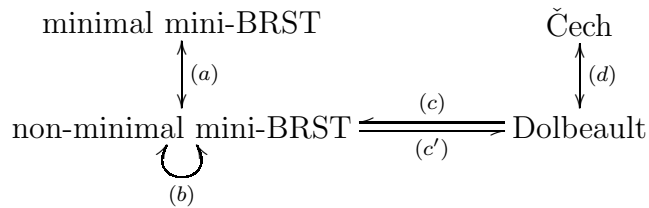
\begin{align*}
\xymatrix{
{ \txt{minimal mini-BRST}\ar@{<->}[d]^-{(a)}  } && { \txt{\v{C}ech}\ar@{<->}[d]^-{(d)} }\\
{ \txt{non-minimal mini-BRST}\ar@^{<-}@<.3ex>[rr]^-{(c)}\ar@_{->}@<-.3ex>[rr]_-{(c')}\ar@{<->}@(dl,dr)_-{(b)}  } && {\txt{Dolbeault}} \\
  &&
}
\end{align*}
\begin{description}
\item[\quad$(a)$] Adding/removing (unconstrained) non-minimal quartet under $\deldol=-r\overline{\omega}$
\item[\quad$(b)$] Different choice of cohomology representatives
\item[\quad$(c)$] Embedding to ``extrinsic'' space of free fields
\item[\quad$(c')$] Restriction to ``intrinsic'' (or gauge invariant) operators on $X_{10}$
\item[\quad$(d)$] Standard \v{C}ech-Dolbeault mapping (partition of unity)
\end{description}
We will come back to the explanation of this diagram in section~\ref{sec:SymPart},
but in short, steps $(a)$ and $(c)$ can be used to
embed the minimal mini-BRST and Dolbeault cohomologies in the
non-minimal mini-BRST cohomology,
and both are then simply different choices of the
cohomology representatives (step $(b)$).
The biggest conceptual step is step $(c)$,
where a connection between free fields (ghost-for-ghost)
and constrained fields (\v{C}ech/Dolbeault) has to be made.
For the pure spinor model, an added technical difficulty arises
due to the infinite number of ghosts on the ghost-for-ghost side.
Nevertheless, one can at least formally state the mapping between the
two regimes,
and establish the equivalence of the cohomologies at each ghost number
(i.e. not just in the ghost number $0$ sector)~\cite{Toymodels}:
\begin{align}
  H^{n}(D) = H^{n}(\delta)\quad (\text{$\delta=\delcech$ or $\deldol$}) \,.
\end{align}
Moreover, as mentioned above, we shall argue that only $H^{0}$ and $H^{3}$ are non-empty,
and that there is a one-to-one mapping between the two.

One of the basic tools for studying cohomologies is to compute their partition function (or character).
In particular, this is convenient for exposing some discrete symmetries
such as the mapping between $H^{0}$ and $H^{3}$.
So we will spend the next several sections
computing the partition functions of various cohomologies.
Eventually, we shall argue that only $H^0$ and $H^{3}$ are non-empty
and that they together form a space that precisely
reproduces the correct superstring spectrum.

%%%%%%%%%%%%%%%%%%%%%%%%%%%%%%%%%%%%%%%%%%%%%%%%%%%%%%%%%%%%%%%%
\section{Naive partition function of pure spinors and missing states}
\label{sec:partPS}

In the last subsection, we explained various cohomologies
that might be used to describe the operators of the pure spinor sector.
Eventually, it will be argued that they are classically all equivalent.
In particular, all have structures that can be summarized by
two discrete symmetries of their partition functions,
the ``field-antifield symmetry'' and the ``$\ast$-conjugation symmetry''.
The former is essential for being able to define the spacetime amplitudes,
and the latter is responsible for the symmetry between
gauge invariant operators (i.e. of the zeroth cohomology $H^{0}$)
and the operators that are not globally defined on the pure spinor space
(which turn out to live only in the third cohomology $H^{3}$).
Also, it is only when the contribution from the latter is taken into account
that the total Hilbert space exhibits the field-antifield symmetry.

To explain what we have just stated, we here compute the partition function
of the globally defined gauge invariant operators by explicitly constructing them
at lower Virasoro levels.
It turns out that, starting from level $2$,
the space $H^{0}$ by itself lacks some operators for having the field-antifield symmetry.
The missing states turn out to be {\em fermionic} and hence are
naturally described by higher cohomologies of odd degrees.

\subsection{Preliminaries}

\subsubsection{Definition of the partition function}
\label{subsec:defpartfn}

The characters of the states we wish to keep track of
are their statistics, weights (Virasoro levels), $t$-charge (measured by $J_t=-\omega\lambda-p\theta$)
and the Lorentz spin.
The Lorentz spin of a state can be labeled by five integers
which we denote by
\begin{align}
\begin{split}
\mu&=(a_1 a_2 a_3 a_4 a_5)\quad\text{Dynkin basis} \,, \\
 &= {1\over2}[\mu_1\mu_2\mu_3\mu_4\mu_5]\quad\text{``five sign'' basis} \,.
\end{split}
\end{align}
Introducing formal variables $(q,t,\vec{\sigma})$ for each quantum numbers,
we define the partition function (character) as
\begin{align}
\begin{split}
  Z(q,t,\vec{\sigma}) &= \Tr (-1)^{F}q^{L_{0}}t^{J_{0}}\mathe^{\mu\cdot \sigma} \\
  &= \sum_{h\ge0}Z_{h}(t,\vec{\sigma})q^{h} \,.
\end{split}
\end{align}
The trace is taken over the various cohomologies explained above,
and we will show that all lead to the same result.
Characters of the basic operators $\omega$ and $\lambda$ are
\begin{align}
\begin{split}
h(\omega,\lambda) &= (1,0)\,,\quad t(\omega,\lambda) = (-1,1)\,, \\
\mu(\omega) &= \mathe^{(00010)} = \mathe^{{1\over2}(\pm \sigma_1 \pm \sigma_2\pm\sigma_3\pm\sigma_4\pm\sigma_5)} \quad\text{\small(odd \# of $-$'s)}\,, \\
\mu(\lambda) &= \mathe^{(00001)} = \mathe^{{1\over2}(\pm \sigma_1 \pm \sigma_2\pm\sigma_3\pm\sigma_4\pm\sigma_5)} \quad\text{\small(even \# of $-$'s)}\,.
\end{split}
\end{align}
The relation between the Dynkin basis and
the ``five sign basis'' can be found in appendix~\ref{app:Dynkin}.

Sometimes, it is convenient to ignore the spin characters
and concentrate on the dimensions of the Hilbert space
\begin{align}
\begin{split}
  Z(q,t) &= \Tr (-1)^{F}q^{L_{0}}t^{J_{0}} \\
    &= \sum_{h\ge0}Z_{h}(t)q^{h} \\
    &= \sum_{h\ge0, n}N_{h,n}q^{h}t^{n} \,.
  \end{split}
\end{align}
One might wish to keep track of the ghost number (or $g$-charge),
but the computation of $Z(q,t)$ is considerably easier than the
computation of $Z(q,t,g)$ as we explain shortly.

A list of partition functions at lower levels can be found
in appendix~\ref{app:characters}.

\subsubsection{Cohomology via partition function}

In section~\ref{sec:lightcone}, we will relate the partition function of pure spinors
to that of the cohomology of the physical BRST operator $Q = \int\lambda^{\alpha}d_{\alpha}$.
Let us explain the basic idea behind this, which is also useful for the
computation of the partition function itself.

Let $\cal O$ be a fermionic nilpotent operator that commutes with $L_{0}$, $J_{0}$
and the Lorentz charge, let $\calH$ be the
cohomology of $\cal O$, and let $\calF$
be the Hilbert space in which the cohomology of $\cal O$
is computed.
Then it can be shown that the traces over $\calH$ and $\calF$ coincide:
\begin{align}
 \label{eq:HFequiv}
  \Tr_{\calH}(-1)^{F}q^{L_{0}}t^{J_{0}}\mathe^{\mu\cdot \sigma}
= \Tr_{\calF}(-1)^{F}q^{L_{0}}t^{J_{0}}\mathe^{\mu\cdot \sigma}  \,.
\end{align}
To show this, first split $\calH$ and $\calF$ to even and odd parts:
\begin{align}
  \calH = \calH_{{e}}\oplus \calH_{o}\,,\quad
  \calF = \calF_{{e}}\oplus \calF_{{o}}\,.
\end{align}
(In our case, fermion numbers will be carried by $\theta$'s and the fermionic BRST ghosts.)
Then, since
\begin{align}
\begin{split}
  \calF_{e} &= \calZ_{e} + \calF_{e}/\calZ_{e}
   = (\calH_{e} + \calB_{e}) + \calB_{o}\,,\\
  (\calZ &= \mathop{\mathrm{Ker}} {\cal O}\,,\quad \calB= \mathop{\mathrm{Im}} {\cal O})\,,
\end{split}
\end{align}
and similar for $e\leftrightarrow o$,
the trace over $\calB_{e}$ and $\calB_{o}$ do not contribute
to the right-hand side of (\ref{eq:HFequiv})
due to the factor $(-1)^{F}$.

All the cohomology operators $\delcech$, $\deldol$, and $D$
satisfy the criterion we stated for general $\cal O$.
Thus, although we defined the partition function as the trace over the cohomology
in the previous subsection, it could have been the trace over the space in which
the cohomology is computed.
Below, we use the formula~(\ref{eq:HFequiv}) freely
when computing the partition functions.

\bigskip
We will also use the formula~(\ref{eq:HFequiv}) in section~\ref{sec:lightcone}
when we relate the partition function of pure spinors to
the cohomology of the physical BRST operator $Q= \int\lambda^{\alpha}d_{\alpha}$.
Although $Q$ does not commute with $J_{0}$,
we will argue that one can twist the $t$-charge using the Lorentz current
so that the twisted charge $0$ piece of $Q$ has the same cohomology as $Q$
(except for the on-shell condition $L_{0}=0$).
Then the cohomology of $Q$ can be read off from the twisted partition function.
It will be shown in section~\ref{sec:lightcone} that the cohomology
thus obtained precisely reproduces the lightcone spectrum of the superstring.

\bigskip
Finally, note that our partition function remains the same for
the classical and quantum cohomologies. Although
some classical cohomology elements may not be in the quantum cohomology,
such elements will drop out in the form of doublets, $f\overset{\cal O}{\to}g$.
Hence, due to the factor of $(-1)^{F}$, they do not affect the partition function.
(For the \v{C}ech operator $\delcech$, the fermion number operator $F$
counts the order of cochains.)

\subsection{Counting of gauge invariant polynomials and the missing states}
\label{sec:gaugeinvs}

Now, let us actually construct the elements of $H^{0}$ at lower Virasoro levels.
The states we construct are polynomials of $\omega$, $\lambda$ and their derivatives,
and are invariant under the ``gauge transformation''
\begin{align}
  \delta_{\Lambda}\omega_{\alpha} &= \Lambda^{\mu}(\gamma_{\mu}\lambda)_{\alpha} \,.
\end{align}
In the language of curved $\beta\gamma$ theory
these correspond to globally defined operators.
Basic invariants with a single $\omega$ are
the $\lambda$-charge\footnote{
We call $J$ the $\lambda$-charge current to distinguish
it from the $t$-charge current $J_{t}=-\omega\lambda-p\theta$.
} and Lorentz currents,
and the energy-momentum tensor for the pure spinor sector
\begin{align}
  J &= -\omega\lambda\,,\quad
  N^{\mu\nu} = -{1\over2}\omega\gamma^{\mu\nu}\lambda \,,\quad
  T = -\omega\del\lambda\,.
\end{align}
Of course, arbitrary products of these operators are again gauge invariant.
Starting from level $2$, there will be certain gauge invariant polynomials with negative $\lambda$-charge
meaning that the number of $\omega$'s is strictly larger than that of $\lambda$'s.
These, however, are perfectly normal gauge invariant operators
and should not be confused with the ``missing states''
alluded to at the beginning of this section.

The true missing states, which first appear at level $2$, are fermionic
and are crucial for reproducing the massive spectrum of the superstring.
The purpose of this section is to show that the Hilbert space of ``naive'' gauge invariants
lacks field-antifield symmetry and hence is not the appropriate
Hilbert space in the pure spinor formalism.
Later, we shall explain how the missing states fit into
the higher degree cohomologies of the \v{C}ech/Dolbeault or ghost-for-ghost descriptions.

Descriptions of gauge invariants at levels $0$ and $1$ can also be
found in references~\cite{Berkovits:2005hy}\cite{Grassi:2005jz}.

\subsubsection{Level $0$ gauge invariants}

At the lowest level, the Hilbert space is spanned by non-vanishing polynomials of $\lambda$.
Due to the pure spinor constraint, $\lambda$'s can only appear in the ``pure spinor representations''
\begin{align}
  \lambda^{\lpar \alpha_{1}}\lambda^{\alpha_{2}}\cdots \lambda^{\alpha_{n}\rpar}
 &= \dynkin(0000n)t^{n}\,,\quad(n\ge0)\,.
\end{align}
Here, we also indicated the $t$-charge of the state,
and the symbol $\lpar \alpha_{1}\alpha_{2}\cdots \alpha_{n}\rpar$
signifies the ``spinorial $\gamma$-traceless condition'', which means that the expression
is zero when any two indices $\alpha_{i}\alpha_{j}$ are contracted using $\gamma^{\mu}_{\alpha_{i}\alpha_{j}}$.
Since the pure spinor representations have dimensions
\begin{align}
\dim\dynkin(0000n)
&= {(n+7)(n+6)(n+5)^2(n+4)^2(n+3)^2(n+2)(n+1) \over 7\cdot6\cdot5^2\cdot4^2\cdot3^2\cdot2} \,,
\end{align}
the level $0$ partition function is easily found to be~\cite{Berkovits:2005hy}
  \begin{align}
   \label{eqn:z0dim}
    Z_{0}(t) &= {1-10t^2+16t^3-16t^5+10t^6-t^8 \over (1-t)^{16} } = {(1+t)(1+4t+t^{2}) \over (1-t)^{11}}\,.
  \end{align}

\subsubsection{Field-antifield symmetry}

Before proceeding to the next level, let us explain an important symmetry
possessed by the zero-mode partition function.
Looking at~(\ref{eqn:z0dim}),
one immediately notices that $Z_{0}(t)$ has the following symmetry:
\begin{align}
  \label{eqn:PSFAFsym}
  Z_{0}(t) &= -t^{-8}Z_{0}(1/t) \,.
\end{align}
As we shall explain shortly, this symmetry is related
to {\em field-antifield symmetry} in the pure spinor superstring.
The symmetry is important for having a non-degenerate inner product on the physical states and
the value $-8$ is related to the $\lambda$-charge anomaly
of the pure spinor system~\cite{Berkovits:2005hy}.

In order to explain how the field-antifield symmetry is related
to the inner product structure of pure spinor superstring,
let us compute the total weight $0$ partition function for the
pure spinor superstring, by including the contribution
from $\theta^{\alpha}$. (The momenta $\omega_{\alpha}$, $p_{\alpha}$ and $\del x^{\mu}$ do not affect the
weight $0$ partition function,
and we ignore the zero modes $k^{\mu}$ of $\del x^{\mu}$ as usual.)
Assigning $t$-charge $1$ to $\theta^{\alpha}$, the partition function for
$\theta^{\alpha}$ is easily computed and reads
\begin{align}
  Z_{\theta,0}(t) &= \Tr_{\theta}(-1)^{F}t^{J_{0}} = (1-t)^{16} \,.
\end{align}
Hence, the total weight $0$ partition function is
\begin{align}
\begin{split}
\mathbf{Z}_{0}(t) &= Z_{\lambda,0}(t)Z_{\theta,0}(t) \\
 &= 1-10t^2+16t^3-16t^5+10t^6-t^8 \,.
\end{split}
\end{align}
Now $\mathbf{Z}_{0}(t)$ is nothing but the partition function for
the cohomology of $Q_{0}=\int \lambda^{\alpha}p_{\alpha}$ carrying $t$-charge zero.
For the massless sector, the cohomology of $Q_{0}$
coincides with the zero-momentum
cohomology of $Q= \int\lambda^{\alpha}d_{\alpha}$.
The cohomology representatives can be explicitly identified as follows:
\begin{align}
\begin{split}
  t^{0}:&\quad 1 \,, \\
  -10t^{2}:&\quad (\lambda\gamma^{\mu}\theta) \,, \\
  16t^{3}:&\quad  (\lambda\gamma^{\mu}\theta)(\gamma_{\mu}\theta)_{\alpha} \,, \\
  -16t^{5}:&\quad (\lambda\gamma^{\nu}\theta)(\lambda\gamma^{\rho}\theta)(\gamma_{\nu\rho}\theta)^{\alpha} \,, \\
  10t^{6}:&\quad (\lambda\gamma^{\nu}\theta)(\lambda\gamma^{\rho}\theta)(\theta\gamma_{\mu\nu\rho}\theta) \,,\\
  -t^{8}:&\quad (\lambda\gamma^{\mu}\theta)(\lambda\gamma^{\nu}\theta)(\lambda\gamma^{\rho}\theta)(\theta\gamma_{\mu\nu\rho}\theta) \,.
\end{split}
\end{align}
It is then easy to see that an appropriate inner product $(V,W)$
can be defined on the cohomology using the zero-mode prescription
\begin{align}
  \label{eqn:zeromoderule}
  \langle (\lambda\gamma^{\mu}\theta)(\lambda\gamma^{\nu}\theta)(\lambda\gamma^{\rho}\theta)(\theta\gamma_{\mu\nu\rho}\theta) \rangle &= 1 \,.
\end{align}
Every cohomology element $V$ has its conjugate (antifield) $V_{A}$ such that
\begin{align}
  (V, V_{A}) &= \langle V^{\dagger}V_{A} \rangle = 1 \,,
\end{align}
where $V^{\dagger}$ denotes the BPZ conjugate of $V$~\cite{Belavin:1984vu}.
Since $\lambda^{\alpha}$ has $t$-charge anomaly $-8$ while $\theta^{\alpha}$ has $16$,
the rule~(\ref{eqn:zeromoderule}) precisely saturates the anomaly.
It is analogous to the rule for the bosonic string, $\langle c\del c \del^{2} c \rangle = 1$, and can be derived from functional integration methods
after including an appropriate BRST-invariant measure factor%
~\cite{Berkovits:2004px}.

Below, we shall argue that the partition function of pure spinors
has the field-antifield symmetry~(\ref{eqn:PSFAFsym}) at each Virasoro level,
and therefore all physical states in the pure spinor superstring
appear in field-antifield pairs.

\subsubsection{Level $1$ gauge invariants}

The weight $1$ can be saturated either by one $\omega$ or one $\del\lambda$,
and we wish to count the states that do not vanish due to
the pure spinor constraints
\begin{align}
  \lambda\gamma^{\mu}\lambda = 0\,,\quad  \del(\lambda\gamma^{\mu}\lambda) = 2\lambda\gamma^{\mu}\del \lambda = 0 \,.
\end{align}
For the states with $\omega$, one must also require
invariance under the gauge transformation
$\delta_{\Lambda}\omega_{\alpha} = \Lambda_{\mu}(\gamma^{\mu}\lambda)_{\alpha}$.
For the level $1$ operators, the latter condition implies that
$\omega$ must appear in the form of the gauge invariant currents $J$ and $N^{\mu\nu}$.
Hence, all the possible states with a single $\omega$ are ($n\ge0$)
\begin{align}
\begin{split}
  \omega_{\alpha}\lambda^{\lpar \alpha}\lambda^{\beta_{1}}\cdots \lambda^{\beta_{n}\rpar }&= \dynkin(0000n)t^{n} \,,\\
  \omega_{\alpha}{(\gamma^{\mu\nu})^{\alpha}}_{\beta}\lambda^{\lpar \beta}\lambda^{\beta_{1}}\cdots \lambda^{\beta_{n}\rpar }&= \dynkin(0100n)t^{n} \,.
\end{split}
\end{align}
The states involving $\del \lambda$ are described by ($n\ge0$)
\begin{align}
\begin{split}
 \del \lambda^{\lpar \alpha}\lambda^{\beta_{1}}\cdots \lambda^{\beta_{n}\rpar}
 &= \dynkin(0000,n+1)t^{n+1}\,,\\
 \del \lambda^{\alpha}\gamma^{\mu\nu\rho}_{\alpha\beta}\lambda^{\lpar \beta}\lambda^{\beta_{1}}\cdots \lambda^{\beta_{n}\rpar}
 &= \dynkin(0010n)t^{n+2}\,.
\end{split}
\end{align}
Note that while $\lambda\gamma^{\mu}\del\lambda=0$ due to the pure spinor constraint,
the $3$-form $\lambda\gamma^{\mu\nu\rho}\del\lambda$ is non-vanishing.

Adding up all four contributions, one finds~\cite{Grassi:2005jz}
\begin{align}
Z_{1}(t)
&= { 46 - 144t + 116t^2 + 16t^3 - 16t^5 -116t^6 + 144 t^7 - 46t^8 \over (1-t)^{16} } \\
& \qquad\qquad\qquad\qquad =  \, {2(1+t)(23+20t+23t^{2})\over (1-t)^{11}}\,.
\end{align}
This satisfies the same field-antifield symmetry as $Z_{0}(t)$:
\begin{align}
  Z_{1}(t) &= -t^{-8}Z_{1}(1/t) \,.
\end{align}

\subsubsection{Level $2$ gauge invariants and a missing state}
\label{sec:ginvpoly2}

Explicit constructions of the gauge invariant polynomials
at level $2$ can be obtained using similar methods.
But at this level we encounter several new features.
Most importantly, we will find that the space of gauge invariant polynomials
does not posses the field-antifield symmetry.
This
implies the space has to be augmented by some finite number of terms.
As already hinted in section~\ref{subsec:BRSTCechDolbeault1},
these ``missing states'' correspond to elements of higher degree cohomologies
in \v{C}ech/Dolbeault and ghost-for-ghost descriptions.
(We shall explain this in detail in section~\ref{sec:SymPart}.)
For now, however, let us focus on the space of gauge invariant polynomials
and enumerate them.

First of all, there are polynomials with two $\omega$'s.
One might expect that these $\omega$'s only appear in the form of
$N^{\mu\nu}$ or $J$, but there in fact is
a gauge invariant polynomial with
negative $\lambda$-charge
\begin{align}
 \label{eq:falpha}
  f_{\alpha} &= 3J \omega_{\alpha} + N^{\mu\nu}(\gamma_{\mu\nu}\omega)_{\alpha} \,.
\end{align}
Appearance of $f_{\alpha}$ is interesting, but we stress that it is
a perfectly normal gauge invariant polynomial
and has nothing to do with the ``missing states''.
Of course, $f_{\alpha}$ multiplied by some function of $\lambda$ is again gauge invariant,
but this carries non-negative $\lambda$-charge and can be expressed in terms
of operators constructed from $N^{\mu\nu}$ and $J$.

The states with two $N$'s, two $J$'s, and one $N$ and one $J$ are ($n\ge0$)
\begin{align}
\label{eq:NNJJ}
\begin{split}
N_{\lbra \mu\nu}N_{\rho\sigma \rbra}\lambda^{(n)}
 &=  (\gamma_{\lbra \mu\nu}\omega)_{\alpha_{1}}(\gamma_{\rho\sigma \rbra}\omega)_{\alpha_{2}}
     \lambda^{\lpar \alpha_{1}}\lambda^{\alpha_{2}}\lambda^{\beta_{1}}\cdots \lambda^{\beta_{n}\rpar}  = \dynkin(0200n)t^n\,, \\
N_{[ \mu\nu}N_{\rho\sigma ]}\lambda^{(n)}
 &= (\gamma_{[ \mu\nu}\omega)_{\alpha_{1}}(\gamma_{\rho\sigma ]}\omega)_{\alpha_{2}}\lambda^{\lpar \alpha_{1}}\lambda^{\alpha_{2}}\lambda^{\beta_{1}}\cdots \lambda^{\beta_{n}\rpar}
  = \dynkin(0001,n+1)t^{n} \,, \\
N_{\mu\nu}J \lambda^{(n)}
 &=  (\gamma_{\mu\nu}\omega)_{\alpha_{1}}\omega_{\alpha_{2}}\lambda^{\lpar \alpha_{1}}\lambda^{\alpha_{2}}\lambda^{\beta_{1}}\cdots \lambda^{\beta_{n}\rpar}
  = \dynkin(0100n)t^{n}\,, \\
JJ\lambda^{(n)}
 &=  \omega_{\alpha_{1}}\omega_{\alpha_{2}}\lambda^{\lpar \alpha_{1}}\lambda^{\alpha_{2}}\lambda^{\beta_{1}}\cdots \lambda^{\beta_{n}\rpar}
 = \dynkin(0000n)t^{n}\,.
\end{split}
\end{align}
Here, we left the $\gamma$-traceless conditions implicit,
and the indices in $\lbra \mu\nu,\rho\sigma \rbra$ are traceless, block-symmetric,
and antisymmetric within each blocks.
In fact, the $4$-form piece of $NN\lambda^{(n)}$, $NJ\lambda^{(n)}$ and $JJ\lambda^{(n)}$ can be written as $(n\ge0)$
\begin{align}
\begin{split}
 f_{\alpha}\lambda^{(n)}
&= (3\omega_{\alpha_0}\omega_{\alpha} + (\gamma^{\mu\nu}\omega)_{\alpha_{0}}(\gamma_{\mu\nu}\omega)_{\alpha})\lambda^{\lpar \alpha_{0}}\lambda^{\beta_{1}}\cdots \lambda^{\beta_{n}\rpar} \\
&= \dynkin(00010)\otimes\dynkin(0000n)t^{n-1}\,,
\end{split}
\end{align}
so one must be careful not to double count.

As for the polynomials with a single derivative, the following states are independent
($n\ge0$):
\begin{align}
\label{eq:delNJ}
\begin{split}
  \del N^{\mu\nu}\lambda^{(n)} &= \del(\omega_{\alpha_{0}}\gamma^{\mu\nu\alpha_{0}}{}_{\alpha_1}\lambda^{\lpar \alpha_1})\lambda^{\beta_{1}}\cdots \lambda^{\beta_{n}\rpar}
   = \dynkin(0100n)t^{n}\,, \\
  \del J\lambda^{(n)} &= \del(\omega_{\alpha_1}\lambda^{\lpar \alpha_1})\lambda^{\beta_{1}}\cdots \lambda^{\beta_{n}\rpar}
   = \dynkin(0000n)t^{n}\,, \\
  N^{\mu\nu}\del\lambda \lambda^{(n)}
 &= \bigl((\omega\gamma^{\mu\nu})_{\alpha_{1}}\del \lambda^{\lpar \alpha_{0}}\lambda^{\alpha_{1}}\lambda^{\beta_{1}}\cdots \lambda^{\beta_{n}\rpar}
  + (\text{$\gamma$-traces})\bigr) \\
 &\qquad+
 (\omega\gamma^{\lbra \mu\nu})_{\alpha_{0}}(\del \lambda\gamma^{\rho\sigma\tau\rbra})_{\alpha_{1}}\lambda^{\lpar \alpha_{0}}\lambda^{\alpha_{1}}\lambda^{\beta_{1}}\cdots \lambda^{\beta_{n}\rpar } \\
 &=\bigl(\dynkin(0100,n+1)  + \dynkin(1001n) + \dynkin(0000,n+1)\bigr) t^{n+1}
  + \dynkin(0110n)t^{n+2}\,, \\
J\del\lambda \lambda^{(n)}
 &= \omega_{\alpha_1}\del\lambda^{\lpar \alpha_0}\lambda^{\alpha_1}\lambda^{\beta_{1}}\cdots \lambda^{\beta_{n}\rpar}
  + \omega_{\alpha_0}(\del\lambda\gamma_{\mu\nu\rho})_{\alpha_{1}}\lambda^{\lpar \alpha_{0}}\lambda^{\alpha_{1}}\lambda^{\beta^{1}}\cdots \lambda^{\beta_{n}\rpar} \\
 &= \dynkin(0000,n+1)t^{n+1} + \dynkin(0010n)t^{n+2}\,, \\
T &= \omega_{\alpha}\del \lambda^{\alpha} = \dynkin(00000)t^{0}\,.
\end{split}
\end{align}
Note that $T\lambda^{(n+1)}$ and $\omega_{\alpha_1}\del\lambda^{\lpar \alpha_0}\lambda^{\alpha_1}\lambda^{\beta_{1}}\cdots \lambda^{\beta_{n}\rpar}$ are not independent.

Finally, there are two types of polynomials with two derivatives, $\del^{2}\lambda\lambda^{(n)}$ and
$(\del \lambda)^{2}\lambda^{(n)}$, and some of them are related by the level $2$
pure spinor condition
\begin{align}
  \lambda\gamma^{\mu}\del^{2}\lambda + \del\lambda\gamma^{\mu}\del\lambda = 0 \,.
\end{align}
The independent states are ($n\ge0$)
\begin{align}
\begin{split}
 \label{eq:del2lam}
 \del^{2}\lambda^{\alpha}\lambda^{\lpar \beta_{1}}\cdots \lambda^{\beta_{n}\rpar}
 &= \dynkin(00001)\otimes\dynkin(0000n)t^{n+1}\,,
\\
\del\lambda^{\lpar \alpha_{1}}\del\lambda^{\alpha_{2}}\lambda^{\beta_{1}}\cdots \lambda^{\beta_{n}\rpar}
&= \dynkin(0000,n+2)t^{n+2}\,,
\\
(\del\lambda\gamma^{\mu\nu\rho})_{\beta_{1}}\del\lambda^{\lpar \alpha}\lambda^{\beta_{1}}\lambda^{\beta_{2}}\cdots \lambda^{\beta_{n+1}\rpar}
&= \dynkin(0010,n+1)t^{n+3}\,,
\\
(\del\lambda\gamma^{\lbra \mu\nu\rho})_{\beta_{1}}(\del\lambda\gamma^{\sigma\tau\kappa \rbra})_{\beta_{2}}\lambda^{\lpar \beta_{1}}\lambda^{\beta_{2}}\cdots \lambda^{\beta_{n+2}\rpar}
&= \dynkin(0020n)t^{n+4}\,.
\end{split}
\end{align}

\bigskip
Adding up all the contributions, (\ref{eq:falpha}), (\ref{eq:NNJJ}), (\ref{eq:delNJ}) and (\ref{eq:del2lam}),
one finds
\begin{align}
Z_{2,\text{poly}}(t)
&= {1\over (1-t)^{16} }\Bigl\{
  16t^{-1} + 817 - 3840t + 7794t^2 - 10848t^3 + 12870t^4 - 12032t^5 \nonumber\\
&\qquad +
 8222t^6 - 4896t^7 + 2823t^8 - 1136t^9 + 240t^{10} - 32t^{11} + 2t^{12} \Bigr\} \,.
\end{align}

\paragraph{The missing state}

As already mentioned, $Z_{2,\text{poly}}$ we just computed
does not posses the field-antifield symmetry.
However, one finds that
\begin{align}
Z_{2}(t) &= Z_{2,\text{poly}}(t)-t^{-4}  \nonumber\\
 &= {1\over (1-t)^{16} }\Bigl\{
- t^{-4} + 16t^{-3} - 120t^{-2} + 576t^{-1} -1003 + 528t - 214t^2 + 592t^3 \nonumber\\
&\qquad - 592t^5 + 214t^6 - 528t^7 + 1003t^8 - 576t^9 + 120t^{10} - 16t^{11} + t^{12} \Bigr\}
\end{align}
{\em does} have the desired symmetry
\begin{align}
  Z_{2}(t) &= -t^{-8}Z_{2}(1/t) \,.
\end{align}
Therefore, we expect to have an extra {\em fermionic} singlet with $t$-charge $-4$
at weight $2$.
Because it is fermionic, it cannot be a usual gauge invariant state.
Indeed, it will be identified as an element of higher cohomology
in all four descriptions, 
minimal and non-minimal ghost-for-ghost, \v{C}ech and Dolbeault.

%%%%%%%%%%%%%%%%%%%%%%%%%%%%%%%%%%%%%%%%%%%%%%%%%%%%%%%%%%%%%%%%
\section{Partition function of pure spinors and its symmetries}
\label{sec:PartPS}

In this section, we present two independent methods
for computing the full partition function of pure spinors.
The first method utilizes Chesterman's ghost-for-ghost
description of pure spinors~\cite{Chesterman:2004xt},
while the second method uses a fixed point formula extending the zero mode result of~\cite{Berkovits:2005hy}.
Neither method gives the complete partition function in closed form,
but the partition functions can be computed level by level unambiguously
once one imposes
the requirements of field-antifield and ``$\ast$-conjugation'' symmetries.

We present the ghost-for-ghost method first because the two
symmetries are (formally) manifest in this formalism. However,
since the ghost-for-ghost description of the pure spinor requires
an infinite tower of ghosts-for-ghosts, the expression for the
partition function is ill-defined and one has to invoke an
analytic continuation in order to maintain the two symmetries.
Also, using this method, it is difficult to compute the partition
function keeping the spin dependence of the states.

For the fixed point method, the difficulty in writing a closed formula
arises because the states that depend on inverse powers of $\lambda$
(or $\lambda\overline{\lambda}$ in non-minimal formulation)
do not appear to contribute.
However, the number of such states is finite at any given level,
and they can be recovered by requiring the two discrete symmetries.

This section is organized as follows.
In section~\ref{sec:BRST} (which is accompanied by appendix~\ref{app:reducibility})
we introduce the BRST description of the pure spinors
using an infinite tower of ghosts-for-ghosts.
We then use it in section~\ref{sec:partBRST} to
motivate the form of field-antifield and ``$\ast$-conjugation'' symmetries
and to compute the partition function.
The last section~\ref{sec:partFP} is on the fixed point formulas for the partition function.
If one accepts the two symmetries described in section~\ref{subsubsec:symparts}, section~\ref{sec:partFP} can be read
independently from sections~\ref{sec:BRST} and~\ref{sec:partBRST}.

\subsection{Ghost-for-ghost description of pure spinors}
\label{sec:BRST}

In this section,
we analyze the (infinite) reducibility conditions
for the pure spinor constraint using the BRST formalism.
The resulting BRST operator $D$ was first introduced by
Chesterman~\cite{Chesterman:2004xt}\footnote{%
Note, however, that the spin contents of the ghosts-for-ghosts we derive is
slightly different from the ones proposed by Chesterman~\cite{Chesterman:2004xt}.
(See footnote~\ref{fn:Chesterman}.)}.
As already mentioned, we sometimes refer to $D$ as the mini-BRST operator
to avoid confusion with the physical BRST operator $Q=\int \lambda^{\alpha}d_{\alpha}$,

Chesterman's
ghost-for-ghost construction is designed so that
the ghost number $0$ cohomology $H^{0}(D)$
reproduces the space of gauge invariant functions of the constrained system.
Indeed, the partition functions of $D$-cohomology
in weight $0$ and $1$ sectors
precisely describe the number of gauge invariant objects described above.
However, starting at weight $2$, we find extra cohomology elements
which do not correspond to the naive gauge invariants,
as is expected from the analysis of toy models~\cite{Toymodels}.
We shall claim that those extra states are as important a part of
the Hilbert space of the pure spinor system
as the naive gauge invariants.
We will come back to this issue in section~\ref{sec:SymPart}.

\subsubsection{Reducibility conditions and the ghosts-for-ghosts}
\label{sec:reducibility}

Let us start by constructing a nilpotent operator $\delta$, whose weight $0$ cohomology
is isomorphic to the space of polynomials of $\lambda$, modulo the pure spinor constraint.
For the time being, we shall concentrate on the ``position space'' $\lambda$,
and ignore the ``momentum space'' $\omega$.
Later in section~\ref{sec:BRSTop}, we will
construct the mini-BRST operator $D$ from $\delta$
by extending the action of $\delta$ to the full phase space at the quantum level.

To facilitate the discussion below, we denote the pure spinor constraint as
\begin{align}
  G^{A_{1}} &\equiv \lambda\gamma^{\mu}\lambda=0\,,\quad A_{1}=\dynkin(10000)=\mathbf{10} \,.
\end{align}
Now, following the usual strategy of the ghost-for-ghost construction,
we introduce a fermionic antighost (or $C$-type ghost) $c^{\mu}$ to `kill' the
pure spinor part of $\lambda$ and define the $\delta$-action
\begin{align}
  \delta C^{A_{1}} = G^{A_{1}}\quad\leftrightarrow\quad \delta c^{\mu} = \lambda\gamma^{\mu}\lambda  \,.
\end{align}
Then, a function $f(\lambda)$ proportional to $\lambda\gamma^{\mu}\lambda$ is $\delta$-exact and does not
contribute to the $\delta$-cohomology.
(Later, in order to construct the mini-BRST operator
in the phase space,
we shall include the momentum like $B$-type ghost conjugate to $C$,
which is in the conjugate representation $\overline{A}_{1}$.)

\paragraph{First order reducibility}

However, because the pure spinor constraint is {\em reducible},
this is not the end of the story:
Using the (strong) identity
\begin{align}
  \label{eq:psred1}
  (\lambda\gamma^{\mu}\lambda)(\gamma_{\mu}\lambda)_{\alpha} = 0 \,,
\end{align}
one can construct a $\delta$-closed state
\begin{align}
  c^{\mu}(\gamma_{\mu}\lambda)_{\alpha}\,,
\end{align}
which must be killed by introducing another generation of ghost.
The coefficient $(\gamma_{\mu}\lambda)_{\alpha}$ in~(\ref{eq:psred1}) is called
the ``reducibility coefficient'' and we denote it as
\begin{align}
  G^{A_{1}}R_{A_{1}}^{A_{2}} = 0 \quad\to\quad R_{A_{1}}^{A_{2}}=(\gamma_{\mu}\lambda)_{\underline{\alpha}}\,,\quad
 A_{2} = \dynkin(00010)=\overline{\mathbf{16}} \,.
\end{align}
(For convenience, we put an underline to the index newly appeared.)
The reducibility coefficient $R_{A_{1}}^{A_{2}}$ is ``complete'',
meaning there are no other (strong) relations for the pure spinor constraint,
independent from~(\ref{eq:psred1}).

Now, in order to eliminate the unwanted $\delta$-closed state,
a second generation ghost must be introduced to kill it cohomologically.
For the case at hand, we introduce a bosonic ghost
\begin{align}
  C^{A_{2}} \quad\leftrightarrow\quad \sigma_{\alpha} \,,
\end{align}
and define the $\delta$-action to trivialize $(\fslash{c}\lambda)_{\alpha}$:
\begin{align}
  \delta C^{A_{2}} = C^{A_{1}}R_{A_{1}}^{A_{2}}\quad\leftrightarrow\quad
  \delta \sigma_{\alpha} = c_{\mu}(\gamma^{\mu}\lambda)_{\alpha} \,.
\end{align}
Then, by definition, the action of $\delta$ is
strongly nilpotent up to this order.

\paragraph{Second order reducibility}

At the next order, the reducibility coefficients are defined by
\begin{align}
  R_{A_{1}}^{A_{2}}R_{A_{2}}^{A_{3}} \approx 0\quad\text{(weak equality)} \,.
\end{align}
As opposed to the first order reducibility~(\ref{eq:psred1}),
weak equality is enough for unwanted $\delta$-closed elements to appear.
A complete reducibility coefficient for the case at hand is\footnote{%
\label{fn:Chesterman}
In~\cite{Chesterman:2004xt}, $R_{A_{2}}^{\bullet_{3}}=\lambda^{\alpha}$ is also
considered as the reducibility coefficient and, correspondingly,
an additional singlet ghost $c$ was introduced.
But, as we explain in appendix~\ref{app:redtech},
there are no $\delta$-closed states associated to this relation
and this additional ghost should not be introduced.}
\begin{align}
  R_{A_{2}}^{A_{3}} &= (\gamma^{\underline{\mu\nu}}\lambda)^{\alpha},\quad A_{3}=\dynkin(01000)=\mathbf{45} \,.
\end{align}
Again, this relation implies the existence of unwanted $\delta$-closed states
of the form $C^{A_{2}}R_{A_{2}}^{A_{3}} + M^{A_{3}}$,
where $M^{A_{3}}$ is some polynomial free of $C^{A_{2}}$.
Explicitly, the following combination is $\delta$-closed
\begin{align}
C^{A_{2}}R_{A_{2}}^{A_{3}} + M^{A_{3}}\quad\leftrightarrow\quad
   \sigma_{\alpha}(\gamma^{\mu\nu}\lambda)^{\alpha} + c^{\mu}c^{\nu}\,.
\end{align}
To kill these, we introduce the third generation fermionic ghosts
\begin{align}
  C^{A_{3}} \quad\leftrightarrow\quad c^{\mu\nu}\,,
\end{align}
and extend the $\delta$-action as
\begin{align}
  \delta C^{A_{3}} = C^{A_{2}}R_{A_{2}}^{A_{3}} + M^{A_{3}} \quad\leftrightarrow\quad
  \delta c^{\mu\nu}  =  \sigma_{\alpha}(\gamma^{\mu\nu}\lambda)^{\alpha} + c^{\mu}c^{\nu}\,.
\end{align}
One can check the strong nilpotency of $\delta$-action up to this order.

\paragraph{Higher order reducibilities}

The analyses of the higher order reducibility conditions are similar.
For the readers interested,
we include the third and fourth order analyses in appendix~\ref{app:reducibility}.
Explicit constructions of the reducibility coefficients $R_{A_{n}}^{A_{n+1}}$
(and hence the mini-BRST operator) soon become tedious at higher orders.
But as we will now explain,
the spin contents of the ghosts-for-ghosts can be easily inferred
without actually doing the reducibility analysis.
Moreover, as we will argue in section~\ref{sec:BRSTop},
the knowledge of spin contents is sufficient for determining
the structure of reducibility coefficients.

\subsubsection{Spins of ghosts-for-ghosts}
\label{sec:spinsofghosts}

Since the ghosts needed in the ghosts-for-ghosts implementation of pure spinor constraints
are all free fields, computation of their partition functions are straightforward.
Demanding that they reproduce the level $0$ spin partition function of pure spinors, $Z_{0}(t,\vec{\sigma})$,
their Lorentz spins can be readily determined.
Let us denote by $A_{n}$ the representations of the $n$th generation $C$-type ghosts.
($A_{n}$'s are not necessarily irreducible.)
By convention, we include a minus sign, $(-1)^{|A_{n}|}$,  in $A_{n}$ if the corresponding ghost is a fermion.
In order to reproduce $Z_{0}(t,\vec{\sigma})$, the $A_{n}$'s must satisfy
\begin{align}
 Z_{0}(t,\vec{\sigma}) &= \prod_{n\ge0}(1-t^{n+1})^{-A_{n}}
  \equiv \prod_{n\ge0}\prod_{\mu\in A_{n}} (1-t^{n+1}\mathe^{\mu\cdot \sigma})^{-(-1)^{|A_{n}|}} \,,
\end{align}
or equivalently (by canceling $(1-t)^{-A_{0}}$ present in both sides),
\begin{align}
 \label{eq:Aneq}
 \mathbf{1}t^{0} -\mathbf{10}t^{2}
 + \overline{\mathbf{16}}t^{3} - \mathbf{16}t^{5}
 + \mathbf{10}t^{6} - \mathbf{1}t^{8}
&= \prod_{n\ge1}(1-t^{n+1})^{-A_{n}} \,.
\end{align}
Now, by expanding both sides in $t$ and comparing the coefficients of $t^{n}$,
the ghost representations $A_{n}$'s can be uniquely determined.
For example, the first few terms of the expansion on the right hand side read
\begin{align}
\begin{split}
  \prod_{n\ge1}(1-t^{n+1})^{-A_{n}}
&= 1 + A_{1}t + (\Sym^{2}A_{1} + A_{2})t^{2}
 + (\Sym^{3}A_{1}+ A_{1}\otimes A_{2} + A_{3})t^{3} \\
&\qquad + (\Sym^{4}A_{1} + \Sym^{2}A_{1}\otimes A_{2} + \Sym^{2}A_{2} + A_{1}\otimes A_{3} + A_{4})t^{4} + \cdots\,,
\end{split}
\end{align}
where the symmetric products $\Sym^{k}$ are understood in the supersense.   
It should be clear that $A_{n}$ is uniquely determined by
the equality at $t^{n}$.
In figure~\ref{fig:An}, we list the
$A_{n}$'s for the first few generations of the BRST ghosts.
(Since it is sometimes convenient to treat $\lambda^{\alpha}$ as the zeroth generation ghost,
we also indicated it in the list.)
\begin{figure}[t]
\centering
\begin{tabular}{M|MMMM}
 \text{gh\#} & \text{$C$-type ghosts} & \text{$t$-charge} & A_{n} & N_{n} \\ \hline
 0& \lambda^{\alpha} & 1&\dynkin(00001) & 16\\ \hline
 1& c^{\mu} & 2&-\dynkin(10000) & -10  \\
 2& \sigma_{\alpha} & 3&\dynkin(00010) & 16\\
 3& c^{\mu\nu}& 4 & -\dynkin(01000) & -45 \\
 4& \sigma^{\mu}_{\alpha}&5 & \dynkin(10010) & 144\\
 5& c^{\mu}_{[\nu\rho]},\;c^{\alpha\beta},\;c^{\mu} & 6& -\bigl(\dynkin(11000)+\dynkin(00020)+\dynkin(10000)\bigr) & -456\\
 6& \sigma^{(\mu\nu)}_{\alpha},\;\sigma^{[\mu\nu]}_{\alpha},\;\sigma^{\mu\alpha},\;\sigma_{\alpha} &7&
 \dynkin(20010)+\dynkin(01010)+\dynkin(10001)+\dynkin(00010) & 1440
\end{tabular}
  \caption{Spin contents of $C$-type BRST ghosts}
  \label{fig:An}
\end{figure}

When the spin contents of the ghosts are not of interest,
one can ignore them in~(\ref{eq:Aneq}) and only keep their dimensions
\begin{align}
N_{n}=\dim A_{n} \,.
\end{align}
In fact, there is a closed formula for $N_{n}$'s given in~\cite{Movshev:2005ei}\cite{Berkovits:2005hy}\footnote{%
Our indexing convention for $N_{n}$ differs from~\cite{Berkovits:2005hy};
we apologize for the inconvenience.}:
\begin{align}
\begin{split}
  N_{0} &= 16\,,\quad N_{1} = -10\,, \\
  N_{n} &= {1\over n+1}\sum_{k\le n}
  (-1)^{k-1}\mu\bigl( (n+1)/k \bigr)\bigl( (2+\sqrt{3})^{k} + (2-\sqrt{3})^{k} \bigr)\,,\quad(n\ge2)
 \,.
\end{split}
\end{align}
Here, $\mu(n)$ is the M\"obius function defined as
\begin{align}
  \mu(n) &= \begin{cases}
    1\,, & \text{if $n=1$}\,, \\
    (-1)^{k}\,, & \text{if $n$ is a product of $k$ distinct primes}\,, \\
    0\,, & \text{otherwise}\,.
  \end{cases}
\end{align}

Below, we will need some moments of $N_{n}$'s
that can be computed using the Mellin transformation of the M\"obius function and
an analytic continuation~\cite{Berkovits:2005hy}.
Some relevant formulas thus obtained are\footnote{See~\cite{crazyformulas}
for some recent mathematical attempts to give meaning to these
manipulations.}
\begin{align}
\label{eq:Nmoments}
 \begin{split}
  \sum_{n\ge0}N_{n} &= 11\,,\quad \sum_{n\ge0}(n+1)N_{n} = 8\,,\quad \sum_{n\ge0}(n+1)^2 N_{n} = 4\,, \\
  \sum_{n\ge0}(n+1)^4 N_{n} &= -4\,,\quad \sum_{n\ge0}(n+1)^6 N_{n} = 4\,,\\
   \sum_{n\ge0}(n+1)^{8}N_{n}&={68\over3}\,,\quad \sum_{n\ge0}(n+1)^{10}N_{n} = -396 \,, \\
\sum_{k\ge1}N_{k} &= -5\,,\quad \sum_{k\ge1} kN_{k} = -3\,,\quad \sum_{k\ge1}k^2 N_{k} = -1\,.
\end{split}
\end{align}
Of these formulas, the first line has clear physical interpretations:
\begin{align}
\begin{split}
  c_{tot} &= 2\sum_{n\ge0}N_{n} = 22\,,\quad{\text{total central charge ($\Hat{T}\Hat{T}$)}}\,, \\
  a_{tot} &= -\sum_{n\ge0}(n+1)N_{n} = -8\,,\quad\text{total $\lambda$-charge anomaly ($\Hat{J}\Hat{T}$)}\,, \\
  k_{tot} &= -\sum_{n\ge0}(n+1)^2N_{n} = -4\,,\quad\text{total $U(1)$-charge anomaly ($\Hat{J}\Hat{J}$)}\,.
\end{split}
\end{align}
$\Hat{J}$ and $\Hat{T}$, which will be defined later,
are the total $\lambda$-charge current and total energy-momentum tensors
for the ghost extended system.

\subsubsection{Chesterman's mini-BRST operator}
\label{sec:BRSTop}

Let us go back to the ghost-for-ghost
program and implement the free field resolution
in the phase space at the fully quantum level.
That is, we shall construct a nilpotent mini-BRST $D$ such that
\begin{align}
 \delta C &= [D, C\}\,.
\end{align}
Using our result on the ghost-for-ghost implementation of the pure spinor constraint
(section~\ref{sec:reducibility} and appendix~\ref{app:reducibility}),
construction of $D$ is straightforward.
First, introduce the $B$-type ghosts conjugate to the $C$-type ghosts.
They carry the conjugate Lorentz representations $\overline{A}_{n}$ and satisfy
the free field operator product expansions
\begin{align}
 b^{\mu}(z)c_{\nu}(w) = {{\delta}^{\mu}_{\nu}\over z-w}\,,\quad
 \rho^{\alpha}(z)\sigma_{\beta}(w) = {-\delta^{\alpha}{}_{\beta} \over z-w}\,,\quad
 b^{\mu\nu}(z)c_{\rho\sigma}(w) = {\delta^{\mu\nu}_{\rho\sigma} \over z-w}\,,\quad\cdots.
\end{align}
Then, for the pure spinor system,
the BRST operator $D$ can be written schematically as
(recall $C_{0}=\lambda$ by convention)
\begin{align}
\begin{split}
 D &= \sum_{n\ge0}\sum_{k=0}^{n} B_{n+1}C_{n}C_{n-k} \\
   &= \sum_{n\ge0}B_{A_{n+1}}(C^{A_{n}}R_{A_{n}}^{A_{n+1}} + M^{A_{n+1}}) \,
 \end{split}
\end{align}
where $M^{A_{n}}$ is discussed in appendix \ref{app:reducibility}.
It is convenient to split $D$ in terms of the resolution degree, or the ``$C$-charge''.
Using the result from appendix~\ref{app:reducibility}, the first
several terms in $D$ are found to be
\begin{align}
\begin{split}
D &= D_{0} + D_{1} + D_{2} + D_{3} + D_{4} + \cdots\,, \\
D_{0} &= b^{\mu}(\lambda\gamma_{\mu}\lambda)\,,\quad
D_{1} = - \rho^{\alpha}c^{\mu}(\gamma_{\mu}\lambda)_{\alpha}\,,\quad
D_{2} = {1\over2}b^{\mu\nu}\bigl((\sigma\gamma_{\mu\nu}\lambda) + c_{\mu}c_{\nu} \bigr)\,, \\
D_{3} &= -\rho^{\mu\alpha}\bigl(c^{\rho\sigma}\eta_{\mu\rho}(\gamma_{\sigma}\lambda)_{\alpha} + {3\over2}c_{\mu}\sigma_{\alpha} \bigr)\,, \\
D_{4} &= {1\over2}b_{\mu,\nu\rho}\bigl( \sigma^{\mu}_{\alpha}(\gamma^{\nu\rho}\lambda)^{\alpha} + {5\over3}c^{\mu}c^{\nu\rho} \bigr)
 + b_{\mu}\bigl( \sigma_{\nu\alpha}(\gamma^{\nu \mu}\lambda)^{\alpha} - {4\over3}c_{\nu}c^{\nu \mu}\bigr) \\
&\qquad
  + b^{\alpha\beta}\sigma_{\mu\alpha}(\gamma^{\mu}\lambda)_{\beta} \,.
\end{split}
\end{align}
Since $D$ is only linear in the $B$-type ghosts,
quantum nilpotency of $D$ follows from that of the $\delta$-action.

In fact, as announced earlier, there is a simple way to specify the form
of $D$ without actually doing the reducibility analysis.
First, for the pure spinor constraint,
the reducibility coefficients $R_{A_{n}}^{A_{n+1}}$'s are linear in $\lambda$ and $M^{A_{n+1}}$'s are
quadratic in $C$-type ghosts.
Then, since there is only one way to construct a Lorentz singlet
from $\lambda^{\alpha}$ and two arbitrary representations $\overline{A}_{n+1}$ and $A_{n}$,
the tensor structure in $R_{A_{n}}^{A_{n+1}}$ is uniquely fixed up to a scale.
The choice of this scale is a matter of convention
and the appropriate choice of $M^{A_{n+1}}$ follows from the nilpotency of $D$.

We hope our discussion in this section convinced the reader
that the ghost-for-ghost mini-BRST operator $D$
is an object much tamer than might be expected,
and we now turn to the analysis of its cohomology.

\subsubsection{Mini-BRST cohomology versus gauge invariant polynomials}
\label{sec:Dcohom}

To initiate the analysis of the mini-BRST (or ghost-for-ghost) cohomology $H^{\ast}(D)$,
we first explain how the gauge invariant polynomials
described in section~\ref{sec:gaugeinvs} are translated to the ghost-for-ghost language.
As is expected from the general theory of the BRST formalism,
we find them in the ghost number $0$ cohomology $H^{0}(D)$.
However, 
we also claim that there should be non-trivial cohomology elements
of higher ghost numbers that do not correspond to naive gauge invariants.

\paragraph{Basic gauge invariant currents and their composites}

In the ghost-for-ghost language,
$\lambda$-charge and Lorentz currents, and the energy-momentum tensor
of the pure spinors are extended to include the ghost contributions as
\begin{align}
  \hat{J} &= \sum_{n\ge0}(n+1)j_{n}\,,\quad
  \Hat{N}^{\mu\nu} = \sum_{n\ge0}N^{\mu\nu}_{n}\,,\quad
  \Hat{T} = \sum_{n\ge0}T_{n}\,.
\end{align}
We note in passing that the BRST ghost number is measured by
\begin{align}
  J_{g} &= -\sum_{n\ge1}n j_{n}\,.
\end{align}
In these formulas, $j_{n}$, $N_{n}^{\mu\nu}$ and $T_{n}$ denote
the $U(1)$ current, Lorentz current,
and energy-momentum tensor for the $n$th generation ghost:
\begin{align}
\begin{split}
  j_{0} &= -\omega_{\alpha}\lambda^{\alpha}\,,\quad j_{1}= -b^{\mu}c_{\mu}\,,\quad j_{2} = -\rho^{\alpha}\sigma_{\alpha}\,,\quad\cdots\,, \\
  N_{0}^{\mu\nu} &= -{1\over2}(\omega\gamma^{\mu\nu}\lambda)\,,\quad
  N_{1}^{\mu\nu} = -2b^{[\mu}c^{\nu]}\,,\quad
  N_{2}^{\mu\nu} = -{1\over2}(\rho\gamma^{\mu\nu}\sigma)\,,\quad\cdots\,. \\
  T_{0} &= -\omega_{\alpha}\del\lambda^{\alpha}\,,\quad
  T_{1} = -b^{\mu}\del c_{\mu}\,,\quad
  T_{2} = -\rho^{\alpha}\del \sigma_{\alpha}\,,\quad\cdots\,.
\end{split}
\end{align}
Note that the $U(1)$-currents are normalized as
\begin{align}
  j_{n} &= -B_{n}C_{n}\quad\to\quad j_{n}B_{n} = -B_{n},\quad j_{n}C_{n}=+C_{n} \,.
\end{align}

It is easy to see that $(\Hat{J},\Hat{N}_{\mu\nu},\Hat{T})$ are $D$-closed and,
corresponding to the fact that the original gauge invariant currents were
not weakly vanishing, these basic $D$-closed currents
are not $D$-exact.

\paragraph{Gauge invariants with negative $\lambda$-charges}

As we saw in section~\ref{sec:ginvpoly2}, there are certain gauge invariant
polynomials in which $\omega$'s do not appear in the form of basic invariants $J$, $N^{\mu\nu}$ or $T$.
A typical example of these is
\begin{align}
  f_{\alpha} &= 3J\omega_{\alpha} + N^{\mu\nu}(\gamma_{\mu\nu}\omega)_{\alpha}\,.
\end{align}
Naively, one would expect that it will be described in the ghost-for-ghost language as
\begin{align}
  \Hat{f}_{\alpha} &\overset{?}{=} 3\Hat{J}\omega_{\alpha} + \Hat{N}^{\mu\nu}(\gamma_{\mu\nu}\omega)_{\alpha} \,.
\end{align}
This guess, however, turns out to be wrong.
Roughly speaking, the ghost contributions in $\Hat{f}_{\alpha}$ has to be doubled,
because $\Hat{J}\omega$ and $\Hat{N} \omega$ are quadratic in $\omega$
while being linear in the ghosts.

The correct expression for the $\Hat{f}_{\alpha}$ can be determined systematically
level by level in the $C$-charge defined by $(B_{n},C_{n})=(0,n)$.
Denoting by $\Hat{f}_{\alpha,n}$ the $C$-charge $n$ piece of $\Hat{f}_{\alpha}$,
the condition
\begin{align}
  D \hat{f}_{\alpha} = 0
\end{align}
leads classically to a set of master equations
\begin{align}
\begin{split}
\text{level $0$}\colon\quad 0&=  [D_{0}\,, \Hat{f}_{\alpha,0}+\Hat{f}_{\alpha,1}]_{0} \,,\\
\text{level $1$}\colon\quad0&=  [D_{0}+D_{1}\,, \Hat{f}_{\alpha,0}+\Hat{f}_{\alpha,1}+\Hat{f}_{\alpha,2}]_{1}\,,  \\
\vdots\qquad\quad & \\
\text{level $n$}\colon\quad0&=  [D_{0}+D_{1}+\cdots +D_{n}\,, \Hat{f}_{0} + \Hat{f}_{\alpha,1}+\Hat{f}_{\alpha,2}+\cdots +\Hat{f}_{\alpha,n+1}]_{n} \,, \\
\vdots\qquad\quad &
\end{split}
\end{align}
which can be used to fix $\Hat{f}_{\alpha,n}$ level by level.
The notation $[D\,,f]_{n}$ suggests that only the pieces up to $C$-charge $n$ are
kept after taking the commutator.

For example, the first several terms of $\Hat{f}_{\alpha}$ in the
$C$-charge decomposition are
\begin{align}
\begin{split}
\Hat{f}_{\alpha,0} &= f_{\alpha} = 3\cdot 1j_{0}\omega_{\alpha} + N_{0}^{\mu\nu}(\gamma_{\mu\nu}\omega)_{\alpha}\,, \\
\Hat{f}_{\alpha,1} &= 2\bigl( 3\cdot 2j_{1}\omega_{\alpha} + N_{1}^{\mu\nu}(\gamma_{\mu\nu}\omega)_{\alpha}  \bigr)\,, \\
\Hat{f}_{\alpha,2} &= 2\bigl( 3\cdot 3j_{2}\omega_{\alpha} + N_{2}^{\mu\nu}(\gamma_{\mu\nu}\omega)_{\alpha}  \bigr) + 8b^{\mu}b^{\nu}(\gamma_{\mu\nu}\sigma)_{\alpha}\,.
\end{split}
\end{align}
and it is straightforward to check $\Hat{f}_{\alpha}$ satisfies the (classical) master equations
up to level~$1$.
The general form of $\Hat{f}_{\alpha}$ would be
\begin{align}
\Hat{f}_{\alpha}
 &= 2\bigl(3\Hat{J}\omega_{\alpha} + \Hat{N}^{\mu\nu}(\gamma_{\mu\nu}\omega)_{\alpha} \bigr) - f_{\alpha}
  + \sum_{n\ge2}\sum_{k=1}^{n-1} a_{n}B_{k}B_{n-k}C_{n} \,.
\end{align}
where the $BBC$ terms are needed because of the $BCC$ terms in $D
$.

Quantum mechanically, things become more complicated because multiple contractions
do not respect the levels of the classical master equation.
However, one can try to find the quantum improvement terms after
working out the classical expressions and see if it remains in the cohomology.
For $\Hat{f}_{\alpha}$ at hand, the quantum correction should be of the form
\begin{align}
  \label{eq:quantumfalpha}
\Hat{f}_{\alpha}  &= \Hat{f}_{\alpha,\text{cl}} + A\del \omega_{\alpha} \,,
\end{align}
with $A$ being some number.
For the models studied in~\cite{Toymodels},
the states analogous to $\Hat{f}_{\alpha}$ drop out from the
quantum mini-BRST cohomology
while they survive in the \v{C}ech/Dolbeault cohomology,
and this leads to a quantum discrepancy of the two methods
that is invisible from the partition function.
However, for the case of pure spinors,
the partition function indicates the existence of the 
(classical {\em and} quantum) operator carrying the same charges
as $\Hat{f}_{\alpha}$ does.
This then implies that there should be a way to
define $\Hat{f}_{\alpha}$ quantum mechanically
both in curved $\beta\gamma$ and ghost-for-ghost methods.

\bigskip
Of course, the procedure described here applies also to the basic gauge invariant
polynomials.
But for them, it is much easier to guess the correct results
than to systematically work out what they get mapped into.

\paragraph{States in the higher cohomology}
\label{subsubsec:toyhighercohom}

The elements of $D$-cohomology we have been considering up to now
are all living in $H^{0}(D)$,
including $\Hat{f}_{\alpha}$ with negative $\lambda$-charge.
That is, they all correspond to some gauge invariant polynomials studied in section~\ref{sec:gaugeinvs}.
To understand why the higher cohomologies are expected to be non-empty,
let us study the higher cohomology of the mini-BRST operator
considered in~\cite{Toymodels}:
\begin{align}
\label{ddd}
  D &= \int b \lambda^{i}\lambda^{i} \quad (i=1, \ldots , N)\,.
\end{align}
Clearly, the operator $b$ is in the $D$-cohomology.
One also finds $b\omega_{i}$, $b\omega_{\lpar i}\omega_{j\rpar}=b(\omega_{i}\omega_{j}-{1\over N}\delta_{ij}\omega^2)$ etc.
to be in the cohomology and, in general, there is a one-to-one mapping between
the gauge invariant operators (elements of $H^{0}(D)$)
and those with a single $b$ (elements of $H^{1}(D)$).
Roughly speaking, the correspondence is given by
\begin{center}
\begin{tabular}{MMM}
  H^{0}(D) && H^{1}(D) \\ \hline
  f(\omega,\lambda;\del) &\leftrightarrow& b f(\lambda,\omega;\del) \\
  \mathbf{1} & & b \\
  \lambda^{i},\,\lambda^{\lpar i}\lambda^{j \rpar} && b\omega_{i},\,b\omega_{\lpar i}\omega_{j \rpar} \\
   \vdots && \vdots
\end{tabular}
\end{center}
where the above table
indicates that the roles of $\lambda^{i}$ and $\omega_{i}$ are
swapped in $H^{0}(D)$ and $H^{1}(D)$.
The precise correspondence can be established by constructing a inner product
that couples $H^{0}(D)$ and $H^{1}(D)$.
We shall refer to the symmetry between $H^{0}(D)$ and $H^{1}(D)$
as the $\ast$-conjugation symmetry, as in ~\cite{Toymodels}.

The symmetry can be seen at the level of partition function as follows.
The correspondence above (for the toy models) relates the states at
\begin{align}
  q^{m}t^{n}g^{0}\quad\leftrightarrow\quad q^{1+n+m}t^{-2-n}g^{1} \,.
\end{align}
The factor $q^{1}t^{-2}g^{1}$ on the right hand side corresponds to $b$, and
the trade-off ${\lambda}^{i} \leftrightarrow {\omega}_{i}$ corresponds to the switch $t \leftrightarrow q/t$, since ${\omega}_{i}$ have conformal weight one and the $t$-charge opposite of that of ${\lambda}^{i}$.
Hence, the partition function should behave under the $\ast$-conjugation as
\begin{align}
\label{eq:ToyZStarconj}
  Z(q,t) &= -q^{1}t^{-2}Z(q,q/t) \,.
\end{align}
In the next section, we start the study of the partition function of pure spinors.
The partition function will be found to have a covariance property very similar
to~(\ref{eq:ToyZStarconj}).
The only difference is that the prefactor will be $-q^{2}t^{-4}$ instead of $-q^{1}t^{-2}$.
Then, in the coming sections, we shall identify the operator responsible for this factor
(i.e. the
generalization of the operator $b$ in (\ref{ddd}))
as an element of the third cohomology $H^{3}(D)$
(instead of $H^{1}(D)$).

\subsection{Partition function of the mini-BRST cohomology}
\label{sec:partBRST}

In the previous section, we resolved the pure spinor constraint using the infinite chain of
free-field ghosts, and constructed the BRST operator $D$.
Since $D$ carries $t$-charge $0$,
the partition function of its cohomology is equal to that
of the total Hilbert space of (now unconstrained) pure spinors and the ghosts.
Therefore, the full partition function of pure spinors can be formally written as
\begin{align}
\label{eq:ZghFull}
\begin{split}
Z(q,t,\vec{\sigma})
&= \prod_{n\ge0}\Bigl(\prod_{h\ge0}(1-q^{h}t^{n+1})^{-A_{n}}
  \prod_{h\ge1}(1-q^{h}t^{-n-1})^{-\overline{A}_{n}}\Bigr) \\
&= Z_{0}(t,\vec{\sigma})
   \prod_{n\ge0}\prod_{h\ge1}
(1-q^{h}t^{n+1})^{-A_{n}}(1-q^{h}t^{-n-1})^{-\overline{A}_{n}} \,.
\end{split}
\end{align}
In the second line, we factored out the zero-mode contributions
which, by definition, reproduces $Z_{0}(t,\vec{\sigma})$.
$\overline{A}_{k}$ signifies the conjugate representation of $A_{k}$. For example, the chiral and antichiral spinors are conjugate to each other, ${\overline{S_{+}}} = S_{-}$.

It may seem difficult to extract useful information from this formal expression. 
In fact, on the contrary, once the moments of $N_{k}$'s are known,
the two important symmetries of the partition function%
---the field-antifield symmetry and the $\ast$-conjugation symmetry---%
can be easily deduced from~(\ref{eq:ZghFull}).
Also, by expanding in $q$, and employing some analytic continuations, one can obtain from~(\ref{eq:ZghFull}) a well-defined expression
at each Virasoro level.
We shall demonstrate this
by computing the partition function of the first and second mass levels.

\subsubsection{Symmetries of the partition function}
\label{subsubsec:symparts}

Elementary calculations show that $Z(q,t)$ defined in~(\ref{eq:ZghFull}) has the following symmetries
(we drop the spin dependence for simplicity):
\begin{align}
\label{eq:ZghFA}
\begin{split}
\text{field-antifield} \\
\text{symmetry:}
\end{split}
&\begin{split}
Z(q,t)
 &= \prod_{n\ge0}\Bigl( (-1)^{N_{n}} t^{-(n+1)N_{n}} \Bigr) Z(q,1/t) \\
 &= -t^{-8}Z(q,1/t) \,, \\
\end{split}
\\
\label{eq:ZghPF}
\begin{split}
\text{$\ast$-conjugation} \\
\text{symmetry:}
\end{split}
&\begin{split}
Z(q,t)
  &= \prod_{n\ge0} \Bigl( (-1)^{-nN_{n}}q^{-{1\over2}n(n+1)N_{n}}t^{n(n+1)N_{n}} \Bigr)  Z(q,q/t) \\
  &= -q^{2}t^{-4}Z(q,q/t) \,.
\end{split}
\end{align}
From these two symmetries, one also finds
\begin{align}
\label{eq:ZghSF}
\begin{split}
Z(q,t) &= \prod_{n\ge0} \Bigl( (-1)^{(n+1)N_{n}}
    q^{-{1\over2}n(n+1)N_{n}}t^{(n+1)^{2}N_{n}} \Bigr)  Z(q,qt) \\
 &= q^{2}t^{-4}Z(q,qt) \,.
\end{split}
\end{align}
Imposing those symmetries on the formal expression for $Z(q,t)$ means
that one has made an analytical continuation
\begin{align}
  Z(q,t) &= \prod_{n\ge0} (-\mathi q^{-1/12}\eta(q)^{-1}\sqrt{t^{n}}\,\vartheta_{11}(q,t^{n}) )^{-N_{n}} \,,
\end{align}
where the elliptic functions are defined as
\begin{align}
 \vartheta_{11}(q,t) &= \mathi\sum_{n=-\infty}^{\infty}(-1)^{n}q^{{1\over2}(n-1/2)^{2}}t^{n-1/2} \\
   &= -\mathi q^{1/12}\eta(q)(t^{1/2}-t^{-1/2})\prod_{h\ge1}(1-q^{h}t)(1-q^{h}t^{-1}) \,, \\
\eta(q) &= q^{1/24}\prod_{h\ge1}(1-q^{h}) \,.
\end{align}
The symmetries above follow from the well-known
transformation properties of the theta function:
\begin{align}
  \vartheta_{11}(q,t) &= -\vartheta_{11}(q,1/t) = -q^{1/2}t\,\vartheta_{11}(q,qt) \,.
\end{align}

\paragraph{$\ast$-conjugation symmetry and the higher cohomology}

As in the case of the toy models (see section~\ref{subsubsec:toyhighercohom} and \cite{Toymodels}),
the $\ast$-conjugation symmetry suggests that there are non-trivial fermionic elements
in the higher $D$-cohomology.
The element with charges $-q^{2}t^{-4}$ generalizing the state $b$ of the toy models
is of particular importance.
Unfortunately, the construction of this state in the BRST framework is not straightforward,
obstructed by the complexity of the infinite ghosts-for-ghosts.
However, the state has a particularly nice interpretation
in the \v{C}ech/Dolbeault cohomologies.
In fact, it turns out to be nothing but the tail term $b_{3}$ of the composite reparameterization $b$-ghost.
Hence, it carries ghost number $3$,
and we expect from the $\ast$-conjugation symmetry
that there is a one-to-one mapping between $H^{0}(D)$ and $H^{3}(D)$.
We shall come back to this issue in section~\ref{sec:SymPart}.

\paragraph{Remark on the modular property}

Modular properties of the total partition function
$\mathbf{Z}(q,t)=Z_{x,p\theta}(q,t)Z_{\omega\lambda}(q,t)$ of the pure
spinor superstring can in principle be studied using the expression given
here. If one defines
$q=\mathe^{2\pi\mathi \tau}$
and transforms to the cylinder coordinate, one expects the contribution to the
partition function
$\calZ(q,t)=q^{-c_{tot}/24}\mathbf{Z}(q,t)$ of fields and
antifields to be separately invariant under $\tau \to -1/\tau$.
However, to verify this symmetry one first needs to decide how to
separate the contributions of the spacetime fields and
antifields. In section~\ref{sec:lightcone}, we shall argue that the
correct identification involves the lightcone boost charges as well as
the $t$-charge: after twisting the $t$-charge current as
$\Tilde{J} = J_{t} + N^{+-}_{\omega\lambda}+N^{+-}_{p\theta} + 2N^{+-}_{x}$,
all the physical fields and their antifields appear at
$\Tilde{t}^{2}$ and $\Tilde{t}^{6}$ respectively, and their partition
functions
are separately invariant under $\tau\to -1/\tau$.

\subsubsection{Partition functions of non-zero modes}

Our full partition function represents the cohomology of
the mini-BRST operator $D$.
In particular, by expanding in $q$ and $t$
\begin{align}
  Z(q,t) &= \sum_{m\ge0,\,n}N_{m,n}q^{m}t^{n} \,,
\end{align}
one gets the number of cohomology elements with weight $m$ and $t$-charge $n$.
Now, we wish to show that $N_{m,n}$ can be determined inductively
using the two symmetries and the initial data
\begin{align}
 N_{0,n} &=
  \begin{cases}
   0\,, &(n<0)\,, \\ \dim\dynkin(0000n)\,, &(n\ge0) \,.
 \end{cases}
\end{align}
Let us demonstrate this by showing how the level $1$
and $2$ partition functions
\begin{align}
  Z_{1}(t) &= \sum_{n}N_{1,n}t^{n} \,,\quad
  Z_{2}(t) = \sum_{n}N_{2,n}t^{n} \,,
\end{align}
are uniquely obtained.

\paragraph{Level $1$ partition function}

The field-antifield symmetry at level $0$ implies that the polynomial
\begin{align}
\begin{split}
 f_{0}(n) &= {(n+7)(n+6)(n+5)^2(n+4)^2(n+3)^2(n+2)(n+1) \over 7\cdot6\cdot5^2\cdot4^2\cdot3^2\cdot2} \\
 &\bigl(= N_{0,n} \quad\text{for $n\ge0$}\bigr) \,,
\end{split}
\end{align}
possesses the property
\begin{align}
  f_{0}(n) &= f_{0}(-n-8) \,,
\end{align}
which follows from the Serre duality
on the space of projective pure spinors
$$
H^{i}({\cal X}_{10}, {\cal O}(n)) = H^{10-i}
({\cal X}_{10} , K_{{\cal X}_{10}} \otimes {\cal O}(-n) )^{*} \,,
$$
and the relation 
$$
K_{{\cal X}_{10}} \approx {\cal O}(8) \,.
$$
Then, from the expression~(\ref{eq:ZghFull}) for the full partition function,
one finds
\begin{align}
\begin{split}
Z_{1}(t) &= {\del \over \del q}Z(q,t)|_{q=0} \\
 &= Z_{0}(t)
    \sum_{k\ge0}(t^{k+1}+t^{-k-1})N_{k}  \\
 &= \sum_{k\ge0}\bigl( \sum_{n\ge k+1} f_0(n-k-1)N_{k}t^{n} + \sum_{n\ge-k-1}f_0(n+k+1)N_{k}t^{n} \bigr) \,.
\end{split}
\end{align}
This expression is not quite well defined as it contains infinite series
both in $t$ and $1/t$.
Since what we wish to have is the series in $t$,
an analytic continuation must be performed to throw away the series in $1/t$.
We perform the analytic continuation in a manner
such that the coefficients of $t^{n}$ in
\begin{align}
  \label{eq:Z1sum}
  Z_{1}(t) &= \sum_{n}f_{1}(n)t^{n}\,,
\end{align}
respect the field-antifield symmetry.
As will be shown shortly, the correct prescription is to simply
discard the $1/t^{2}$ and higher poles in $t$:
\begin{align}
 Z_{1}(t)
  &= \sum_{n\ge-1}t^{n}\sum_{k\ge0}\bigl(f_{0}(n-k-1)+f_{0}(n+k+1) \bigr) N_{k} \,.
\end{align}

Now, since $f_0(k)$ is a polynomial of order $10$, the sum over $k$
in the last expression can be performed
if one knows the (even) moments of $N_{k}$ up to $\sum_{k}(k+1)^{10}N_{k}$.
We have already listed those moments in~(\ref{eq:Nmoments})
and the result of the sum is a geometric series of the form~(\ref{eq:Z1sum}) with
\begin{align}
f_{1}(n) &=
{(n+1) (n+2) (n+3) (n+4)^2 (n+5) (n+6) (n+7) (11 n^2+88 n+345) \over 2^{5}\cdot 3^3\cdot 5^2\cdot7} \,.
\end{align}
The sum over $n$ can then be readily done and therefore the
partition function (which takes into account the number of states)
at level $1$ is:
\begin{align}
\begin{split}
Z_{1}(t) &= \sum_{n\ge0}f_{1}(n)t^{n} \\
   &= {46 - 144t + 116t^2 + 16t^3 - 16t^5 - 116t^6 + 144t^7 - 46t^8 \over (1-t)^{16} } \,,
 \end{split}
\\
N_{1,n} &=
  \begin{cases}
    0\,, & (n<0)\,, \\
    f_{1}(n)\,, & (n\ge0)\,.
  \end{cases}
\end{align}
Note that $N_{0,n}$ and $N_{1,n}$ are consistent both with
the field-antifield and the $\ast$-conjugation symmetries.
Also, the form of $f_{1}(n)$ is consistent with our earlier result:
\begin{align}
 f_{1}(n)
 &=
 \begin{cases}
  \dim\bigl(\dynkin(0100n)+\dynkin(0000n)\bigr)\,, & (n=0)\,, \\
  \dim\bigl(\dynkin(0100n)+2(0000n)\bigr)\,, & (n=1)\,, \\
  \dim\bigl(\dynkin(0100n)+2\dynkin(0000n)+(0010,n-2)\bigr)\,, & (n\ge2)\,.
\end{cases}
\end{align}

In general, there is no reason why the result obtained using the prescription above
also respects the $\ast$-conjugation symmetry.
That is, some (finite) number of states might be missing,
which are implied by the lower level partition function and the $\ast$-conjugation symmetry:
\begin{align}
  N_{m,n} &= N_{m+n+2,-n-4} \,.
\end{align}
This will be the case for level $2$ and higher.
For these cases, we simply add the missing pieces together with
their antifields so that the number of states is consistent
with both symmetries.

\paragraph{Level $2$ partition function}

The computation of the weight $2$ partition function $Z_{2}$ can be
done in a similar manner.
That is, it can be determined from the two symmetries
and the lower dimensional ones $Z_{0}$ and $Z_{1}$.
First,
\begin{align}
\begin{split}
Z_{2}(t) &= {1\over2}{\del^{2} \over \del q^{2}}Z(q,t)|_{q=0} \\
 &= Z_{0}(t)\Bigl\{\sum_{n\ge1}N_{n}(t^{n+1}+t^{-n-1})
  +{1\over2}\bigl(\sum_{n\ge0} N_{n}(t^{n+1}+t^{-n-1}) \bigr)^{2} \\
&\qquad\qquad +{1\over2}\sum_{n\ge0}N_{n}(t^{2(n+1)}+t^{-2(n+1)})\Bigr\} \\
&=  Z_{1}(t) + {1\over2}Z_{1}(t)\sum_{n\ge0}N_{n}(t^{n+1}+t^{-n-1})
+ {1\over2}Z_{0}(t)\sum_{n\ge0}N_{n}(t^{2(n+1)}+t^{-2(n+1)}) \,.
\end{split}
\end{align}
Again, by an appropriate analytic continuation,
the
series in $1/t$ can be discarded in such a way that the coefficient $f_{2}(t)$ in
\begin{align}
 \label{eq:Z2naive}
  Z_{2}(t)&=\sum_{n\ge-1}f_2(n)t^{n} \,
\end{align}
contains the field-antifield symmetry.
As before, this can be achieved by simply discarding $1/t^{2}$ and higher poles
giving the result
\begin{align}
\begin{split}
  f_{2}(n)
&= {1\over 2^6\cdot 3^3 \cdot 5 \cdot 7}
   { (2 + n)(3 + n)(5 + n)(6 + n)}\\
&\qquad \times {(360528 + 580664n + 321543n^2 + 90400n^3 + 14450n^4 +
   1320n^5 + 55n^6)} \,.
\end{split}
\end{align}

One can check that the polynomial $f_{2}(n)$ obtained here coincides
with our earlier result obtained by counting the number of gauge invariant polynomials, i.e.
$\sum_{n\ge-1}f_{2}(n)t^{n} = Z_{2,\text{poly}}(t)$.
This means that,
even though $f_{2}$ is consistent with the field-antifield symmetry, $f_{2}(n)  =  f_{2}(-n-8)$,
the result after the summation of the geometric series~(\ref{eq:Z2naive}) is not.
More concretely, one finds
\begin{align}
  Z_{2,\text{poly}}(t) + t^{-8}Z_{2,\text{poly}}(1/t) = 2/t^{4}\,.
\end{align}
Also, the definition~(\ref{eq:Z2naive}) is inconsistent with the
$\ast$-conjugation symmetry because
\begin{align}
  N_{2,-4}\overset{?}{=} 0 \quad\ne\quad -N_{0,0}=-1\,.
\end{align}
As expected from the general analytic continuation of the full
partition function $Z(q,t)$, those two failures are related.
Indeed, both can be remedied at the same time by adding the
term\footnote{Note that this is equal to ${1\over 2} \times f_{2}(-4)t^{-4}$} $-{1\over t^4}$ to (\ref{eq:Z2naive}) so that
\begin{align}
 \label{eq:Z2correct}
  Z_{2}(t)&=-{1\over t^{4}} + \sum_{n\ge-1}f_2(n)t^{n} \nonumber\\
  &= {1\over (1-t)^{16}}\bigl\{-t^{-4} + 16t^{-3} - 120t^{-2} + 576t^{-1}  -1003  + 528t - 214t^2 + 592t^3 \nonumber\\
&\qquad - 592t^5 + 214t^6 - 528t^7 + 1003t^8 - 576t^9 + 120t^{10} - 16t^{11} + t^{12}\bigr\} \,.
\end{align}
It is easy to see the number of states implied by the modified
partition function~(\ref{eq:Z2correct}),
\begin{align}
  N_{2,n} &= \begin{cases}
    0\,, & (n<-4)\,, \\
   -1\,, & (n=-4)\,, \\
   f_{2}(n)\,, &(n>-4) \,,
  \end{cases}
\end{align}
is now consistent with both field-antifield symmetry and $\ast$-conjugation symmetry.

\subsection{Fixed point formulas for the full partition function}
\label{sec:partFP}

In the previous section, we presented a formula
for the full partition function of pure spinors using an infinite tower of ghosts-for-ghosts.
The formula is natural and convenient for motivating the two important
symmetries of the partition function,
i.e. the field-antifield symmetry and the $\ast$-conjugation symmetry.
Also, we were able to compute the partition function level by level,
respecting those two symmetries.
However, the computation using (\ref{eq:ZghFull}) is not easy
if one wishes to keep the spin information of the states.
We here present a very simple fixed point formula for the
partition function including the spin character,
extending the zero mode formula given in~\cite{Berkovits:2005hy}.
Although our formulas~(\ref{eq:ZFPfull1}) and~(\ref{eq:ZFPfull2}) miss some finite number of states
at each Virasoro level, these missing states
can be recovered by imposing the two symmetries as in the case of
the ghost-for-ghost method.

It may seem troubling that the fixed point formulae we present below ``miss''
some states, but in fact this is not so hard to explain. 

\subsubsection{Some remarks on the fixed point formulae}

A classic example of the fixed point formula is Weyl formula for the character ${\chi}_{\mu}$ of an irreducible representation of a simple Lie group $G$:
\begin{equation}
{\chi}_{\mu} (g) = {\Tr}_{R_{\mu}} \left( g \right) \,, \ g \in G \,.
\label{char}
\end{equation} 
Such a representation $R_{\mu}$ can be realized by geometric quantization of a coadjoint orbit
${\calO}_{\mu} \approx G/T$, 
\begin{equation}
R_{\mu} = H^{0} \left( {\calO}_{\mu}, L_{\mu} \right)
\label{coh}
\end{equation}
where $L_{\mu}$ is a holomorphic line bundle over ${\calO}_{\mu}$ whose first Chern class $c_{1}\left( L_{\mu} \right)$ coincides with the symplectic form on ${\calO}_{\mu}$, determined, in turn, by the dominant weight ${\mu} \in P_{+}$. The character can be written as a sum over the fixed points $w$ of the action of the element $g \in G$ on ${\calO}_{\mu}$,
each point contributing essentially
a character of the Hilbert space
obtained by quantizing the tangent space $T_{w}{\calO}_{\mu}$:
\begin{equation}
{\chi}_{\mu}(g) = \sum_{w} 
{e_{\mu, w}(g) \over {\rm det}_{T_{w}{\calO}_{\mu}} \left( 1 - g \right)}
\label{charw}
\end{equation}
where $e_{\mu , w}(g) \in {\mathbb{C}}^{*}$ is the eigenvalue of the action of $g$ on the fiber of the line bundle $L_{\mu}\vert_{w}$ over $w$. 

The formula~(\ref{charw}) can be also understood as follows. 
Let us cover ${\calO}_{\mu}$ by the open neighbourhoods  $U_{w}$ of the fixed points $w$, which 
are invariant with respect to the action of the maximal torus $T_{g} \subset G$ containing $g$. 
Then the intersections $U_{w_{1}} \cap U_{w_{2}} \cap \cdots \cap U_{w_{k}}$
are also $T_{g}$-invariant. The character, being additive, can be written as follows:
\begin{align}
{\Tr}_{H^{0}} (g) &= \sum_{i} (-1)^{i}{\Tr}_{H^{i}}(g) \qquad \text{\small (positivity of $\mu$)} \\
\label{inclusion}
\begin{split}
 &= \sum_{w} {\Tr}_{U_{w}}(g) 
  - \sum_{w_{1} < w_{2}} {\Tr}_{U_{w_{1}}\cap U_{{w_{2}}}}(g) + \cdots \\
 &\qquad + (-1)^{k-1} \sum_{w_{1} < w_{2} < \cdots < w_{k}} {\Tr}_{U_{w_{1}}\cap U_{w_{2}}\cdots \cap U_{w_{k}}}(g) + \cdots \,.
\end{split}\text{\small (inclusion-exclusion)}
\end{align}
On the one hand, the $l$-fold intersections $U_{w_{1}} \cap U_{w_{2}} \cap \ldots \cap U_{w_{l}}$
contain $\left( {\mathbb{C}}^{*} \right)^{\times l-1}$, and, therefore, contribute zero to the character, using the definition
\begin{equation}
\sum_{p\in {\bf Z}} z^{p} \sim 0\ , \ 
z\neq 1 \,.
\label{vanishch}
\end{equation}
On the other hand, set-theoretically, the exclusions-inclusions of the sets
$U_{w}$ and their intersections,
as in~(\ref{charw}):
\begin{align}
&
\left( \cup_{w} U_{w} \right) \backslash \left( \cup_{w_{1} < w_{2}} U_{w_{1}} \cap U_{w_{2}} \right) \cup \left( \cup_{w_{1} < w_{2}< w_{3}} U_{w_{1}} \cap U_{w_{2}} \cap U_{w_{3}} \right) \cdots = 
 \cup_{w} \{ w \} 
\label{wpoints}
\end{align}
i.e. we are left with the fixed points only, or, rather, their infinitesimal neighbourhoods. 

These well-known considerations 
\cite{Pressley:1988qk} stumble immediately once we replace the compact orbit ${\calO}_{\mu}$ by the space $X_{10}$ of pure spinors. As we stated many times before
the space $X_{10}$ is the total
space of the ${\mathbb{C}}^{*}$-bundle over
the space ${\calX}_{10}$ of projective
pure spinors. Thus, the group $SO(10) \times U(1)$ acts on $X_{10}$
without fixed points. Therefore
the character must vanish, for the simple reason~(\ref{vanishch}). 

In reality, however, we manage to get a non-vanishing character, both the zero modes~\cite{Berkovits:2005hy} and the non-zero ones, as we shall do below. 

The resolution of the apparent paradox is the fact that in a sense the total character, taking into account both positive and negative
degrees of ${\lambda}_{+}$, counts both fields and anti-fields of the spacetime theory, which cancel each other.

The creative part of the computation is to separate the vanishing character into the two 
opposite contributions, one coming
from the fields, the other coming from the antifields. 
Picking up the contributions from the fields, one obtains a non-vanishing character.
Moreover, after coupling to the fermionic matter sector $(p_{\alpha},\theta^{\alpha})$,
one can introduce another definition of the fields and antifields
using the norm $\langle \lambda^{3} \theta^{5}\rangle = 1$.
Of course, this new definition differs from the one expected
from the pure spinor sector alone, but it has been 
customarily used in the pure spinor formalism (and in the present paper), 
and is known to lead to the correct spacetime amplitudes.

Thus, the true fields and anti-fields in the pure spinor superstring
are constructed only after including the contributions of the
$(p_{\alpha}, {\theta}^{\alpha})$ and $x^{\mu}$ sector. Since these
contributions also carry the $t$-charge (and there are different
ways to attribute the $t$-charge
to the $(p_{\alpha}, {\theta}^{\alpha}, x^{\mu})$-fields, to be explored below), the total $t$-charge of both fields and anti-fields
may become both positive
and negative. For example, at the
level of zero modes we only have
positive $t$-charge states.

\subsubsection{Review of fixed point formula for level $0$ partition function}

For convenience, we briefly review the fixed point formula
for the zero mode partition function~\cite{Berkovits:2005hy}.
(See also~\cite{Nekrasov:2005wg}.)

\paragraph{Geometric preliminary}

Let us begin by refining our description of the space of pure spinors
\begin{align}
\begin{split}
  X_{10} &= \{ \lambda^{\alpha} \;|\: \lambda^{\alpha}\gamma^{\mu}_{\alpha\beta}\lambda^{\beta} = 0\,,\;\lambda\ne0 \}  \\
   &= (\text{$\mathbb{C}^{\ast}$-bundle over $\calX_{10}$})\,, \quad\bigl( \calX_{10}=SO(10)/U(5) \bigr)\,.
\end{split}
\end{align}
With the removal of the origin understood, $X_{10}$ can be covered by 16 patches,
where in each patch at least one component of $\lambda$ is non-vanishing.
It is convenient to use the ``five sign'' notation to
describe the components of $\lambda$.
(See section~\ref{subsec:defpartfn} and appendix~\ref{app:Dynkin} for explanations.)
In this notation, the character of $16$ components are
\begin{align}
\lambda_{\pm\pm\pm\pm\pm} = \mathe^{{1\over2}(\pm \sigma_{1}\pm \sigma_{2}\pm \sigma_{3}\pm \sigma_{4}\pm \sigma_{5})}\quad\text{\small(even number of $-$'s)}\,,
\end{align}
and $X_{10}$ can be covered by $16$ patches
\begin{align}
 U_{\pm\pm\pm\pm\pm} &= \{\lambda\in X_{10} \;|\; \lambda_{\pm\pm\pm\pm\pm}\ne0 \}\,.
\end{align}
In a given patch, a pure spinor can be parameterized using eleven parameters $(g,u_{ab})$
which are in the $(\mathbf{1},\mathbf{10})$ of $U(5)$.
$u_{ab}$ is the ``angular'' coordinate parameterizing the base $\calX_{10}$
and $g$ is the coordinate for the fiber $\mathbb{C}^{\ast}$.
For example in $U_{+++++}$,
\begin{align}
\begin{split}
  \lambda &= (\lambda_{+},\lambda_{ab},\lambda^{a})
  = (g,gu_{ab},{1\over8}g\epsilon^{abcde}u_{bc}u_{de})\,.
\end{split}
\end{align}
where
\begin{align}
\begin{split}
  \lambda_{+} &= \lambda_{+++++} \ne 0\,, \\
  \lambda_{ab} &= \{ \lambda_{+++--} \text{ and permutations } \}\,, \\
  \lambda^{a} &= \{ \lambda_{+----} \text{ and permutations } \}\,.
\end{split}
\end{align}
The characters of $g$ and $u_{ab}$ in this patch are
\begin{align}
  g = \mathe^{{1\over2}(\sigma_{1}+\sigma_{2}+\sigma_{3}+\sigma_{4}+\sigma_{5})}\,,\quad
  u_{ab} = \mathe^{-(\sigma_{a}+\sigma_{b})}\,,\;(1\le a < b \le 5)\,.
\end{align}
In other patches $U_{\epsilon(+++++)}$, the characters will be
\begin{align}
  g_{\epsilon} = \mathe^{{1\over2}\epsilon(\sigma_{1}+\sigma_{2}+\sigma_{3}+\sigma_{4}+\sigma_{5})} = \mathe^{{1\over2}\epsilon\cdot \sigma}\,,\quad
  u_{\epsilon,ab} = \mathe^{-\epsilon(\sigma_{a}+\sigma_{b})} = \mathe^{-(\epsilon_{a}\sigma_{a}+\epsilon_{b}\sigma_{b})}\,,
\end{align}
where $\epsilon$ acts by even number of sign changes.

\paragraph{The fixed point formula}

By constructing the symmetry generators explicitly,
one finds the action of $N_{\mu\nu}$ on $X_{10}$ which commutes
with the $J$-rescaling of the $\mathbb{C}^{\ast}$-fiber.
A generic action of the maximal torus of $SO(10)$ has $16$ fixed points
which are nothing but the ``origins'' ($u_{ab}=0$) of $16$ patches.
The spin character of pure spinors can then be written as a sum of
the contributions at the fixed points~\cite{Nekrasov:2005wg}\cite{Pressley:1988qk}
\begin{align}
\label{eq:ZFP0}
Z_{0}(t,\vec{\sigma})
&= \sum_{\epsilon=1}^{16}{1 \over 1-t\mathe^{{1\over2}\epsilon\cdot \sigma}}
  \prod_{(ab)=1}^{10}{1\over 1-\mathe^{-(\epsilon_{a}\sigma_{a}+\epsilon_{b}\sigma_{b})} } \,
\end{align}
where we use the notation of~\cite{Berkovits:2005hy}.
The sum over $\epsilon$ describes the sum over $16$ fixed points.
The first term of the summand is the character of the non-vanishing component $g_{\epsilon}$,
and the second term is the character of the rest $u_{\epsilon,ab}$,
both at a given fixed point $\epsilon$.
Summation over the fixed points in~(\ref{eq:ZFP0}) is
straightforward and one gets~\cite{Berkovits:2005hy}
\begin{align}
Z_{0}(t,\vec{\sigma}) &=
 {\mathbf{1} - \mathbf{10}t^{2} + \overline{\mathbf{16}}t^{3}
  - \mathbf{16}t^{5} + \mathbf{10}t^{6} - \mathbf{1}t^{8} \over (1-t)^{\mathbf{16}} } \,.
\end{align}

At this point we can make more
explicit the somewhat  abstract
discussion at the end of the
previous section. 

Had we not known that each coordinate patch $U_{A}$ should only contribute the positive
powers of $t$, i.e. the functions
which are regular at ${\lambda}_{+} = 0$, despite the fact that the 
${\lambda}_{+}$ is not allowed to vanish, the full character would have been zero. Indeed, instead
of the 
\begin{align}
{1 \over 1-t\mathe^{{1\over2}\epsilon\cdot \sigma}}
\end{align}
factor in ~(\ref{eq:ZFP0}) we would have had
\begin{align}
{1 \over 1-t\mathe^{{1\over2}\epsilon\cdot \sigma}}
 + 
{t^{-1}\mathe^{-{1\over2}\epsilon\cdot \sigma} \over 1-t^{-1}\mathe^{-{1\over2}\epsilon\cdot \sigma}} = 0 \,.
\end{align}
It is thus with this experience that we approach the full partition function. We shall take the following
point of view. We shall trust the
fixed point formula for sufficiently
large positive $t$-charge and then
use the $*$-conjugation and field-antifield symmetries to fix the rest. The result will be identical to
what we had before using ghosts-for-ghosts.

\subsubsection{Fixed point formulas for the full partition function}

Now, let us introduce two ways to extend the zero-mode fixed point formula~(\ref{eq:ZFP0}).
The two utilize different parameterizations for the non-zero modes
and lead to partition functions that differ by a
finite number of terms at each level.
Also, both miss a finite number of terms with respect to the fully symmetric partition function.
However, the missing states can be unambiguously recovered by imposing
the field-antifield and $\ast$-conjugation symmetries,
and the two formulas then give the same symmetric partition function.

The first way of extending is to simply include the non-zero modes of $\lambda$
at each patch $(g_{\epsilon},u_{\epsilon,ab})\in U_{\epsilon}$
together with the modes of their conjugates $(h_{\epsilon},v^{ab}_{\epsilon})$:
\begin{align}
\label{eq:ZFP}
Z(q,t,\vec{\sigma})
&= \sum_{\epsilon=1}^{16}Z_{\epsilon}(q,t,\vec{\sigma})\,, \\
\label{eq:ZFPfull1}
\begin{split}
Z_{\epsilon}(q,t,\vec{\sigma})
&=  \prod_{h\ge0}
   {1\over 1-q^{h}t\mathe^{{1\over2}\epsilon\cdot \sigma }}
   \prod_{(ab)=1}^{10}{1\over 1-q^{h}\mathe^{-(\epsilon_{a}\sigma_{a}+\epsilon_{b}\sigma_{b})} }  \\
 &\qquad
   \times \prod_{h\ge1}
   {1\over 1-q^{h}t^{-1}\mathe^{-{1\over2}\epsilon\cdot \sigma } }
   \prod_{(ab)=1}^{10}{1\over 1-q^{h}\mathe^{\epsilon_{a}\sigma_{a}+\epsilon_{b}\sigma_{b}} } \,.
 \end{split}
\end{align}
The first line represents the modes of $(g,u_{ab})$
and the second line represents the modes of $(h,v^{ab})$.

To obtain another way of parameterizing the non-zero modes,
one observes that the constraints for the non-zero modes are
essentially linear $\lambda_{0}\gamma^{\mu}\lambda_{-h}+\cdots =0$
($\lambda_{-h}\sim \del^{h}\lambda$) while the
constraint for the zero mode is quadratic $\lambda_{0}\gamma^{\mu}\lambda_{0}=0$.
Therefore, the $11$ components of non-zero modes $\lambda_{-h}$
(and their conjugates) can be thought as carrying different characters from the zero-mode $\lambda_{0}$,
and the contribution from a fixed point is
\begin{align}
\label{eq:ZFPfull2}
\begin{split}
Z_{\epsilon}(q,t,\vec{\sigma})
&= {1 \over 1-t\mathe^{{1\over2}\epsilon\cdot \sigma}}
  \prod_{(ab)=1}^{10}{1\over 1-\mathe^{-(\epsilon_{a}\sigma_{a}+\epsilon_{b}\sigma_{b})} } \\
 &\qquad
   \times \prod_{h\ge1}
   {1\over 1-q^{h}t\mathe^{{1\over2}\epsilon\cdot \sigma }}
   \prod_{(ab)=1}^{10}{1\over 1-q^{h}t\mathe^{{1\over2}\epsilon\cdot \sigma -(\epsilon_{a}\sigma_{a}+\epsilon_{b}\sigma_{b})} }  \\
 &\qquad
   \times \prod_{h\ge1}
   {1\over 1-q^{h}t^{-1}\mathe^{-{1\over2}\epsilon\cdot \sigma } }
   \prod_{(ab)=1}^{10}{1\over 1-q^{h}t^{-1}\mathe^{-{1\over2}\epsilon\cdot \sigma+(\epsilon_{a}\sigma_{a}+\epsilon_{b}\sigma_{b})} } \,.
 \end{split}
\end{align}
The second line describes the contributions of the $\lambda$ non-zero modes
and the third line describes the contributions of the $\omega$ non-zero modes.
As mentioned above, it carries essentially the same information
as the first formula~(\ref{eq:ZFPfull1}).

By expanding either~(\ref{eq:ZFPfull1}) or~(\ref{eq:ZFPfull2}) in $q$, the
level $h$ partition function with spin information
is expressed in a simple form for all $h\ge1$.
The summation over $16$ fixed points is straightforward,
and one gets a result of the form
\begin{align}
 \label{eq:ZhSpinresult}
  Z_{h}(t,\vec{\sigma}) &= {P'_{h}(t,\vec{\sigma}) \over (1-t)^{\mathbf{16}} } \,,
\end{align}
where $P'_{h}(t,\vec{\sigma})$ is some polynomial in $t$ with coefficients
taking values in the representations of $SO(10)$,
and $(1-t)^{\mathbf{16}}=\prod_{\mu\in\mathbf{16}}(1-t\mathe^{\mu\cdot \sigma})$.
We put a prime on $P'_{h}(t,\vec{\sigma})$ as it lacks field-antifield symmetry
as of yet:
\begin{align}
  P'_{h}(t,\vec{\sigma}) \ne  P'_{h}(1/t,-\vec{\sigma}) \,.
\end{align}
We now turn to our results on $P'_{h}(t,\vec{\sigma})$
and explain how to improve them so that they respect the
field-antifield and $\ast$-conjugation symmetries.

\subsubsection{Partition functions for non-zero modes with spin character}

Although the summation over $16$ fixed points is straightforward,
it is not obvious how
to combine local $U(5)$ characters into $SO(10)$ characters in a simple
manner.
A convenient computational trick is to utilize the Weyl character formula
to take care of the combinatorics.
To do this, one first augments the factor for the
zero-mode character $\prod_{(ab)=1}^{10}(1-\mathe^{-(\epsilon_a\sigma_a+\epsilon_b\sigma_b)})^{-1}$
representing the $10$ ``positive roots of $SO(10)/U(5)$''
by the character of the remaining $10$ positive roots of $SO(10)$,
i.e. those of $U(5)$, $\prod_{(ab)=1}^{10}(1-\mathe^{-(\epsilon_a\sigma_a-\epsilon_b\sigma_b)})^{-1}$,
and then extends the summation over
the $16$ fixed points $\epsilon$ to $1920$ elements of the $SO(10)$ Weyl group $W$.
Using the first parameterization of~(\ref{eq:ZFPfull1}),
$1920$ ``local'' contributions are given by
\begin{align}
\begin{split}
Z_{w}(q,t,\vec{\sigma})
&= {1 \over 1-t\mathe^{{1\over2}w\cdot \sigma}}
  \prod_{(ab)=1}^{10}{1\over (1-\mathe^{-(w_{a}\sigma_{a}+w_{b}\sigma_{b})}) (1-\mathe^{-(w_{a}\sigma_{a}-w_{b}\sigma_{b})}) } \\
&\qquad
 \times \prod_{h\ge1}
   {1\over 1-q^{h}t\mathe^{{1\over2}w\cdot \sigma }}
   \prod_{(ab)=1}^{10}{1\over 1-q^{h}\mathe^{-(w_{a}\sigma_{a}+w_{b}\sigma_{b})} }  \\
 &\qquad
   \times \prod_{h\ge1}
   {1\over 1-q^{h}t^{-1}\mathe^{-{1\over2}w\cdot \sigma } }
   \prod_{(ab)=1}^{10}{1\over 1-q^{h}\mathe^{w_{a}\sigma_{a}+w_{b}\sigma_{b}} } \,.
\end{split}
\end{align}
(Using the second parameterization of~(\ref{eq:ZFPfull2}), the
formula is the same
except for the last two lines representing non-zero modes.)
An element $w\in W$ acts on the five-sign basis by permutations and an even number of sign changes.
The two modifications ``cancel'' each other and simply gives $\sum_{w}Z_{w}=\sum_{\epsilon}Z_{\epsilon}$.

Now multiplying $\mathe^{w\cdot \rho}$ (where $\rho$ is the half sum of positive roots)
to both the numerator and denominator of $Z_{w}(q,t,\vec{\sigma})$,
and denoting the $SO(10)$ Weyl denominator by $\mathe^{\rho}R$,
the sum over $w$ reads
\begin{align}
\begin{split}
Z(q,t,\vec{\sigma})
= \sum_{w}^{1920}Z_{w}(q,t,\vec{\sigma})
&= {1\over (1- t)^{\mathbf{16}}}{1\over \mathe^{\rho}R}
 \sum_{w} (-1)^{w}\mathe^{w\rho}\prod_{\epsilon\ne1}^{15}(1- t\mathe^{{1\over2}w(\epsilon\cdot\sigma)}) \\
&\qquad \times\prod_{h\ge1}\bigl\{ \text{non-zero modes} \bigr\} \,.
\end{split}
\end{align}
Using the Weyl character formula, the summation over $w\in W$ is readily done
leading to the expressions of the form~(\ref{eq:ZhSpinresult}).
This trick also explains why one gets $SO(10)$ representations as the coefficients of $t$.

\paragraph{Level $1$}

Using the computational trick just mentioned at this level,
the second parameterization of~(\ref{eq:ZFPfull2}) yields
\begin{align}
\label{eq:Z1spinsecond}
\begin{split}
Z_{1,\text{2nd}}(t,\vec{\sigma})
 = {1\over (1-t)^{\mathbf{16}}}
 &\bigl\{
  (\mathbf{45}+\mathbf{1})t^{0}
  - \mathbf{\overline{144}}t^{1}
  + (\mathbf{126}^{-}-\mathbf{10})t^{2}
  + \mathbf{\overline{16}}t^{3} \\
&\qquad
  - \mathbf{16}t^{5}
  - (\mathbf{126}^{+}-\mathbf{10})t^{6}
  + \mathbf{144}t^{7}
  -(\mathbf{45}+\mathbf{1})t^{8}
 \bigr\} \,,
\end{split}
\end{align}
while the first parameterization~(\ref{eq:ZFPfull1}) yields
\begin{align}
  Z_{1,\text{1st}} = Z_{1,\text{2nd}} - \mathbf{1} \,.
\end{align}
The singlet missing from $Z_{1,\text{1st}}$ is the gauge invariant current $J=-\omega\lambda$,
and the only way to make $Z_{1,\text{1st}}$ consistent with
field-antifield symmetry and $\ast$-conjugation symmetry up to this level
is to add $\mathbf{1}$ to it.
So we conclude
\begin{align}
 Z_{1}(t,\vec{\sigma}) = Z_{1,\text{2nd}}(t,\vec{\sigma}) =
 Z_{1,\text{1st}}(t,\vec{\sigma}) + \mathbf{1}
 \,,
\end{align}
where $Z_{1,\text{2nd}}$ obtained from our second parameterization
is defined in~(\ref{eq:Z1spinsecond}).

\paragraph{Higher levels}

An important point to notice
is that although $Z_{\text{2nd}}$ reproduces the fully symmetric
partition function at level 1, neither $Z_{\text{1st}}$ nor
$Z_{\text{2nd}}$ reproduce
the fully symmetric partition function at higher levels.
In particular, they both miss the fermionic singlet at $-q^{2}t^{-4}$ discussed above,
and (a part of) analogous states at higher levels.
Also, both $Z_{\text{1st}}$ and
$Z_{\text{2nd}}$ miss some gauge-invariant operators.
For example, at level $3$, the numerator $P'_{3}(t,\vec{\sigma})$ in
$Z_{\text{2nd}}$
starts as
\begin{align}
  P'_{3}(t,\vec{\sigma}) &= -\mathbf{10}t^{-2} + (\mathbf{144}+\mathbf{560}+3\cdot\overline{\mathbf{16}})t^{-1} + \cdots\,,
\end{align}
and the correction required includes both bosonic and fermionic operators.
Nevertheless, at least up to the fifth level,
the difference between the $t$-expansions of the fully symmetric partition function
and the result from the fixed point formulas is always finite.
Therefore, the fixed point result can be unambiguously improved to the symmetric one
using the method described for the ghost-for-ghost partition function.
(A list of the improved numerator $P_{h}(t,\vec{\sigma})$ can be found in appendix~\ref{app:spincharacters}.)

\paragraph{Towards fixed point formula for the fully symmetric partition function}

Since we use the field-antifield and $\ast$-conjugation symmetries
as guiding principles for computing the complete partition function,
it will be useful to build them into the fixed point formula itself.
Although we do not have an answer to this problem at the present time,
organizing the complete partition function into 
a character of $\widehat{SO}(10)$ affine Lie algebra seems to be
promising. Note that it is probably not going to work for the pure spinor sector only, since the level of the $SO(10)$ current algebra is negative. Together with the $(p_{\alpha}, {\theta}^{\alpha})$ contribution one can expect a reasonable expression. 

This should also be useful for extending our result to
all mass levels.
However, we leave the study of these issues for future research,
and we now turn to an explanation of 
the symmetries of the partition function.

%%%%%%%%%%%%%%%%%%%%%%%%%%%%%%%%%%%%%%%%%%%%%%%%%%%%%%%%%%%%%%%%
\section{Structure of pure spinor cohomology}
\label{sec:SymPart}

In this section, we explain the structure of the Hilbert space of the pure spinor system.
In particular, we will give a ``microscopic'' explanation of the states
which do not correspond to the usual gauge invariant polynomials.
As mentioned repeatedly, those ``missing'' states carry ghost number $3$
and are essential for the partition function to have the symmetries
\begin{align}
 \label{eq:Zsymsagain}
  Z(q,t) &= -t^{-8}Z(q,1/t)  = -q^{2}t^{-4}Z(q,q/t) \,.
\end{align}
Since the structure of the pure spinor cohomology is surprisingly similar
to that of simpler models analyzed in~\cite{Toymodels},
we will first briefly review the result obtained there.

\subsection{A brief summary of the toy models}

\subsubsection{Partition function and its symmetry}

In~\cite{Toymodels}, the curved $\beta\gamma$-systems
with a single quadratic constraint
\begin{align}
  \lambda^{i}\lambda^{i} = 0 \,,\quad (i=1, \ldots ,  N)\,.
\end{align}
are analyzed in detail. 
Unlike the pure spinor constraint, the constraint $\lambda\lambda=0$ is irreducible for $N\ge2$.
Therefore, the BRST approach is very effective in this case.
Only a single pair of fermionic ghosts $(b,c)$ has to be introduced
and the mini-BRST operator is given by
\begin{align}
  D &= \int b\lambda\lambda \,.
\end{align}
The partition function of the $D$-cohomology is hence
\begin{align}
 Z(q,t) = {1-t^2\over (1-t)^{N}} \prod_{h\ge1}{(1-q^{h}t^{2})(1-q^{h}t^{-2}) \over (1-q^{h}t)^{N}(1-q^{h}t^{-1})^{N}}\,,
\end{align}
where the charges of the fields are defined as
\begin{align}
  h(b,c,\omega_i,\lambda^{i}) = (1,0,1,0) \,,\quad t(b,c,\omega_{i},\lambda^{i}) = (-2,2,-1,1) \,.
\end{align}
Clearly, $Z(q,t)$ satisfies the symmetries analogous to~(\ref{eq:Zsymsagain}):
\begin{align}
\text{field-antifield symmetry}\colon&\quad
 Z(q,t) = -(-t)^{2-N}Z(q,1/t)\,, \\
\text{``$\ast$-conjugation'' symmetry}\colon&\quad
 Z(q,t)= -q^{1}t^{-2}Z(q,q/t) \,.
\end{align}
Note that these symmetries naturally appear in pairs,
as they are related to the double periodicity of the theta function.
(In other words, failure of one type implies the failure of the other.)

Let us now turn to explain the ``microscopic'' origin of the symmetries.
Both symmetries reflect certain discrete symmetries of the cohomology
and can be understood in terms of inner products that pair the elements of cohomologies.
(In particular, the inner product for the $\ast$-conjugation
pairs $H^{0}(D)$ and $H^{1}(D)$.)

\subsubsection{Field-antifield symmetry}

Importance of the first symmetry, $Z(q,t)=-(-t)^{2-N}Z(q,1/t)$, is best appreciated
when the system is coupled to $N$ fermionic $bc$ systems $(p_{i},\theta^{i})$.
Then, the total partition function,
\begin{align}
  \mathbf{Z}(q,t) &= (1-t^{2})\prod_{h\ge1}(1-q^{h}t^{2})(1-q^{h}t^{-2}) \,,
\end{align}
satisfying
\begin{align}
  \mathbf{Z}(q,t) &= -t^{2}\mathbf{Z}(q,1/t) \,,
\end{align}
represents the cohomology of the ``physical'' BRST operator $Q=\int \lambda^{i}p_{i}$
(or more appropriately that of $\Hat{Q}=Q+D$).
The space on which $\Hat{Q}$ acts can be split into three pieces
\begin{align}
  \calF &= \calF_{\omega\lambda}\otimes \calF_{bc}\otimes \calF_{p\theta}
 =  \begin{cases}  \calH \colon &\text{$\Hat{Q}$-cohomology}\,, \\
  \calA \colon& \text{$\Hat{Q}$-non-closed}\,, \\
  \calB \colon& \text{$\Hat{Q}$-exact}\,.
\end{cases}
\end{align}
Now, define a non-degenerate inner product $(V,W)$ on $\calF$ using the standard BPZ conjugation
$\bra<V|=\ket|V>^{\dagger}$
\begin{align}
  (V,W) &= \braket<V|W> = \lim_{z\to\infty,w\to0}z^{2L_{0}}\bra<0|V(z)W(w)\ket|0> \,,
\end{align}
with the basic overlap
\begin{align}
  \bra<0|(\lambda^{i}\theta^{i}-c)_{0}\ket|0> = 1\,.
\end{align}
It is easy to see that the inner product couples $\calA$ with $\calB$,
and $\calH$ with itself.
(There can be the coupling $(\calA,\calH)$
but this can be set to zero by an appropriate choice of the cohomology
representatives.)
Then, whenever there is a cohomology element $V\in\calH$,
its antifield $V_{A}$ satisfying $(V,V_A)=1$ also represents a cohomology.
For example, the cohomology $\mathbf{1}$ is paired with $\lambda\theta-c$.
In general, $V$ and $V_A$ are related by the transformation $t\leftrightarrow 1/t$
but with the $t$-charge anomaly $-t^2$ which comes
from the $t$-charge of $\lambda\theta-c$.

\subsubsection{Paring of cohomology via $\ast$-conjugation symmetry}

The other symmetry $Z(q,t)=-q^{1}t^{-2}Z(q,q/t)$ can be understood in a similar manner~\cite{Toymodels}.
The relevant inner product $\langle V, W\rangle$ can be defined as the overlap
\begin{align}
\langle V,W\rangle = \llangle V|W\rangle\quad \text{with}\quad \llangle0|b_{-1}|0\rangle = 1\,.
\end{align}
Here, $\llangle V|=\ket|V>^{\ast}$ refers to a certain definition of
the BPZ conjugation
and $b=b_{-1}$ accounts for the prefactor $(-q^{1}t^{-2})$.
For the same reason as before, the inner product induces a non-degenerate paring between
$H^{k}(D)$ and $H^{1-k}(D)$
where the charges of a pair is related by
\begin{align}
  q^{m}t^{n}g^{k} \quad\leftrightarrow\quad -q^{1+m+n}t^{-2-n}g^{1-k} \,.
\end{align}
Since $H^{k}(D)=0$ for $k<0$ (more or less by construction),
the pairing implies that $H^{\ast}(D)$ is non-empty only at
$H^{0}(D)$ and $H^{1}(D))$.

\subsubsection{Mini-BRST and \v{C}ech/Dolbeault cohomologies}

To conclude our brief summary of the toy models,
let us explain how the mini-BRST and \v{C}ech/Dolbeault languages are related.
For definiteness, we concentrate on the relation between
the (minimal) mini-BRST cohomology $H^{\ast}(D)$ and the Dolbeault cohomology $H^{\ast}(\deldol)$.
The idea is to use the non-minimal mini-BRST cohomology $H^{\ast}(D+\deldol)$
to bridge between the two,
as the following figure indicates:
\begin{align*}
\xymatrix{
   & \calF^{0}(\deldol)\ar@{}[l]|{\cdots}\ar@{<.>}[d]\ar@{->}[r]^{\deldol} & \calF^{1}(\deldol)\ar@{<.>}[d]|-{(c)} \ar@{}[r]|{\cdots} & &\\
\calF^{0}(D)\ar@{<.>}[r]\ar@{->}[d]_{Q}\ar@{}[u]|{\vdots}
  & \calF^{0,0}\ar@{->}[d]_{Q}\ar@{->}[r]^{\deldol} &\calF^{1,0}\ar@{->}[d]^{Q}\ar@{->}[r]& & \\
\calF^{1}(D)\ar@{<.>}[r]|-{(a)}\ar@{}[d]|{\vdots} & \calF^{0,1}\ar@{->}[d]\ar@{->}[r]_{\deldol} \ar@{<.>}[ur]|-{(b)}
    &\calF^{1,1}\ar@{->}[d]\ar@{->}[r] & & \\
  &&&&
}
\end{align*}
In the figure, we put the $D$-cohomology on the left-most column
and the $\deldol$-cohomology on the top row.
Both cohomologies can be embedded
in the non-minimal mini-BRST cohomology of $D+\deldol$
as indicated by the arrows $(a)$ and $(c)$.
$(D+\deldol)$-cohomology is
graded by the sum of BRST ghost number and the Dolbeault form degree (number of $r$'s)
which runs diagonally from north-west to south-east.
The cohomologies of $D$ and $\deldol$ then simply
correspond to the different choices of cohomology representatives of
$(D+\deldol)$-cohomology (arrow $(b)$).

Step $(a)$ is simply an adding of the non-minimal quartet $(\overline{\omega},\overline{\lambda},s,r)$.
Since the non-minimal variables are unconstrained in the
ghost-for-ghost framework,
this does not affect the cohomology.
In particular, any representative of $D$-cohomology
also serves as a representative of $(D+\deldol)$-cohomology.

However, non-minimal variables can be used to obtain
different representatives for $(D+\deldol)$-cohomologies,
by replacing $b$ in terms of $r$ (step $(b)$).
In fact, (ignoring the terms proportional to the constraint $(\lambda\lambda)$ and the $c$ ghost)
one can choose a representative in $\calF^{m,0}$ ($m=0$ or $1$),
where all the cohomology degrees are carried by $r$
instead of $b$.

Finally, note that the cohomology representative in $\calF^{m,0}$
is necessarily $D$-closed and free of
non-minimal conjugates $\overline{\omega}_{\alpha}$ and $s_{\alpha}$.
Therefore, the object is ``gauge invariant'' (both in minimal and non-minimal senses)
and has intrinsic meaning on the constrained cone $\lambda\lambda=0$.
This gives the identification $(c)$.

\paragraph{Mapping of $b\in H^{1}(D)$:}

All the essential points of the mapping above can be seen by
computing the element of $H^{1}(\deldol)$
that corresponds to $b\in H^{1}(D)$ of the minimal mini-BRST cohomology.
First, $b$ is clearly in the non-minimal cohomology of $D+\deldol$.
But since
\begin{align}
\label{eq:toybeqD}
  b &= D\Bigl( {\overline{\lambda}\omega \over 2\lambda\overline{\lambda} } \Bigr) \,,
\end{align}
it is equivalent to
\begin{align}
\label{eq:toybdol}
 b\simeq  -\deldol \Bigl( {\overline{\lambda}\omega \over 2\lambda\overline{\lambda} } \Bigr)
  = { (\lambda r)(\overline{\lambda}\omega) - (\lambda\overline{\lambda})(r \omega) \over 2(\lambda\overline{\lambda})^{2}}  \,.
\end{align}
Because the last expression is $b$-independent
and $D$-closed (i.e. gauge invariant), it can be expressed in terms of
the local coordinates on $\lambda\lambda=0$.
Let us denote this ``intrinsic'' expression by $\Bar{b}$.
It is of course $\deldol$-closed.
In fact, it also appears $\deldol$-exact at first sight,
but since $(\lambda\overline{\lambda})^{-1}\overline{\lambda}\omega$ is gauge {\em non}-invariant,
there is nothing that can trivialize $\Bar{b}$ on the intrinsic geometry of $\lambda\lambda=0$.
Hence, $\Bar{b}$ represents a cohomology element of $H^{1}(\deldol)$.

\subsection{Mini-BRST and \v{C}ech/Dolbeault cohomologies for pure spinor}

The relation between mini-BRST and \v{C}ech/Dolbeault cohomologies
should be the same for the pure spinor system,
but explicit identifications of cohomology elements are not straightforward
due to the infinite number of BRST ghosts.
Let us nevertheless explain what one should expect for the structure of
pure spinor cohomology,
in view of the analysis in the previous subsection.

Recall that the full partition function satisfies:
\begin{align}
\text{field-antifield symmetry}\colon&\quad Z(q,t) = -t^{-8}Z(q,1/t) \,, \\
\text{``$\ast$-conjugation'' symmetry}\colon&\quad Z(q,t) = -q^{2}t^{-4}Z(q,q/t) \,.
\end{align}
The field-antifield symmetry implies that one can define an inner product
$(V,W)$ for the coupled system $(\omega,\lambda;p,\theta)$ with the overlap defined
by using a weight $t^{8}$ operator ${\lambda}^{3}{\theta}^{5}$~\cite{Berkovits:2000fe}
\begin{align}
  \langle0| (\lambda\gamma^{\mu}\theta)(\lambda\gamma^{\nu}\theta)(\lambda\gamma^{\rho}\theta)(\theta\gamma_{\mu\nu\rho}\theta) |0\rangle = 1 \,.
\end{align}
The other symmetry suggests
that there is an analog of the $\ast$-conjugation operation and
the $b$ operator of the toy model such that the inner product
$\langle V,W\rangle=\llangle V|W\rangle$ defined by
\begin{align}
\llangle V| = |V\rangle^{\ast}\quad \text{with}\quad\llangle0|b|0\rangle = 0
\end{align}
induces a pairing of the cohomology.
We claim that the $b$ operator is nothing but the tail element $b_{3}$ of
the reparameterization $b$-ghost carrying charges $-q^{2}t^{-4}g^{3}$.

Then, the inner product $\langle V,W\rangle$ pairs
the cohomologies at ghost number $k$ with those at ghost number $3-k$:
\begin{align}
  H^{k}(\delta) \quad\leftrightarrow\quad  H^{3-k}(\delta) \,.
\end{align}
($\delta$ here denotes either \v{C}ech, Dolbeault, minimal or non-minimal mini-BRST operators.)
In particular, there is a one-to-one mapping between $H^{0}(\delta)$ and $H^{3}(\delta)$
as has been repeatedly announced.
Cohomologies at negative ghost numbers $H^{n}(\delta)$ ($n<0$),
and hence $H^{n}(\delta)$ ($n>3$), should be empty.
As for the remaining pair $H^{1}(\delta)$ and $H^{2}(\delta)$,
we conjecture that they are also empty.
(One piece of evidence
for this conjecture is that
$H^{0}(\delta)$ and $H^{3}(\delta)$ are sufficient to reproduce
the superstring spectrum.)
Leaving proofs of these conjectures as a future problem,
let us see how the operator $b_3$ at $-q^{2}t^{-4}g^{3}$
can be described in various cohomologies.

\subsection{The operator $b_3$}

Under the $\ast$-conjugation symmetry, the operator $\mathbf{1}$ in $H^{0}(\delta)$ is mapped
to a fermionic singlet at $-q^{2}t^{-4}$,
which we first encountered in section~\ref{sec:ginvpoly2} and called the missing state.
This missing state will be identified with $b_{3}$.

\subsubsection{\v{C}ech/Dolbeault descriptions}

The identification of $b_{3}$ is easier
in the \v{C}ech/Dolbeault cohomologies than in the mini-BRST cohomology (this is different from the toy models).
Indeed, it can be identified as the ``tail term'' of the reparameterization $b$-ghost
of the pure spinor formalism.
Recall that in the non-minimal pure spinor formalism the $b$-ghost is written as~\cite{Berkovits:2004px,Berkovits:2005bt}\cite{Oda:2005sd,Oda:2007ak}
\begin{align}
\label{eq:nmbghost}
\begin{split}
  b &= b_{0} + b_{1} + b_{2} + b_{3}\,, \\
 b_{0} &= -s^{\alpha}\del\overline{\lambda}_{\alpha} + {\overline{\lambda}_{\alpha}G^{\alpha} \over (\lambda\overline{\lambda}) }
  = -s^{\alpha}\del\overline{\lambda}_{\alpha}
 -{\overline{\lambda}_{\alpha}\{\pi^{\mu}(\gamma_{\mu}d)^{\alpha} - N_{\mu\nu}(\gamma^{\mu\nu}\del\theta)^{\alpha} - J\del\theta^{\alpha} - {1\over4}\del^{2}\theta^{\alpha} \}
   \over4(\lambda\overline{\lambda})} \,,\\
 b_{1} &= {\overline{\lambda}_{\alpha}r_{\beta}H^{[\alpha\beta]} \over (\lambda\overline{\lambda})^{2} }
  = {(\overline{\lambda}\gamma^{\mu\nu\rho}r)\{ (d\gamma_{\mu\nu\rho}d)-48N_{\mu\nu}\pi_{\rho} \}
  \over 768(\lambda\overline{\lambda})^{2}}\,,\\
 b_{2} &= {\overline{\lambda}_{\alpha}r_{\beta}r_{\gamma}K^{[\alpha\beta\gamma]} \over  (\lambda\overline{\lambda}) }
  = -{(r\gamma^{\mu\nu\rho}r)N_{\mu\nu}(\overline{\lambda}\gamma_{\rho}d)
   \over64(\lambda\overline{\lambda})^{3} }\,, \\
 b_{3} &= {\overline{\lambda}_{\alpha}r_{\beta}r_{\gamma}r_{\delta}L^{[\alpha\beta\gamma\delta]} \over (\lambda\overline{\lambda})^{4} }
   = {(r\gamma^{\mu\nu\rho}r)(\overline{\lambda}\gamma_{\sigma\tau\rho}r)N_{\mu\nu}N^{\sigma\tau}
   \over 512(\lambda\overline{\lambda})^{4} }\,,
\end{split}
\end{align}
and satisfies
\begin{align}
\{Q\,, b_{0}\} = T\,,\quad
\{\deldol\,, b_{i} \} + \{Q\,, b_{i+1}\} = 0\,,\;(i=0, 1, 2)\,,\quad
\{\deldol\,, b_{3} \} =0\,.
\end{align}
Being the tail of the $b$-ghost, $b_{3}$ is clearly in the Dolbeault cohomology
of intrinsic, or gauge invariant operators.
It is independent of $(x,p,\theta)$ and carries charges $-q^{2}t^{-4}g^{3}$.
So this is the ``missing state'' we were looking for.

The \v{C}ech description of $b_{3}$ is similar.
It simply corresponds to a $3$-cochain
\begin{align}
b_{3} =   (b^{ABCD}) = {L^{[ABCD]} \over \lambda^{A}\lambda^{B}\lambda^{C}\lambda^{D}} \,,
\end{align}
which can be related to the Dolbeault version using the partition of unity.
Again, it is clearly in the \v{C}ech cohomology of the gauge invariant operators.

We leave the geometrical interpretation of $b_{3}$
and the related construction of inner products
as future projects.

\subsubsection{Mapping to ghost-for-ghost language}

Now, we move on to the identification of $b_3$ in
the mini-BRST cohomology of the ghost-for-ghost language.
It can be obtained from the \v{C}ech/Dolbeault version by a mapping similar to~(\ref{eq:toybdol}).
For ease of notation, we choose to start from the Dolbeault language.
We do not work out the coefficients and the spinor index structures completely.

First, one has to embed $b_{3}$ to the non-minimal mini-BRST cohomology.
This can be achieved by forgetting the pure spinor constraint ($\overline{\lambda}$ is kept constrained)
and adding the ghost contributions so that
\begin{align}
 b_{3} \quad\to\quad D b_{3,0} = \deldol b_{3,0} = 0 \,.
\end{align}
Here, the notation $b_{3,0}$ indicates that it is in $\calF^{3,0}$
(i.e. carries BRST ghost number $0$) and the main part of $b_{3,0}$ would look like
\begin{align}
  b_{3,0} &\sim {(\overline{\lambda}\gamma^{\mu\nu\rho}r)(r\gamma_{\sigma\tau \rho}r)\Hat{N}_{\mu\nu}\Hat{N}^{\sigma\tau} \over (\lambda\overline{\lambda})^{4}} \,.
\end{align}
(Recall that $\Hat{N}^{\sigma\tau}$ is the Lorentz generators for the full system.)

Then, we proceed in the direction toward the minimal mini-BRST cohomology
and try to obtain the expression $b_{0,3}$ in which
the non-minimal variables are absent:
\begin{align*}
\xymatrix{
&  & a_{2,0}\ar@{->}[r]^{-\deldol}\ar@{->}[d]_{D} & b_{3,0}\ar@{.>}[ld] \\
& a_{1,1}\ar@{->}[r]^{-\deldol}\ar@{->}[d]_{D} & b_{2,1}\ar@{.>}[ld]& \\
a_{0,2}\ar@{->}[r]^{-\deldol}\ar@{->}[d]_{D}& b_{1,2}\ar@{.>}[ld]&& \\
b_{0,3}
}
\end{align*}
At each step, one Dolbeault form $r=\targetd\overline{\lambda}$ will be taken off
and the final form $b_{0,3}$ is supposed to be free of all the non-minimal fields,
as in~(\ref{eq:toybeqD}).

\bigskip
To get an idea on how $b_{0,3}$ should look like,
let us do a very rough computation.
First, $b_{3,0}$ is a $\deldol$ of
a gauge {\em non}-invariant ($D$-non-closed) operator $a_{2,0}$:
\begin{align}
a_{2,0}
\sim { (\overline{\lambda}\gamma^{\mu\nu\rho}r)(\omega\gamma_{\rho}r)(\omega\gamma_{\mu\nu}\lambda) \over (\lambda\overline{\lambda})^{3} }+\cdots
\quad\to\quad
-\deldol a_{2,0} = b_{3,0} \,.
\end{align}
Then we get another representative
\begin{align}
b_{2,1} = D a_{2,0}
  \sim { b^{\sigma}(\overline{\lambda}\gamma^{\mu\nu\rho}r)(\lambda\gamma_{\sigma \rho}r)(\omega\gamma_{\mu\nu}\lambda) \over (\lambda\overline{\lambda})^{3} }+ \cdots
 \simeq b_{3,0}\,,
\end{align}
where $b^{\mu}$ is the first generation BRST ghost
contained in the first term of $D=\sum_{k}D_{k}$, i.e. $D_0 = \lambda\fslash{b}\lambda$.
(See section~\ref{sec:reducibility} and appendix~\ref{app:reducibility}
for more details on the mini-BRST operator.)

Similarly, $b_{2,1}$ can be written as
\begin{align}
  a_{1,1}\sim { b_{\rho}(\overline{\lambda}\gamma^{\mu\nu\rho}r)(\omega\gamma_{\mu\nu}\lambda) \over (\lambda\overline{\lambda})^{2} }+ \cdots
\quad\to\quad
  -\deldol a_{1,1} = b_{2,1} \,,
\end{align}
and it leads to
\begin{align}
  b_{1,2} = D a_{1,1} \sim
  { \rho^{\alpha}(\gamma_{\rho}\lambda)_{\alpha}(\overline{\lambda}\gamma^{\mu\nu\rho}r)(\omega\gamma_{\mu\nu}\lambda) \over (\lambda\overline{\lambda})^{2} }+ \cdots \simeq b_{2,1} \simeq b_{3,0}\,.
\end{align}
Here, $\rho^{\alpha}$ is the second generation BRST ghost contained in
$D_{1}=\rho^{\alpha}(\fslash{c}\lambda)_{\alpha}$,
and we dropped the term $D_{0} a_{1,1}\sim (\lambda\overline{\lambda})^{-2}b_{\mu}b_{\nu}(\lambda\gamma_{\rho}\lambda)(\overline{\lambda}\gamma^{\mu\nu\rho}r)$
which presumably is not very important.
Finally, we rewrite $b_{1,2}$ as
\begin{align}
 a_{0,2} &\sim {\rho^{\alpha}(\gamma^{\mu\nu}\overline{\lambda})_{\alpha}(\omega\gamma_{\mu\nu}\lambda) \over (\lambda\overline{\lambda})}\quad\to\quad -\deldol a_{0,2}=b_{1,2} \,,
\end{align}
and obtain from hitting it by $D_{2}=b^{\mu\nu}\bigl(\sigma_{\alpha}(\gamma_{\mu\nu}\lambda)^{\alpha}+c_{\mu}c_{\nu}\bigr)$
\begin{align}
 b_{0,3} &= D a_{0,2} = b^{\mu\nu}(\omega\gamma_{\mu\nu}\lambda) + \cdots \simeq b_{3,0}\,.
\end{align}
This should give an expression for the $-q^{2}t^{-4}g^{3}$ element
in the minimal mini-BRST cohomology.

\bigskip
Now, quite independently from the computation above
(i.e. entirely within the minimal ghost-for-ghost framework),
it can be argued that there is a $D$-closed operator $\Hat{b}$
that starts from
\begin{align}
 \Hat{b} =  b^{\mu\nu}\Hat{N}_{\mu\nu} + \cdots\,.
\end{align}
Since $D$ commutes with the total Lorentz generator $\Hat{N}_{\mu\nu}$,
one finds
\begin{align}
\label{eq:DbN}
  D(b^{\mu\nu}\Hat{N}_{\mu\nu}) = D(b^{\mu\nu})\Hat{N}_{\mu\nu} = \rho^{\mu\alpha}\Hat{N}_{\mu\nu}(\gamma^{\nu}\lambda)_{\alpha} + \cdots
\end{align}
where $\rho^{\mu\alpha}$ is the fourth generation ghost
that came from $D_{3}=\rho^{\mu\alpha}(c_{\mu\nu}(\gamma^{\nu}\lambda)_{\alpha}+c_{\mu}\sigma_{\alpha})$,
and the ellipsis denotes the (less important) contribution
from $D_{k}\ni B_{k+1}C_{3}C_{k-3}$.\footnote{
$(B_{k},C_{k})$ are the $k$th generation ghost; in particular $C_{3}=c_{\mu\nu}$.}

Note that the pure spinor identity
\begin{align}
  2N^{\mu\nu}(\gamma_{\mu}\lambda)_{\alpha} - J(\gamma^{\nu}\lambda)_{\alpha} = -{1\over2}(\omega\gamma^{\mu}\gamma^{\nu})_{\alpha}(\lambda\gamma_{\mu}\lambda) \approx 0
\end{align}
is rephrased in the ghost-for-ghost language as
\begin{align}
  2\Hat{N}^{\mu\nu}(\gamma_{\mu}\lambda)_{\alpha} - \Hat{J}(\gamma^{\nu}\lambda)_{\alpha} = D\Hat{Y}^{\nu}_{\alpha}\,,
\end{align}
for some $\Hat{Y}^{\nu}_{\alpha}=-(1/2)\omega_{\alpha}c^{\nu}+\cdots$.
Since $\rho^{\mu\alpha}$ is $\gamma$-traceless, one then finds
\begin{align}
D(b^{\mu\nu}\Hat{N}_{\mu\nu} + \rho^{\mu\alpha}\Hat{Y}_{\mu\alpha})
&= D(\rho^{\mu\alpha})\Hat{Y}_{\mu\alpha} + \cdots \,,
\end{align}
and is taken back to a situation similar to~(\ref{eq:DbN})
with $B_{3}=b^{\mu\nu}$ replaced by $B_{4}=\rho^{\mu\alpha}$
and $\Hat{N}_{\mu\nu}$ replaced by $\Hat{Y}_{\nu\alpha}$.
Thus the general expectation is
\begin{align}
  \Hat{b} &= \Bigl(\sum_{k\ge3}B_{k}\Hat{N}_{k}\Bigr) + \cdots \,,
\end{align}
where $\Hat{N}_{3}=\Hat{N}_{\mu\nu}$, $\Hat{N}_{4}=\Hat{Y}_{\nu\alpha}$ and so on,
and the ellipsis is responsible for
the corrections from $D_{k}\ni B_{k+1}C_{m}C_{k-m}$.

\subsubsection{Properties of $b_3$ and the remaining missing states}

Having explained that the fermionic singlet with charges $-q^{2}t^{-4}g^{3}$
is nothing but the tail term $b_{3}$ (i.e. the three form piece) of the composite $b$-ghost,
let us turn to a brief discussion of the remaining missing states that we found at higher Virasoro levels
(see~(\ref{eq:app:Zqt})).

Although the number of missing states are finite at a given level,
the $\ast$-conjugation symmetry of the partition function
indicates there are an infinite number
of them, one for every gauge invariant state,
$\lambda^{\lpar \alpha}\lambda^{\beta\rpar}$, $N^{\mu\nu}$, $T$ etc.
In view of our analysis of the toy models, the charges of
``$\ast$-conjugation pairs'' must be related as
\begin{align}
q^{m}t^{n}g^{k}\quad\leftrightarrow\quad -q^{2+m+n}t^{-4-n}g^{3-k}  \,.
\end{align}
In particular, we conjecture that all the missing states are carrying ghost number $3$
and can be constructed by multiplying ghost number $0$ operators
to a single $b_{3}$ (perhaps with derivatives).

One should be able to prove the conjecture by constructing an inner product
that couples the $\ast$-conjugation pairs.
As we explained in section~\ref{subsubsec:toyhighercohom} using toy models,
the inner product should be such that
the cohomologies with $b_{3}$ can be obtained roughly
by swapping the role of $\lambda$ and $\omega$.
Some questions that can be checked explicitly even before constructing the inner product are
whether $b_3\lambda^{\alpha}$ and $b_{3}\del b_3$ are trivial,
and whether operators of the form $b_3 \omega_{\lpar \alpha_{1}}\cdots \omega_{\alpha_{n} \rpar}+\cdots$ are in the cohomology.
In the case of the toy models these questions can be confirmed very easily.
For example, using the BRST method (remember $D=b\lambda^i\lambda^i$)
\begin{align}
  b\lambda^{i} &\propto D(w^{i}) \,,\quad b\del b \propto D(\omega^{i}\omega^{i}) \,,
\end{align}
and it is also easy to show that $b \omega_i$ etc. are in the cohomology.
We believe the answers are also affirmative for the pure spinor case,
but will leave a proof of this to future investigation.

%%%%%%%%%%%%%%%%%%%%%%%%%%%%%%%%%%%%%%%%%%%%%%%%%%%%%%%%%%%%%%%%
\section{Derivation of the lightcone spectrum}
\label{sec:lightcone}

Finally in this section we derive the Green-Schwarz
lightcone spectrum by combining the pure spinor partition function
with those of the physical variables $x^{\mu}$ and $(p_{\alpha},\theta^{\alpha})$.
The lightcone spectrum we are to derive is the Fock space spanned by
the transverse oscillators
\begin{align}
  \alpha^{i}_{-n}\,,\quad S^{a}_{-n}\,,\quad (i\in\mathbf{8}_{v}\,,\; a\in \mathbf{8}_{s}\,,\; n\ge 1)\,,
\end{align}
on the  super-Maxwell ground states
\begin{align}
  \ket|i>+\ket|\Dot{a}> = \mathbf{8}_{v}+\mathbf{8}_{a}\,.
\end{align}
Their partition function is simply
\begin{align}
\begin{split}
  Z_{\text{lc}}(q,\vec{\sigma}) &= \Tr{}_{\text{lc}}(-1)^{F}q^{L_{0}}\mathe^{\mu\cdot \sigma} \\
  &= (\mathbf{8}_{v} - \mathbf{8}_{a})
  \prod_{h\ge1}{(1-q^{h})^{\mathbf{8}_{s}} \over (1-q^{h})^{\mathbf{8}_{v}}} \,.
\end{split}
\end{align}

Now, since the physical BRST operator of the pure spinor formalism
$Q$ (or more precisely $Q+\delta$) contains pieces with {\em non-zero} $t$-charge,
the total partition function of the pure spinor superstring
$\mathbf{Z}(q,t,\vec{\sigma})$ (which includes $x$ and $(p,\theta)$ sectors)
is not directly related to the cohomology of $Q$.
Moreover, $\mathbf{Z}(q,t,\vec{\sigma})$ differs from the lightcone partition function.
However, it will be shown in this section that
if the $t$-charge is twisted appropriately using the lightcone boost charge ($t\to\Tilde{t}$),
$\mathbf{Z}(q,t,\vec{\sigma})$ can be related to the lightcone partition function as
\begin{align}
\label{eq:Zlcguess}
  \Tilde{\mathbf{Z}}(q,\Tilde{t},\vec{\sigma})
\quad\leftrightarrow\quad -\Tilde{t}^{2}Z_{\text{lc}}(q,\vec{\sigma}) + \Tilde{t}^{6}Z_{\text{lc}}(q,\vec{\sigma}) \,.
\end{align}
The first term at $\Tilde{t}^{2}$ represents the usual lightcone spectrum
and the second term at $\Tilde{t}^{6}$ represents the spectrum of the antifields.
If one writes $Q=Q_0 + Q_1 + \cdots$ where $Q_n$ carries $t$-charge $n$, it is
obvious that
the twisted total partition function $\Tilde{\mathbf{Z}}(q,\Tilde{t},\vec{\sigma})$
represents the cohomology of $\Tilde{t}$-charge $0$ piece $Q_{0}$ of $Q$.
One might think that the cohomology of $Q_{0}$ has nothing to do with that of $Q$,
but it will be shown that $Q_{0}$ and $Q$ have the same cohomology,
{\em except} that the on-shell condition ($L_0=0$)
is not implied for the former.

Let us begin by first illustrating the analogous result for
the bosonic string.

\subsection{Lightcone spectrum of bosonic string from covariant partition function}

The BRST operator of the bosonic string takes the form
\begin{align}
\begin{split}
  Q &= cT_{x} + bc\del c  \\
   &= \sum_{n\in\mathbb{Z}} c_{-n}L_{n} - {1\over2}\sum_{m,n\in\mathbb{Z}}(m-n)c_{-m}c_{-n}b_{m+n}\,,
 \end{split}
\end{align}
where we left the normal orderings implicit and the Virasoro operators are given by
\begin{align}
\begin{split}
  L_{0} &= {1\over2}k^2 + \sum_{m\ge1}\alpha_{-m}^{\mu}\alpha_{\mu,m} -1\,, \\
  L_{n} &= {1\over2}\sum_{m\in\mathbb{Z}}\alpha_{n-m}^{\mu}\alpha_{\mu,m} \quad (n\ne0)\,.
\end{split}
\end{align}
Because of the ghost zero-mode oscillators $\{b_0,\,c_0\}=1$,
the cohomology of $Q$ consists of two identical copies of the lightcone spectrum%
---those without $c_0$ (fields) and those with $c_0$ (antifields).
Thus, the partition function defined by $\Tr(-1)^{F}q^{L_{0}}$ vanishes identically
due to field-antifield cancellation.
One way to get a non-zero result is to impose an additional condition $b_{0}=0$
which drops all the antifields from the trace,
but it is difficult to perform an analogous operation in the pure spinor formalism.
Another way to get a non-zero result
is to introduce a charge that distinguishes fields from antifields.
Clearly, the ghost number ($t$-charge) measured by
\begin{align}
  J &= -bc\quad\bigl(\to\quad   t(b,c) = (-1,1)\, \bigr)
\end{align}
does the job.
The (lightcone) partition function would then be
\begin{align}
\label{eq:Zboslc}
 Z_{\text{lc}}(q,t) &= \Tr(-1)^{F}q^{L_{0}}t^{J_{0}} =
  -q^{-1}(t-t^2)\prod_{h\ge1}{1\over (1-q^{h})^{24}} \,,
\end{align}
where the prefactor represents the ground state tachyon ($c=-q^{-1}t$)
and its antifield ($c\del c=q^{-1}t^{2}$).

In obtaining the expression~(\ref{eq:Zboslc}),
we used the well-known fact that
the physical spectrum is spanned by the transverse oscillators $\alpha^{i}_{-n}$ ($i=1, \ldots ,  24$, $n>0$).
Now, let us explain how it can be obtained from the covariant partition function
\begin{align}
\begin{split}
  \mathbf{Z}(q,t,\vec{\sigma}) &= Z_{x}(q,t,\vec{\sigma})Z_{bc}(q,t,\vec{\sigma})\,, \\
  Z_{x} &= \prod_{h\ge1}{1\over(1-q^{h})^{\mathbf{26}}} \,,\\
  Z_{bc} &=\prod_{h\ge2}{(1-q^{h}t^{-1})^{\mathbf{1}}}\prod_{h\ge-1}{(1-q^{h}t)^{\mathbf{1}}} \,.
\end{split}
\end{align}
If the BRST operator $Q$ carried ghost number ($t$-charge) $0$,
the total partition function $\mathbf{Z}(q,t,\vec{\sigma})$ would represent its cohomology.
But since $Q$ carries ghost number $1$, $\mathbf{Z}(q,t,\vec{\sigma})$ is not directly related to the cohomology.
Nevertheless, there is a simple way to obtain the partition function
of cohomology~(\ref{eq:Zboslc}) from $\mathbf{Z}(q,t,\vec{\sigma})$.
The procedure is simple and one only has to twist the $t$-charge by the lightcone boost charge
for the non-zero modes $\alpha^{\pm}_{n'}$
\begin{align}
  J \to \Tilde{J} = J + {1\over2}N^{+-}_{x'} \,.
\end{align}
(The zero-modes $k^{\pm}$ are kept intact.)
Then, the twisted $\Tilde{t}$-charges read
\begin{align}
  \Tilde{t}(k^{\pm},\alpha^{\pm}_{n'},\alpha^{i}_{n},b,c) &= (0,\pm1, 0,-1,1)\,,
\end{align}
and the twisted partition function becomes identical to~(\ref{eq:Zboslc})
representing lightcone fields and antifields:
\begin{align}
\label{eq:twistedZbos}
 \Tilde{\mathbf{Z}}(q,\Tilde{t},\vec{\sigma}) &=
  -q^{-1}(\Tilde{t}-\Tilde{t}^2)\prod_{h\ge1}{1\over (1-q^{h})^{\mathbf{24}}}  \,.
\end{align}
Of course, $\Tilde{\mathbf{Z}}(q,\Tilde{t},\vec{\sigma})$ represents the cohomology of
the $\Tilde{t}$-charge $0$ piece of $Q$,
\begin{align}
\label{eq:bosQdecomp}
\begin{split}
  Q &= Q_{0} + Q_{1} + Q_{2} \,, \\
  Q_{0} &= -{1\over2}k^{+}\sum_{n\ne0}c_{-n}\alpha^{-}_{n} \,,
\end{split}
\end{align}
and not necessarily that of $Q$ itself.
However, as is apparent from~(\ref{eq:bosQdecomp})
the cohomologies of $Q$ and $Q_{0}$ are identical,
except that the on-shell conditions are not implied for the latter.
(Recall that we are not imposing the $b_{0}=0$ condition.)
So the twisted partition function~(\ref{eq:twistedZbos}) in fact
represents the lightcone spectrum but {\em without} the on-shell condition.

In the previous paragraph, we recovered the well-known fact that
the BRST cohomology reproduces the lightcone spectrum~\cite{Kato:1982im}.
Anticipating application to the pure spinor formalism,
let us briefly recall why this is the case.
The crucial points to understand are
(i)~why the non-zero modes of $bc$ ghosts and lightcone oscillators $\alpha^{\pm}_{n}$
form a BRST quartet
and (ii)~how the on-shell condition appears.
An efficient way to understand both at the same time is to use a unitary transformation
that reveals the essential structure of $Q$.
It can be shown that $Q$ can be brought to the form (see for example~\cite{Aisaka:2004ga})
\begin{align}
 \mathe^{R}Q\mathe^{-R} &=  d+d'  \,,
\end{align}
where
\begin{align}
\begin{split}
d  &= Q_{0} = -{1\over2}k^{+}\sum_{n\ne0}c_{-n}\alpha^{-}_{n}\,, \\
d' &= c_{0}L^{\text{tot}}_{0} = c_{0}\bigl({1\over2}k^2 - 1 + \sum_{n\ne0}(\alpha^{\mu}_{-n}\alpha_{\mu,n}+ nc_{-n}b_{n}) \bigr)  \,.
\end{split}
\end{align}
Clearly, $d$ and $d'$ commute and the cohomology of $d+d'$
is the lightcone spectrum ($d=0$) {\em with} the on-shell condition ($d'=0$).
Since $Q_{0}$ is identical to $d$ and does {\em not} contain $d'$,
its cohomology is different from that of $Q$ by the on-shell condition.

One could have tried to define the $\Tilde{t}$-charge so that
\begin{align}
  Q_{0}=d+d' \,.
\end{align}
It can be achieved for example by treating the ghost zero-modes differently
from the ghost non-zero modes
and assigning $\Tilde{t}(b_{0},c_{0})=(0,0)$ instead of $(-1,1)$.
But this makes the partition function vanish due to the
field-antifield cancellation.
We expect this phenomenon to be a general feature
of ``on-shell partition functions'' if one does
{\em not} impose the $b_{0}=0$ condition.
Since it is not straightforward to impose the $b_{0}=0$ condition
in the pure spinor formalism, we will be content with
the ``off-shell partition function'' of the type~(\ref{eq:twistedZbos})
and discuss the on-shell condition separately.

\subsection{Lightcone spectrum from pure spinor partition function}

As explained earlier, physical states in the pure spinor formalism are
defined as the cohomology of $Q+\delta$ in \v{C}ech-Dolbeault framework. 
We now wish to define an operator that is the analog of
$Q_{0} = -(1/2)k^{+}\sum_{n\ne0}c_{-n}\alpha^{-}_{n}$
of the bosonic string
as the twisted $\Tilde{t}$-charge $0$ piece of the physical BRST operator.
To find the appropriate twisting of the $t$-charge current $J_{t}=-\omega\lambda-p\theta$,
let us study the massless vertex operators
and see where the lightcone degrees of freedom reside.

\subsubsection{Twisting of $t$-charge}

The super-Poincar\'{e} covariant vertex operator for the
super-Maxwell fields is given by
\begin{align}
\label{eq:SMvert8}
\begin{split}
 V &= \lambda^{\alpha}A_{\alpha}(x,\theta) \\
   &= \lambda^{\alpha}\psi_{\alpha}(x) + (\lambda\gamma^{\mu}\theta)a_{\mu}(x) + (\lambda\gamma^{\mu}\theta)(\theta\gamma_{\mu})_{\alpha}\chi^{\alpha}(x) + \cdots \\
\end{split}
\end{align}
with $a^{\mu}(x)$ and $\chi^{\alpha}(x)$ the photon and photino wave functions.
(The first term $\lambda^{\alpha}\psi_{\alpha}$ is pure gauge
and the ellipsis involve spacetime derivatives of $a^\mu$ and $\chi^\a$.)
In this form, the lightcone degrees of freedom
$(a^{i},\chi^{\Dot{a}})$ are contained in the terms at $t^{2}$ and $t^{3}$.
But if the $t$-charges of $\lambda$ and $\theta$ are twisted by the lightcone boost charge as
\begin{align}
\label{eq:twist0}
\begin{split}
  J \quad\to\quad &\Tilde{J} = - \omega\lambda - p\theta
 + N^{+-}_{\omega\lambda} + N^{+-}_{p\theta}\,, \\
  &\bigl(  \Tilde{t}(\gamma^{+}\lambda,\gamma^{-}\lambda,\gamma^{+}\theta,\gamma^{-}\theta) = (2,0,2,0) \bigr)\,,
\end{split}
\end{align}
both are brought to $\Tilde{t}^{2}$
\begin{align}
  \Tilde{t}^{2}\colon&\quad (\lambda\gamma^{i}\theta)a_{i}(x)\,,\quad (\lambda\gamma^{-}\theta)(\theta\gamma^{+})_{\Dot{a}}\chi^{\Dot{a}}(x)\,.
\end{align}
Similar analysis shows that the
lightcone degrees of freedom for
the antiphoton and antiphotino are brought to $\Tilde{t}^{6}$,
which explains our expectation~(\ref{eq:Zlcguess}):
\begin{align}
  \Tilde{\mathbf{Z}}(q,\Tilde{t},\vec{\sigma}) = -\Tilde{t}^{2}Z_{\text{lc}}(q,\vec{\sigma}) + \Tilde{t}^{6}Z_{\text{lc}}(q,\vec{\sigma}) \,.
\end{align}

The analysis here does not tell us how the
$t$-charges of (the non-zero modes of) $\delx^{\mu}$ should be twisted,
but it turns out that the appropriate definition of $\Tilde{J}$ is
\begin{align}
  \label{eq:twist}
  \Tilde{J} &= -\omega\lambda-p\theta
 + K\,, \quad K = N^{+-}_{\omega\lambda} + N^{+-}_{p\theta} + 2N^{+-}_{x'} \,.
\end{align}
Note that we twisted the lightcone coordinates $\del x^{\pm}$ twice as much as others,
and we indicate by the prime in $N^{+-}_{x'}$
that the zero-mode of $\del x^{\pm}=k^{\pm}$ are kept intact.
In our convention, the boost charge $K$ of the basic operators are
\begin{align}
\begin{split}
  &{K}(\gamma^{\pm}\omega,\gamma^{\pm}\lambda) = (\pm1,\pm1)\,,\quad {K}(\gamma^{\pm}p,\gamma^{\pm}\theta) = (\pm1,\pm1)\,,\\
  &{K}(k^{\pm},\delx^{\prime \pm},\delx^{i}) = (0,\pm4,0)\,.
\end{split}
\end{align}

It will now be argued that the $\Tilde{t}$-charge $0$ piece of the physical BRST operator
plays a role analogous to the $Q_{0} = -(1/2)k^{+}\sum_{n\ne0}c_{-n}\alpha^{-}_{n}$
of the bosonic string.
As a first step, let us see how the total partition function $\mathbf{Z}(q,t,\vec{\sigma})$
is twisted at several lower mass levels.

\subsubsection{Massless states}

It is easy to see that the twisted partition function for the zero modes
$\Tilde{\mathbf{Z}}_{0}(\Tilde{t},\vec{\sigma})$
represents the lightcone super-Maxwell ground state.
The twisted partition function can be easily computed
from the original spin partition function:
\begin{align}
\label{eq:Z0lc}
\begin{split}
  \mathbf{Z}_{0}(t,\vec{\sigma})
 &= \mathbf{1} - \mathbf{10}t^{2} + \overline{\mathbf{16}}t^{3}
  - \mathbf{16}t^{5} + \mathbf{10}t^{6} - \mathbf{1}t^{8} \,, \\
\to\quad
  \Tilde{\mathbf{Z}}_{0}(\Tilde{t},\vec{\sigma})
 &= -(\mathbf{8}_{v} - \mathbf{8}_{a})\Tilde{t}^{2}
  + (\mathbf{8}_{v} - \mathbf{8}_{a})\Tilde{t}^{6}\,.
\end{split}
\end{align}

At the level of vertex operators, this formula can be understood as follows.
Covariant vertex operators for the super-Maxwell antifields $(a^{\ast},\chi_{\alpha}^{\ast})$, for the
ghost $c$, and for the antighost $c^{\ast}$ are similar to that of the
super-Maxwell field~(\ref{eq:SMvert8})
but have different numbers of $\lambda$:
\begin{align}
\begin{split}
V^{\ast}
 &= \lambda^{\alpha}\lambda^{\beta}A_{\alpha\beta}(x,\theta) \\
 &= \cdots +(\lambda\gamma_{\nu}\theta)(\lambda\gamma_{\rho}\theta)(\gamma^{\nu\rho}\theta)^{\alpha}\chi^{\ast}_{\alpha}(x) +(\lambda\gamma_{\nu}\theta)(\lambda\gamma_{\rho}\theta)(\theta\gamma^{\mu\nu\rho}\theta)a^{\ast}_{\mu}(x) + \cdots \,
\\
U &= A(x,\theta)
  = \mathbf{1}c(x) + \cdots \,,
\\
U^{\ast} &= \lambda^{\alpha}\lambda^{\beta}\lambda^{\gamma}A_{\alpha\beta\gamma}(x,\theta)
  = \cdots + (\lambda\gamma^{\mu}\theta)(\lambda\gamma^{\nu}\theta)(\lambda\gamma^{\rho}\theta)(\theta\gamma_{\mu\nu\rho}\theta)c^{\ast}(x) + \cdots \,.
\end{split}
\end{align}
The terms of the covariant partition function $\mathbf{Z}_{0}$ corresponds to
the (vertex operators of) component fields as
\begin{align}
\mathbf{1} - \mathbf{10}t^{2} + \overline{\mathbf{16}}t^{3}
  - \mathbf{16}t^{5} + \mathbf{10}t^{6} - \mathbf{1}t^{8}
\quad\leftrightarrow\quad
(c,a_{\mu},\chi^{\alpha},\chi^{\ast}_{\alpha},a^{\ast}_{\mu},c^{\ast}) \,.
\end{align}
Under the twisting~(\ref{eq:twist0}) one finds that
only the lightcone degrees of freedom survives as in
\begin{center}
\begin{tabular}{r|MMMMMMMMM}
   & \Tilde{t}^{0} & \Tilde{t}^{1} & \Tilde{t}^{2} & \Tilde{t}^{3} & \Tilde{t}^{4} & \Tilde{t}^{5} & \Tilde{t}^{6} & \Tilde{t}^{7} & \Tilde{t}^{8} \\ \hline
$\mathbf{1}$ & c  \\
$-\mathbf{10}t^{2}$ & a^{+} & & a^{i} & & a^{-} \\
$\mathbf{16}t^{3}$ & && \chi^{\Dot{a}} && \chi^{a} \\
$-\mathbf{16}t^{5}$ & & &&& \chi^{\ast}_{a} && \chi^{\ast}_{\Dot{a}} \\
$\mathbf{10}t^{6}$ &&& &&a^{\ast +} & & a^{\ast i} && a^{\ast -}\\
$-\mathbf{1}t^{8}$ &&&&&&&&& c^{\ast}
\end{tabular}
\end{center}
Spurious degrees of freedom and the ghosts ($a^{\pm}$, $\chi^{a}$, $c$ etc.)
are brought outside $\Tilde{t}^{2,6}$ and get canceled by components
with the opposite statistics.

\subsubsection{First massive states}

The lightcone partition function at level $1$ can be derived in a similar
manner.
A new feature here is the appearance of non-zero modes of $x^{\mu}$
which have to be twisted twice as much~(\ref{eq:twist}).
The total partition function before the twisting is
\begin{align}
\begin{split}
\mathbf{Z}_{1}(t,\vec{\sigma})
&= Z_{\omega\lambda,1}Z_{p\theta,0} + Z_{\omega\lambda,0}Z_{p\theta,1}  + Z_{\omega\lambda,0}Z_{p\theta,0}Z_{x,1} \\
&= \bigl(
 (\mathbf{45}+\mathbf{1})
 - \mathbf{144}t
 + (\overline{\mathbf{126}}-\mathbf{10})t^2
 + \overline{\mathbf{16}}t^{3} \nonumber\\
&\qquad\qquad
 - \mathbf{16}t^5
 - (\mathbf{126}-\mathbf{10})t^6
 + \mathbf{144}t^7
 - (\mathbf{45}+\mathbf{1})t^8
\bigr)_{\omega\lambda} \\
&\quad + \bigl(
  (\mathbf{1}-\mathbf{10}t^2+\overline{\mathbf{16}}t^3-\mathbf{16}t^5+\mathbf{10}t^6-\mathbf{1}t^8)_{\lambda}
\otimes(\mathbf{10}_{x} - \overline{\mathbf{16}}_{p}t^{-1} - \mathbf{16}_{\theta}t)
\bigr) \,,
\end{split}
\end{align}
and after the twisting~(\ref{eq:twist}), it becomes
\begin{align}
\Tilde{\mathbf{Z}}_{1}(\Tilde{t},\vec{\sigma})
&=
  (\mathbf{8}_{v}-\mathbf{8}_{a})\Tilde{t}^{-2}
 +(-\mathbf{56}_{va}+\mathbf{35}_{aa}+\mathbf{28}-\mathbf{8}_{s}+\mathbf{1})\Tilde{t}^{0}
 +(-\mathbf{56}_{vs}+\mathbf{56}_{sa})\Tilde{t}^{2} \nonumber\\
&\qquad -(-\mathbf{56}_{vs}+\mathbf{56}_{sa})\Tilde{t}^{6}
 -(-\mathbf{56}_{va}+\mathbf{35}_{aa}+\mathbf{28}-\mathbf{8}_{s}+\mathbf{1})\Tilde{t}^{9}
  -(\mathbf{8}_{v}-\mathbf{8}_{a})\Tilde{t}^{10} \nonumber\\
&\qquad + \Tilde{Z}_{0}(\Tilde{t},\vec{\sigma})
  \otimes\bigl((\mathbf{1}_{x}- \mathbf{8}_{p,a})\Tilde{t}^{-2}
    + (\mathbf{8}_{p,s}+\mathbf{8}_{\theta,a})\Tilde{t}^{0}
   + (\mathbf{1}_{x}+\mathbf{8}_{\theta,s})\Tilde{t}^{2} \bigr) \,.
\end{align}
A little algebra shows that again only the terms at $\Tilde{t}^{2,6}$ survives:
\begin{align}
\begin{split}
\Tilde{\mathbf{Z}}_{1}(\Tilde{t},\vec{\sigma})
&= -\Tilde{t}^{2}\bigl(\mathbf{35}+\mathbf{28}+\mathbf{1}+\mathbf{56}_{sa}+\mathbf{8}_{v}
 -\mathbf{56}_{va}-\mathbf{8}_{s}-\mathbf{56}_{vs}-\mathbf{8}_{a} \bigr) \\
&\qquad  + \Tilde{t}^{6}\bigl(\mathbf{35}+\mathbf{28}+\mathbf{1}+\mathbf{56}_{sa}+\mathbf{8}_{v}
 -\mathbf{56}_{va}-\mathbf{8}_{s}-\mathbf{56}_{vs}-\mathbf{8}_{a} \bigr) \,.
\end{split}
\end{align}
The cancellation among spurious states and ghosts
occurs as indicated in figure~\ref{fig:firstmassive}.
\begin{figure}[p]
  \centering
\begin{small}
\begin{tabular}{M|MMMMMMMMM}
 & \Tilde{t}^{-4} & \Tilde{t}^{-2} & \Tilde{t}^{0} & \Tilde{t}^{2} & \Tilde{t}^{4} & \Tilde{t}^{6} & \Tilde{t}^8 & \Tilde{t}^{10} & \Tilde{t}^{12}\\ \hline
-\overline{\mathbf{16}}t^{-1} & & -\mathbf{8}_{a} & -\mathbf{8}_{s} &&&
\\
\mathbf{45}t^{0} && \mathbf{8}_{v} & \mathbf{28}+\mathbf{1} & \mathbf{8}_{v}  &&&
\\
\mathbf{10}_{x}t^{0} & \mathbf{1} & & \mathbf{8}_{v} && \mathbf{1}  &&
\\
\mathbf{1}t^0 && & \mathbf{1}  &&&&
\\
-\mathbf{120}t^2 && & -\mathbf{28} & \underline{-\mathbf{56}_{sa}-\mathbf{8}_{v}} & -\mathbf{28}
\\
-\mathbf{100}_{x}t^2 & -\mathbf{1} & -\mathbf{8}_{v} &-\mathbf{1}&\underline{-\mathbf{35}} &-\mathbf{1}& -\mathbf{8}_{v}  &-\mathbf{1}
\\
& & & -\mathbf{8}_{v} & \underline{-\mathbf{28}} & -\mathbf{8}_{v}
\\
&& && \underline{-\mathbf{1}} &&&
\\
-2\cdot\mathbf{10}t^2 && & -2\cdot\mathbf{1} & -2\cdot\mathbf{8}_{v} & -2\cdot\mathbf{1}
\\
\mathbf{144}t^3 && & \mathbf{8}_{s} &  \underline{\mathbf{56}_{vs}+\mathbf{8}_{a}} &
  \mathbf{56}_{va}+\mathbf{8}_{s} & \mathbf{8}_{a}
\\
\overline{\mathbf{144}}_{x}t^3 & & \mathbf{8}_{a} & & \underline{\mathbf{56}_{va}+\mathbf{8}_{s}} &
  \mathbf{56}_{vs}+\mathbf{8}_{a} && \mathbf{8}_{s}
\\
\mathbf{16}_{x}t^3 && & \mathbf{8}_{s} &&& \mathbf{8}_{a}
\\
2\cdot\overline{\mathbf{16}}t^{3} && && 2\cdot\mathbf{8}_{a} & 2\cdot\mathbf{8}_{s}
\\
& &&&& \star
\\
-2\cdot\mathbf{16}t^{5} && & && -2\cdot\mathbf{8}_{s} & -2\cdot\mathbf{8}_{a}
\\
-\overline{\mathbf{16}}_{x}t^5 &&  & &-\mathbf{8}_{a}&& & -\mathbf{8}_{s}
\\
-\mathbf{144}_{x}t^5  && &  -\mathbf{8}_{s} && -\mathbf{56}_{vs}-\mathbf{8}_{a} &
  \underline{-\mathbf{56}_{va}-\mathbf{8}_{s}} && -\mathbf{8}_{a}
\\
-\overline{\mathbf{144}}t^5 && &  & -\mathbf{8}_{a} & -\mathbf{56}_{va}-\mathbf{8}_{s} &
  \underline{-\mathbf{56}_{vs}-\mathbf{8}_{a}} & -\mathbf{8}_{s}
\\
2\cdot\mathbf{10}t^6 &&&& & 2\cdot\mathbf{1} & 2\cdot\mathbf{8}_{v} & 2\cdot\mathbf{1}
\\
\mathbf{100}_{x}t^6 && &&&& \underline{\mathbf{1}}
\\
 && && &\mathbf{8}_{v} & \underline{\mathbf{28}} &\mathbf{8}_{v}
\\
 &&&\mathbf{1} & \mathbf{8}_{v}&\mathbf{1} & \underline{\mathbf{35}}& \mathbf{1} & \mathbf{8}_{v} & \mathbf{1}
\\
\mathbf{120}t^6 && &&& \mathbf{28} & \underline{\mathbf{56}_{sa}+\mathbf{8}_{v}} & \mathbf{28}
\\
-\mathbf{1}t^{8} && &&&&  & -\mathbf{1} &
\\
-\mathbf{10}_{x}t^{8} && &&& -\mathbf{1} && -\mathbf{8}_{v} && -\mathbf{1}
\\
-\mathbf{45}t^{8} && &&&& -\mathbf{8}_{v} & -\mathbf{28}-\mathbf{1} & -\mathbf{8}_{v}
\\
\mathbf{16}t^{9} &&&&&&& \mathbf{8}_{s} & \mathbf{8}_{a}
\\ \hline
& 0 & 0 & 0 &  -(\mathbf{35},\mathbf{28},\mathbf{1})  & 0 &  (\mathbf{35},\mathbf{28},\mathbf{1})& 0 &0 & 0
\\
&& & & -(\mathbf{56}_{sa},\mathbf{8}_{v})  && (\mathbf{56}_{sa},\mathbf{8}_{v})
\\
&&&& (\mathbf{56}_{va},\mathbf{8}_{s}) &&-(\mathbf{56}_{va},\mathbf{8}_{s})
\\
&&&& (\mathbf{56}_{vs},\mathbf{8}_{a}) &&-(\mathbf{56}_{vs},\mathbf{8}_{a})
\end{tabular}
\caption{Lightcone first massive states from level $1$ twisted character}
\label{fig:firstmassive}
\end{small}
\end{figure}

\subsubsection{Higher massive states}

The very same twisting procedure leads to the
lightcone spectrum for the higher massive states.
The computations are straightforward once
the spin partition functions $Z_{h}(t,\vec{\sigma})$ of the pure spinor are obtained.
We list the latter up to level $h=5$ in appendix~\ref{app:spincharacters},
so the interested reader can readily check the emergence of the lightcone spectrum.

\subsection{$Q_{0}$-cohomology and absence of on-shell condition}

Finally, let us study the relation between the cohomologies of $Q_{0}$ and $Q$
(or more precisely that of $Q_{0}+\delta$ and $Q+\delta$
where $\delta$ is either the \v{C}ech or Dolbeault operator).
It will be argued that $Q_{0}$ is an analog of
$k^{+}\sum_{n\ne0}\alpha^{-}_{n}c_{-n}$ of the bosonic string,
and in particular that it does not imply the on-shell condition.

Under the twisted $\Tilde{t}$-charge, $Q$ splits into three pieces
\begin{align}
\begin{split}
  Q &= Q_{0} + Q_{2} + Q_{4}\,, \\
  Q_{0} &= \lambda^{\alpha}d'_{\alpha}\,,\quad
 \bigl(  d'_{a} = p_{a} + k^{+}\theta_{a}\,,\;\; d'_{\Dot{a}} = p_{\Dot{a}} + \theta_{\Dot{a}}\del x^{\prime -} \bigr)\,, \\
  Q_{2} &= (\lambda^{\Dot{a}}\theta^{\Dot{a}})k^{-} + (\lambda\gamma^{i}\theta)\delx^{i} \,, \\
  Q_{4} &= (\lambda^{a}\theta^{a})\delx^{\prime +} - {1\over2}(\lambda\gamma^{\mu}\theta)(\theta\gamma_{\mu}\del\theta)\,,
\end{split}
\end{align}
where the notation $\delx^{\prime\pm}$ signifies the omission of zero-modes $k^{\pm}$.
(The \v{C}ech or Dolbeault operators $\delta$ also carries $\Tilde{t}$-charge $0$
and we implicitly include it in $Q_{0}$.)
$Q_{0}$ is certainly nilpotent,
but since it only contains the $k^{+}$ component of the momentum,
setting $Q_{0}=0$ cannot imply the on-shell condition.

In order to see that $Q_{0}$ indeed works as $k^{+}\sum_{n\ne0}\alpha^{-}_{n}c_{-n}$ of
the bosonic string, we study its cohomology directly,
by employing the method utilized in~\cite{Berkovits:2000nn}
to derive the (on-shell) lightcone spectrum from $Q$.

\subsubsection{Ghost-for-ghost method with an $SO(8)$ parameterization of pure spinor}

In section~\ref{sec:reducibility}, we analyzed the reducibility conditions
of the pure spinor constraint in an $SO(10)$ covariant manner.
As was noted in~\cite{Berkovits:2000nn}, there is a simpler version of this analysis
if one breaks the covariance down to $SO(8)$.

First, parameterize $SO(8)$ antichiral and chiral components of $\lambda^{\alpha}$ as
\begin{align}
  \lambda^{\Dot{a}} = s^{\Dot{a}}\,,\quad \lambda^{a}= (\fbslash{v}s)^{a} = v^i \gamma_{i}^{a\Dot{b}}s^{\Dot{b}} \,.
\label{eq:param}
\end{align}
$\lambda^{\alpha}$ satisfies the pure spinor condition
provided $s^{\Dot{a}}$ is constrained to be {\em null}, $s^{\Dot{a}}s^{\Dot{a}}=0$.
However, half of $v^{i}$ is spurious because of the gauge invariance
\begin{align}
  \delta_{\Lambda}v^{i} &= \Lambda^{a}(\gamma^{i}s)^{a}\quad\to\quad \delta_{\Lambda}\lambda^{a} = \Lambda^{a}(ss) = 0 \,.
\end{align}
Repeating the BRST construction (section~\ref{sec:BRST}) in an $SO(8)$ covariant manner,
one obtains a chain of free-field ghosts-for-ghosts\footnote{%
We departed from~\cite{Berkovits:2000nn} in notation to match the notation of the present paper.
In~\cite{Berkovits:2000nn}, the initial parameterization was chosen oppositely (i.e. $\lambda^{a}=s^a$)
and the ghosts $(b_{2n-1},c_{2n-1},\rho_{2n},\sigma_{2n})_{n\ge1}$ were denoted by $(u_n,t_n,w_n,v_n)$.}
\begin{align}
  (B_n,C_n) &\colon\quad  (b_1^{a},c_1^a)\,,\;\;(\rho_2^{i},\sigma_2^{i})\,,\;\;(b_3^{a},c_3^a)\,,\;\;(\rho_4^{i},\sigma_4^{i})\,,\;\;\cdots\,,
\end{align}
where as before $(b^{a}_{2n-1},c^{a}_{2n-1})$ are fermionic and $(\rho^i_{2n},\sigma^i_{2n})$ are bosonic.
Introducing a fermionic ghost pair $(b,c)$ for the remaining constraint $s^{\Dot{a}}s^{\Dot{a}}=0$
(and denoting the conjugate to $s^{\Dot{a}}$ and $v^{i}$ by $t^{\Dot{a}}$ and $w^{i}$),
the mini-BRST operator reads~\cite{Berkovits:2000nn}
\begin{align}
  D &= \int \bigl(  bs^{\Dot{a}}s^{\Dot{a}} + s^{\Dot{a}}\calG_{\text{gh}}^{\Dot{a}} + c\calT_{\text{gh}}  \bigr)\,,
\end{align}
where
\begin{align}
\begin{split}
\calG_{\text{gh}}^{\Dot{a}} &=  -(\fbslash{w}c_{1})^{\Dot{a}} + (b_1\fbslash{\sigma}_2)^{\Dot{a}} - (\fbslash{\rho}_2c_3)^{\Dot{a}} + \cdots\,,  \\
  \calT_{\text{gh}} &= (w^{i}\sigma^{i}_{2}) + (b_1c_3) + (\rho_{2}^{i}\sigma_{4}^{i}) + \cdots \,.
\end{split}
\end{align}
Using a regularization $1-1+1-\cdots = \lim_{x\to1}(1+x)^{-1} =1/2$
familiar in covariant treatments of the $\kappa$-symmetry, it is straightforward to check
that the combined system of $(t^{\Dot{a}},s^{\Dot{a}};w^i,v^i;B_n,C_n,b,c)$ has the desired central charge $22$.
Moreover, one can construct a set of generators for the full $SO(10)$ Lorentz current algebra (with appropriate level $-3$),
under which $D$ and the physical BRST operator $Q'$ (to be defined shortly) are
invariant~\cite{Berkovits:2000nn,Berkovits:2001mx}.

\bigskip
In~\cite{Berkovits:2000nn}, the $SO(8)$ mini-BRST operator $D$ was used to construct
the ghost extended physical BRST operator $Q'=D+\int \lambda^{\alpha}d_{\alpha}+\cdots$, whose cohomology is equivalent to
that of $Q=\int \lambda^{\alpha}d_{\alpha}$.
The operator can be written in the same form as $D$,
\begin{align}
 \label{eq:QprimeTot}
  Q' &= \int \bigl(  bs^{\Dot{a}}s^{\Dot{a}} + s^{\Dot{a}}\calG^{\Dot{a}} + 2c\calT   \bigr) \,,
\end{align}
provided one defines
\begin{align}
\begin{split}
  \calG^{\Dot{a}} &= d^{\Dot{a}} + (\fbslash{v}d)^{\Dot{a}} + \calG^{\Dot{a}}_{\text{gh}} \,,\\
  \calT &= -(\pi^{-}+ 2v^i\pi^{i}+ v^2\pi^{+}) + 2c^{a}_1 d^{a}+ \calT_{\text{gh}} \,, \\
\end{split}
\end{align}
with $\pi^{\mu}$ being the superinvariant momentum $\delx^{\mu} - \theta\gamma^{\mu}\del\theta$.
The combinations $(\calG^{\Dot{a}},\calT)$ and $(\calG^{\Dot{a}}_{\text{gh}},\calT_{\text{gh}})$
satisfy the same algebra
\begin{align}
\calG^{\Dot{a}}(z)\calG^{\Dot{b}}(w) &= {-2\delta^{\Dot{a}\Dot{b}} \calT \over z-w} \,,\quad
\calG^{\Dot{a}}(z)\calT(w) = \calT(z)\calT(w) = \text{regular} \,.
\end{align}
This algebra appears repeatedly in the pure spinor formalism,
and is related to the algebra generated by the first-class
part of the Green-Schwarz-Siegel constraint $d_{\alpha}$~\cite{Siegel:1985xj}.

Now that the ghost extended physical BRST operator~(\ref{eq:QprimeTot}) is written entirely
in terms of free fields, the analysis of its cohomology is straightforward
as explained in~\cite{Berkovits:2000nn}.
Let us apply the argument to the case at hand,
where the full operator $Q$ is replaced by its $\Tilde{t}$-charge $0$ piece $Q_{0}$.

\subsubsection{Lightcone ``off-shell'' spectrum from $Q_{0}$-cohomology}

By coupling the $SO(8)$ mini-BRST operator $D$ to $Q_{0}$,
one concludes that the cohomology of $Q_{0}$
is equivalent to that of the $\tilde t$-charge 0 contribution to $Q'$
which is
\begin{align}
\begin{split}
  Q'_0 &=  \int\bigl(  bs^{\Dot{a}}s^{\Dot{a}} + s^{\Dot{a}}\calG^{\Dot{a}}_{0} + 2c\calT_{0}  \bigr) \,, \\
\end{split}
\end{align}
where
\begin{align}
\begin{split}
\calG^{\Dot{a}}_0 &= d^{\prime \Dot{a}} +(\fbslash{v}d')^{\Dot{a}} + \calG_{\text{gh}}^{\Dot{a}} \,,\\
\calT_0 &= -{1\over2}\delx^{\prime -} + v^{2}k^{+} + 2c_1^{a}d^{\prime a} + \calT_{\text{gh}} \,.
\end{split}
\end{align}

To study the cohomology of $Q'_0$, it is convenient to introduce the grading defined by
\begin{align}
 l(p_{\Dot{a}},\theta_{\Dot{a}},\delx^{\prime\pm}) = (1,-1,\pm1)\,.
\end{align}
Under the $l$-grading, $Q'_0$ splits to
\begin{align}
\begin{split}
 Q'_0 &= Q'_{0,1} + Q'_{0,0} \,,\\
 Q'_{0,1} &= \text{\Large$\smallint$}( s^{\Dot{a}}p_{\Dot{a}} - c'\delx^{\prime -} )\,,\\
Q'_{0,0} &= (\text{rest}) \\
  &= \text{\Large$\smallint$}( bs^{\Dot{a}}s^{\Dot{a}} + s^{\Dot{a}}\theta_{\Dot{a}}\delx^{\prime -}
   + s^{\Dot{a}}(\fbslash{v}d')^{\Dot{a}} + s^{\Dot{a}}\calG^{\Dot{a}}_{\text{gh}}
   + 2cv^2 k^+
   + 4c(c_1^a d^{\prime a})
   + 2c\calT_{\text{gh}}
  ) \,.
  \end{split}
\end{align}
It immediately follows that two quartets
\begin{align}
(p_{\Dot{a}},\theta_{\Dot{a}},t^{\Dot{a}},s^{\Dot{a}})\,,\quad
(\delx^{\prime \pm}, b',c')\,,
\end{align}
decouple from the cohomology.
Furthermore, the conditions implied by $Q'_{0}$ on the remaining fields
\begin{align}
 (k^{\pm}, \delx^{i})\,,\quad(p_{a},\theta^{a})\,,\quad (b_0,c_0)\,,\quad(w^i,v^i)\,,\quad (B_n,C_n)\,,
\end{align}
are the cohomology condition of
\begin{align}
\begin{split}
 Q'' &= c_{0}\text{\Large$\smallint$}(k^{+}v^2 + 2(c_1 d') + \calT_{\text{gh}}) \\
  &= c_{0}(k^{+}v^2 + 2(d' c_1 ) + w^{i}\sigma_{2}^{i} + b^{a}_1c_3^{a} + \cdots)_0 \,.
\end{split}
\end{align}
Remembering $d'_a= p_a+k^{+}\theta_a$,
it then follows that the cohomology of $Q'$ (and hence that of $Q_0$) is
spanned by
\begin{align}
  \delx^{i}\,,\quad (p_a-k^{+}\theta_{a})\,,
\end{align}
on the super-Maxwell ground states $(1,\lambda\gamma^{\mu}\theta,(\lambda\gamma^{\mu}\theta)(\gamma_{\mu}\theta)_{\alpha},\cdots,\lambda^{(3)}\theta^{(5)})$
(with appropriate BRST ghost extensions).

If the full operator $Q$ was used in place of $Q_{0}$
one would find $c_{0}k^{-}$ (among other terms) in the final form of $Q''$,
and this leads to the on-shell condition~\cite{Berkovits:2000nn}.
Summing up, we have learned that the physical BRST operator $Q$ of the pure spinor formalism
contains a piece $Q_{0}$ which plays an analogous role as $k^{+}\sum_{n\ne0} \alpha^{-}_{n}c_{-n}$
of the bosonic string,
and the role of the rest of $Q$ is to impose the on-shell condition
on the ``off-shell'' lightcone spectrum.
This was what we wanted to explain.

\subsubsection{Equivalence with light-cone partition function}
\label{subsubsec:equiv}

In this subsection, we shall argue that after performing the
twist of~(\ref{eq:twist}), the ghost modes of $(\omega,\lambda)$ cancel
against the longitudinal modes of $(p,\theta,x)$ such that only
the light-cone degrees of freedom contribute to the partition
function. Since the light-cone partition function is modular
invariant, the complete partition function which is twisted with
respect to ~(\ref{eq:twist}) must also be modular invariant. 

To show that after performing the 
twist of~(\ref{eq:twist}), the ghost modes of $(\omega,\lambda)$ cancel
against the longitudinal modes of $(p,\theta,x)$, it will
be convenient to parameterize $\lambda^\alpha$ in terms of
$\lambda^{\dot a}$ and $\lambda^a$ as in~(\ref{eq:param}). 
Since $\lambda^{\dot a}$ satisfies the constraint
$\lambda^{\dot a}\lambda^{\dot a}=0$, it can be described by 8 unconstrained
bosons $s^{\dot a}$ together with a fermion $c$ which replaces the 
constraint. Note that $s^{\dot a}$ carries $+2$ $\tilde t$-charge
and $c$ carries $+4$ $\tilde t$-charge. Their conjugate momenta,
which will be called $t_{\dot a}$ and $b$, carry opposite $\tilde t$-charge
and carry conformal weight $+1$.

The variables $(t_{\Dot{a}},s^{\dot a})$ have the same $SO(1,1)\times
SO(8)$ Lorentz spin and $\tilde t$-charge as $(\gamma^- p,\gamma^+\theta)$
and have the opposite statistics. So the partition function for
$(t_{{\dot a}},s^{\dot a})$ cancels the partition function for 
$(\gamma^- p,\gamma^+\theta)$. Similarly, $(b,c)$ have the same
$SO(1,1)\times
SO(8)$ Lorentz spin and $\tilde t$-charge as $(\del x'^+,\partial x'^-)$
and have the opposite statistics. So just as in bosonic string
theory, the partition function for the
$(b,c)$ ghosts cancels the partition function for the longitudinal
$(\del x'^+,\partial x'^-)$ variables.

Finally, one has the remaining $\lambda^a$ ghost variables which
satisfy the constraint $s^{\dot b} \gamma_i^{a \dot b} \lambda^a=0$.
This constraint implies that only four of the eight $\lambda^a$
variables are independent and since
a null $s^{\dot a}$ breaks $SO(8)$ to $U(4)$, one can write this constraint
as $\lambda^A=0$ where $\lambda^a$ has been decomposed into
its $U(4)$ components as $\lambda^a=(\lambda^A,\lambda^{\overline A})$
for $A=1$ to 4. 
Under this decomposition of $SO(8)$ into $U(4)$, 
the variables $\gamma^-\theta$ and $\gamma^+ p$ decompose
as $(\theta^A,\theta^{\overline A})$ and $(p_A, p_{\overline A})$.
And since $(\omega_{\overline A},\lambda^{\overline A})$ carries
the same $SO(1,1)\times U(4)$ Lorentz spin and $\tilde t$-charge
as
$(p_{\overline A}, \theta^{\overline A})$ and has the opposite statistics,
their partition functions cancel out.

So after cancelling out these partition functions of the ghost and
longitudinal matter variables, the only remaining
contribution comes from the light-cone variables $(p_{A},\theta^A,x^{i})$
which produce the standard modular invariant
light-cone partition function of the superstring.

%%%%%%%%%%%%%%%%%%%%%%%%%%%%%%%%%%%%%%%%%%%%%%%%%%%%%%%%%%%%%%%%
\section{Summary and future applications}
\label{sec:summary}

In this paper, we computed the partition function for pure spinors
up to the fifth mass level using both the ghost-for-ghost method
and the fixed-point method. After including the partition function
for the matter variables, we showed agreement with the light-cone superstring
spectrum up to the fifth mass level.

The main surprise in the computation is the appearance of fermionic
states in the pure spinor partition function starting at the second
mass level. These fermionic states
all correspond to three-forms on the pure spinor space, and
are related to a term in the $b$ ghost
in the pure spinor formalism. Based on the symmetry properties
of the pure spinor partition function, we conjecture there
is a one-to-one correspondence between these fermionic states
and the usual bosonic states which are associated to gauge-invariant
polynomials in $(\lambda^\a,\w_\a)$.

There are several possible applications of these results for
amplitude computations and for superstring field theory. Using
the RNS formalism, scattering amplitudes can be computed either using
conformal field theory techniques or using the operator method. Although
conformal field theory techniques are more convenient for multiloop
amplitudes, the operator method is convenient for one-loop computations
where one expresses the amplitude as a trace over states in the Hilbert
space.

In this paper, the pure spinor partition function was only computed
up to the fifth mass level, but it might be possible to extend our
results and construct an explicit formula for the complete pure spinor
partition function. One could then use the operator method in
the pure spinor formalism, which might simplify the
computations of one-loop amplitudes.

Another possible application of our results concerns the role of the
$b$ ghost in computing multiloop scattering amplitudes. As discussed in~\cite{Berkovits:2005bt}\cite{Berkovits:2006vi},
there are subtleties in computing $g$-loop amplitudes when the $3g-3$
$b$ ghosts contribute terms which diverge as fast as $(\lambda\lb)^{-11}$
when $\lambda^\a\to 0$. Since the integral
$\int \targetd^{11}\lambda ~\targetd^{11}\lb (\lambda\lb)^{-11}$ diverges near $\lambda^\a=0$,
the functional integral over the pure spinor ghosts
needs to be regularized when there are terms which diverge
as fast as $(\lambda\lb)^{-11}$. A consistent BRST-invariant
regularization for multiloop amplitudes was defined in~\cite{Berkovits:2006vi}, however,
this regularization was complicated and not very practical for computations.

In this paper, it was argued that the only states in the cohomology of
the pure spinor Hilbert space correspond to functions which are either
zero-forms or three-forms. These operators are either regular when
$\lambda^\a\to 0$, or diverge as $(\lambda\lb)^{-3}$. When multiplying $3g-3$  $b$
ghosts, one can in general get terms which diverge as fast as $(\lambda\lb)^{-9g+9}$.
However, this cohomology argument implies that
there must exist an operator $\Lambda(z_1, ...,z_{3g-3})$ such that,
after ignoring total derivatives in the Teichm\"uller parameters,
$$b(z_1)b(z_2)... b(z_{3g-3}) - [Q,\Lambda(z_1,...,z_{3g-3})]$$
diverges no faster than $(\lambda\lb)^{-3}$.
(Note that
$[Q, b(z_1) b(z_2) ... b(z_{3g-3})]$ is proportional to
total derivatives in the Teichm\"uller parameters, so the
cohomology argument can only be used if one can ignore these total
derivatives.)

Since BRST-trivial operators do not affect on-shell scattering amplitudes,
one can use $[Q,\Lambda(z_1,...,z_{3g-3})]$ to remove the dangerous
divergences when $(\lambda\lb)\to 0$. Even though the construction of $\Lambda$
may be complicated,
this is an alternative BRST-invariant regularization
method for $(\lambda\lb)\to 0$ divergences
which may be more efficient for computations than the regularization
method described in~\cite{Berkovits:2006vi}.

A third possible application of these results is for superstring
field theory. In~\cite{Berkovits:2005bt}, a cubic open superstring field theory action
was constructed using the pure spinor formalism. However, the correct
definition of the Hilbert space was unclear because of the possibility
of states diverging when $(\lambda\lb)\to 0$. Using the results of this paper,
one now knows that the Hilbert space must at least allow states which
diverge as $(\lambda\lb)^{-3}$ in order to reproduce the correct massive
spectrum. But it is an open question if one can consistently define
a multiplication rule for string fields in such a manner that states
diverging like $(\lambda\lb)^{-11}$ are never produced.
Note that in string field theory,
one cannot use BRST-trivial operators to remove these
dangerous states since off-shell string fields are not
necessarily BRST-closed.

%%%%%%%%%%%%%%%%%%%%%%%%%%%%%%%%%%%%%%%%%%%%%%%%%%%%%%%%%%%%%%%%
\section*{Acknowledgment}

Part of the research was done while NB and NN were visiting
Princeton University in the spring of 2007 and while NB was
visiting IHES in the winter of 2007. 
The research of YA was supported by FAPESP grant 06/59970-5.
The research of EAA was supported by FAPESP grant 04/09584-6.
The research of NB was partially supported by CNPq grant 300256/94-9 and FAPESP grant
04/11426-0. The research of NN
was partially supported by European RTN
under the contract 005104 "ForcesUniverse", by  {\it l'Agence Nationale de la Recherche} under the grants
ANR-06-BLAN-3$\_$137168 and ANR-05-BLAN-0029-01, by the Russian Foundation for Basic Research through the grants
RFFI 06-02-17382 and NSh-8065.2006.2, and by DARPA
through its Program ``Fundamental Advances in Theoretical
Mathematics''. NN thanks E.~Frenkel, A.~Polyakov and S.~Shatashvili for discussions.

%%%%%%%%%%%%%%%%%%%%%%%%%%%%%%%%%%%%%%%%%%%%%%%%%%%%%%%%%%%%%%%%
\section*{Appendix}
\appendix
\setcounter{equation}{0}
\def\thesection{\Alph{section}}
\renewcommand{\theequation}{\Alph{section}.\arabic{equation}}

%%%%%%%%%%%%%%%%%%%%%%%%%%%%%%%%%%%%%%%%%%%%%%%%%%%%%%%%%%%%%%%%
\section{$SO(10)$ conventions and formulas}
\label{app:conventions}

\subsection{Dynkin labels}
\label{app:Dynkin}

As is well known, all the irreducible representations
of $SO(10)$ can be labeled by five integers called Dynkin labels.
Those are nothing but the highest weights of the representations
in an appropriate basis.
In our convention,
\begin{align}
\begin{split}
&\text{vector:}\quad  \dynkin(10000)= \mathbf{10}\,, \\
&\text{$2$-form:}\quad  \dynkin(01000)= \mathbf{45}\,,\\
&\text{$3$-form:}\quad  \dynkin(00100)= \mathbf{120}\,, \\
&\text{antichiral spinor}:\quad \dynkin(00010)= \overline{\mathbf{16}}\,, \\
&\text{chiral spinor}:\quad \dynkin(00001)= \mathbf{16}\,.
\end{split}
\end{align}

When computing the partition functions, it is sometimes more convenient
to introduce an orthogonal basis for the Cartan subalgebra, $e_{a}$ ($a=1, \ldots , 5$)
such that the fundamental roots are
\begin{align}
  e_{1} - e_{2},\quad  e_{2} - e_{3},\quad  e_{3} - e_{4},\quad e_{4}\pm e_{5} \,.
\end{align}
We then denote the character of $e_{a}$ by $\mathe^{\sigma_{a}}$
where $\sigma_{a}$ is a formal variable for bookkeeping.
Also the weight vectors in this basis are denoted by square bracket:
\begin{align}
  \mu &= \sum_{a} \mu_{a}e_{a}\quad\leftrightarrow\quad
 [\mu_1\mu_2\mu_3\mu_4\mu_5]\quad\leftrightarrow\quad
\mathe^{\mu\cdot \sigma}\,.
\end{align}
The components $\mu_{a}$'s take values in half integers and are related to
the (integer valued) Dynkin labels $(a_1a_2a_3a_4a_5)$ by
\begin{align}
\begin{pmatrix}
  \mu_{1} \\  \mu_{2} \\  \mu_{3} \\  \mu_{4} \\  \mu_{5}
\end{pmatrix}
=
 \begin{pmatrix}
  1 & 1 & 1 & 1/2 & 1/2 \\
  0 & 1 & 1 & 1/2 & 1/2 \\
  0 & 0 & 1 & 1/2 & 1/2 \\
  0 & 0 & 0 & 1/2 & 1/2 \\
  0 & 0 & 0 & -1/2 & 1/2
  \end{pmatrix}
\begin{pmatrix}
 a_1 \\ a_2 \\ a_3 \\ a_4 \\ a_5
\end{pmatrix} \,.
\end{align}
We refer to this basis as the ``five sign basis'' because the
weights and characters of chiral spinors are expressed as
\begin{align}
  \mu = {1\over2}[\pm1,\pm1,\pm1,\pm1,\pm1]
\quad\leftrightarrow\quad
 \mathe^{{1\over2}(\pm \sigma_1 \pm \sigma_2\pm\sigma_3\pm\sigma_4\pm\sigma_5)} \,,
\end{align}
with even number of minus signs.

\subsection{Some dimension formulas}

Dimensions of the $SO(10)$ irreducible representations are given by
\begin{align}
&\dim\dynkin(a\,b\,c\,d\,e)\nonumber\\
&= {1\over 2^{4}\cdot 3^{4}\cdot 4^{3}\cdot5^{2}\cdot 6\cdot 7} \Bigl\{
 (a+1)(b+1)(c+1)(d+1)(e+1) \nonumber\\
&\qquad(a+b+2)(b+c+2)(c+d+2)(c+e+2) \nonumber\\
&\qquad (a+b+c+3)(b+c+d+3)(b+c+e+3)(c+d+e+3)   \nonumber\\
&\qquad  (a+b+c+d+4)(a+b+c+e+4)(b+c+d+e+4)(b+2c+d+e+5) \nonumber\\
&\qquad  (a+b+c+d+e+5) (a+b+2c+d+e+6) (a+2b+2c+d+e+7)
\Bigr\}\,.
\end{align}
Of special interest are
the `(chiral) pure spinor representations' $\dynkin(0000n)$,
which have the following dimensions
\begin{align}
\dim\dynkin(0000n)
&= {(n+7)(n+6)(n+5)^2(n+4)^2(n+3)^2(n+2)(n+1) \over 7\cdot6\cdot5^2\cdot4^2\cdot3^2\cdot2} \,.
\end{align}

%%%%%%%%%%%%%%%%%%%%%%%%%%%%%%%%%%%%%%%%%%%%%%%%%%%%%%%%%%%%%%%%
\section{Table of partition functions}
\label{app:characters}

\subsection{Partition functions without spin: number of states}
List of coefficients $N_{m,n}$ present in the expansion
$Z(q,t)=\sum_{m\ge0}\sum_{n}N_{m,n}q^{m}t^{n}$ of the pure spinors
partition function. We include the usual gauge invariant states
($N_{m,n}>0$) as well as the extra states ($N_{m,n}<0$) which are
described using BRST or \v{C}ech/Dolbeault cohomologies (of third
degree).
\begin{align}
\label{eq:app:Zqt}
\begin{tabular}{|c|c|c|c|c|c|c|c|c}
    \hline
$n$ &  $N_{0,n}$    &   $N_{1,n}$  & $N_{2,n}$  &  $N_{3,n}$    &   $N_{4,n}$  & $N_{5,n}$ & $N_{6,n}$ & $\cdots$   \\
    \hline
-8   & 0 & 0 & 0 & 0 & 0 & 0 &  $-$2772  \\
-7   & 0 & 0 & 0 & 0 & 0 & $-$672 & $-$19824   \\
-6   & 0 & 0 & 0 & 0 & $-$126 & $-$4068 & $-$70522 \\
-5   & 0 & 0 & 0 & $-$16 & $-$592 & $-$11408  & $-$153408  \\
-4   & 0 & 0 & $-$1 & $-$46 & $-$1073 & $-$16974 & $-$205373 \\
-3   & 0 & 0 & 0 & $-$16 & $-$592 & $-$11408  & $-$152736 \\
-2   & 0 & 0 & 0 & 0 & 0 & 0  & 0 \\
-1   & 0 & 0 & 16 & 592 & 11408 & 152736  & 1597520 \\
0   & 1 & 46 & 1073 & 16974 & 205373 & 2031130  & 17130386 \\
1   & 16 & 592 & 11408 & 153408 & 1617344 & 14228752  & 108567392 \\
2   & 126 & 4068 & 70522 & 868012 & 8479364 & 69771888  &  501686294 \\
3   & 672 & 19824 & 320304 & 3716208 &  34489920 & 271222800  & 1872478496 \\
4   & 2772 & 76824 & 1180602 & 13125484 &1173525227 & 892615196  & 5979762150 \\
\vdots   & \vdots & \vdots  & \vdots & \vdots &  \vdots & \vdots &
\vdots
\end{tabular}
\end{align}

\subsection{Spin partition functions}
\label{app:spincharacters}

For convenience, we here list the partition functions with spin dependence
up to fifth Virasoro levels.
Partition functions at each level are of the form
\begin{align}
 Z_{h}(t,\vec{\sigma}) = { P_{h}(t,\vec{\sigma}) \over (1-t)^{S} }
 \;\;\;\;\; \text{where} \;\;\;\; (1-t)^{S} \equiv \prod_{\mu\in S}(1-t \mathe^{\mu\cdot \sigma}),\quad S = \dynkin(00001) = \mathbf{16}
\end{align}
and $P_{h}(t,\vec{\sigma})$ is a polynomial of $t$ with coefficients taking values
in the representations of $SO(10)$.
For brevity, we only write the numerator $P_{h}(t)$.
Again, formulas include the extra states in the third cohomology.

\paragraph{Level $0$:}
\begin{align}
  \label{eq:app:Z0spin}
P_{0}(t,\vec{\sigma})
&= \dynkin(00000)_{1}
 - \dynkin(10000)_{10}t^{2}
 + \dynkin(00010)_{16}t^{3} \nonumber\\
&\qquad - \dynkin(00001)_{16}t^{5}
 + \dynkin(10000)_{10}t^{6}
 - \dynkin(00000)_{1}t^{8}
\end{align}

\paragraph{Level $1$:}
\begin{align}
P_{1}(t,\vec{\sigma}) &= \bigl(\dynkin(01000)_{45} + \dynkin(00000)_{1}\bigr)
  - \dynkin(10010)_{144}t^{1}
  + \bigl(\dynkin(00020)_{126}- \dynkin(10000)_{10} \bigr)t^{2}  \nonumber\\
&\qquad  + \dynkin(00010)_{16}t^{3}
  - \dynkin(00001)_{16}t^{5}
  - \bigl(\dynkin(00002)_{126}- \dynkin(10000)_{10} \bigr) t^{6} \nonumber\\
&\qquad  + \dynkin(10001)_{144}t^{7}
  -\bigl(\dynkin(01000)_{45} + \dynkin(00000)_{1}\bigr) t^{8}
\end{align}

\paragraph{Level $2$:}
\begin{align}
&P_{2}(t,\vec{\sigma}) \nonumber\\
&=
   -\dynkin(00000)_{1}t^{-4}
   +\dynkin(00001)_{16}t^{-3}
   -\dynkin(00100)_{120}t^{-2} \nonumber\\&\qquad
   + \bigl(   +\dynkin(01010)_{560} +\dynkin(00010)_{16} \bigr)t^{-1} \nonumber\\&\qquad
   +  \bigl(   -\dynkin(10020)_{1050} +\dynkin(01000)_{45} +2\dynkin(00000)_{1} \bigr)t^{0}
   +  \bigl(   +\dynkin(00030)_{672} -\dynkin(10010)_{144} \bigr)t^{1} \nonumber\\&\qquad
   +  \bigl(   -\dynkin(11000)_{320} +\dynkin(00020)_{126} -2\dynkin(10000)_{10} \bigr)t^{2} \nonumber\\&\qquad
   +  \bigl(   +\dynkin(01010)_{560} +2\dynkin(00010)_{16} \bigr)t^{3} \nonumber\\&\qquad
   +  \bigl(   -\dynkin(01001)_{560} -2\dynkin(00001)_{16} \bigr)t^{5} \nonumber\\&\qquad
   +  \bigl(   +\dynkin(11000)_{320} -\dynkin(00002)_{126} +2\dynkin(10000)_{10} \bigr)t^{6} \nonumber\\&\qquad
   +  \bigl(   -\dynkin(00003)_{672} +\dynkin(10001)_{144} \bigr)t^{7}
   +  \bigl(   +\dynkin(10002)_{1050} -\dynkin(01000)_{45} -2\dynkin(00000)_{1} \bigr)t^{8} \nonumber\\&\qquad
   +  \bigl(   -\dynkin(01001)_{560} -\dynkin(00001)_{16} \bigr)t^{9} \nonumber\\&\qquad
   +     \dynkin(00100)_{120}t^{10}
   -\dynkin(00010)_{16}t^{11}
   + \dynkin(00000)_{1}t^{12}
 \end{align}

\paragraph{Level $3$:}
  \begin{align}
&P_{3}(t,\vec{\sigma}) \nonumber\\
&=
    -\dynkin(00010)_{16}t^{-5}
    +\dynkin(00011)_{210}t^{-4}
    -\dynkin(00110)_{1200} t^{-3}
    +\dynkin(01020)_{3696} t^{-2}\nonumber\\&\qquad
   +  \bigl(   -\dynkin(10030)_{5280} +\dynkin(01010)_{560} +2\dynkin(00010)_{16} \bigr)t^{-1} \nonumber\\&\qquad
   +  \bigl(   +\dynkin(00040)_{2772} -\dynkin(10020)_{1050} +\dynkin(02000)_{770} +3\dynkin(01000)_{45} +3\dynkin(00000)_{1} \bigr)t^{0}\nonumber\\&\qquad
   +  \bigl(   -\dynkin(11010)_{3696} +\dynkin(00030)_{672} -3\dynkin(10010)_{144} \bigr)t^{1}\nonumber\\&\qquad
   +  \bigl(   +\dynkin(01020)_{3696} -2\dynkin(11000)_{320} +3\dynkin(00020)_{126} -3\dynkin(10000)_{10} \bigr)t^{2} \nonumber\\&\qquad
   +  \bigl(   +2\dynkin(01010)_{560} +3\dynkin(00010)_{16} \bigr)t^{3}\nonumber\\&\qquad
   +  \bigl(   -2\dynkin(01001)_{560} -3\dynkin(00001)_{16} \bigr)t^{5}  \nonumber\\&\qquad
   +  \bigl(   -\dynkin(01002)_{3696} +2\dynkin(11000)_{320} -3\dynkin(00002)_{126} +3\dynkin(10000)_{10} \bigr)t^{6}\nonumber\\&\qquad
   +  \bigl(   +\dynkin(11001)_{3696} -\dynkin(00003)_{672} +3\dynkin(10001)_{144} \bigr)t^{7}\nonumber\\&\qquad
   +  \bigl(   -\dynkin(00004)_{2772} +\dynkin(10002)_{1050} -\dynkin(02000)_{770} -3\dynkin(01000)_{45} -3\dynkin(00000)_{1} \bigr)t^{8}\nonumber\\&\qquad
   +  \bigl( +\dynkin(10003)_{5280} -\dynkin(01001)_{560} -2\dynkin(00001)_{16} \bigr)t^{9}\nonumber\\&\qquad
   -\dynkin(01002)_{3696}t^{10}
   +\dynkin(00101)_{1200} t^{11}
   -\dynkin(00011)_{210} t^{12}
   +\dynkin(00001)_{16} t^{13}
  \end{align}

\paragraph{Level $4$:}
\begin{align}
&P_{4}(t,\vec{\sigma}) \nonumber\\
&= -\dynkin(00020)_{126} t^{-6}
    + \bigl(   +\dynkin(00021)_{1440} -\dynkin(00010)_{16} \bigr)t^{-5} \nonumber\\&\qquad
    + \bigl(   -\dynkin(00120)_{6930} +\dynkin(00011)_{210} -\dynkin(00000)_{1} \bigr)t^{-4} \nonumber\\&\qquad
    + \bigl(   +\dynkin(01030)_{17280} -\dynkin(00110)_{1200} +\dynkin(00001)_{16} \bigr)t^{-3} \nonumber\\&\qquad
    + \bigl(   -\dynkin(10040)_{20790} +\dynkin(01020)_{3696} +\dynkin(00020)_{126} -\dynkin(00100)_{120} \bigr)t^{-2}\nonumber\\&\qquad
    + \bigl(   +\dynkin(00050)_{9504} -\dynkin(10030)_{5280} +\dynkin(02010)_{8064} \nonumber\\&\qquad\qquad+\dynkin(10001)_{144} +3\dynkin(01010)_{560} +4\dynkin(00010)_{16} \bigr)t^{-1}\nonumber\\&\qquad
    + \bigl(   -\dynkin(11020)_{23040} +\dynkin(00040)_{2772} -3\dynkin(10020)_{1050} \nonumber\\&\qquad\qquad +2\dynkin(02000)_{770} +5\dynkin(01000)_{45} +6\dynkin(00000)_{1} \bigr)t^{0}\nonumber\\&\qquad
    + \bigl(   +\dynkin(01030)_{17280} -\dynkin(20001)_{720} -2\dynkin(11010)_{3696} +3\dynkin(00030)_{672} -5\dynkin(10010)_{144} \bigr)t^{1}\nonumber\\&\qquad
    + \bigl(   -\dynkin(12000)_{4410} +2\dynkin(01020)_{3696} -4\dynkin(11000)_{320} +5\dynkin(00020)_{126} -6\dynkin(10000)_{10} \bigr)t^{2}\nonumber\\&\qquad
    + \bigl(   +\dynkin(02010)_{8064} +\dynkin(00021)_{1440} +4\dynkin(01010)_{560} +6\dynkin(00010)_{16} \bigr)t^{3}\nonumber\\&\qquad
    + \bigl(   -\dynkin(02001)_{8064} -\dynkin(00012)_{1440} -4\dynkin(01001)_{560} -6\dynkin(00001)_{16} \bigr)t^{5}\nonumber\\&\qquad
    + \bigl(   +\dynkin(12000)_{4410} -2\dynkin(01002)_{3696} +4\dynkin(11000)_{320} -5\dynkin(00002)_{126} +6\dynkin(10000)_{10} \bigr)t^{6}\nonumber\\&\qquad
    + \bigl(   -\dynkin(01003)_{17280} +\dynkin(20010)_{720} +2\dynkin(11001)_{3696} -3\dynkin(00003)_{672} +5\dynkin(10001)_{144} \bigr)t^{7}\nonumber\\&\qquad
    + \bigl(   +\dynkin(11002)_{23040} -\dynkin(00004)_{2772} +3\dynkin(10002)_{1050} \nonumber\\&\qquad\qquad -2\dynkin(02000)_{770} -5\dynkin(01000)_{45} -6\dynkin(00000)_{1} \bigr)t^{8}\nonumber\\&\qquad
    + \bigl(   -\dynkin(00005)_{9504} +\dynkin(10003)_{5280} -\dynkin(02001)_{8064} \nonumber\\&\qquad\qquad -\dynkin(10010)_{144} -3\dynkin(01001)_{560} -4\dynkin(00001)_{16} \bigr)t^{9}\nonumber\\&\qquad
    + \bigl(   +\dynkin(10004)_{20790} -\dynkin(01002)_{3696} -\dynkin(00002)_{126} +\dynkin(00100)_{120} \bigr)t^{10}\nonumber\\&\qquad
    + \bigl(   -\dynkin(01003)_{17280} +\dynkin(00101)_{1200} -\dynkin(00010)_{16} \bigr)t^{11}\nonumber\\&\qquad
    + \bigl(   +\dynkin(00102)_{6930} -\dynkin(00011)_{210} +\dynkin(00000)_{1} \bigr)t^{12}\nonumber\\&\qquad
    + \bigl(   -\dynkin(00012)_{1440} +\dynkin(00001)_{16} \bigr)t^{13}\nonumber\\&\qquad
    + \dynkin(00002)_{126} t^{14}
\end{align}

\paragraph{Level $5$:}
\begin{align}
&P_{5}(t,\vec{\sigma}) \nonumber\\
&=  -\dynkin(00030)_{672} t^{-7}\nonumber\\&\qquad
    + \bigl(   +\dynkin(00031)_{6930} -\dynkin(00020)_{126} -\dynkin(00100)_{120} \bigr)t^{-6}\nonumber\\&\qquad
    + \bigl(   -\dynkin(00130)_{29568} +\dynkin(00021)_{1440} +\dynkin(00101)_{1200} -2\dynkin(00010)_{16} \bigr)t^{-5}\nonumber\\&\qquad
     + \bigl(   +\dynkin(01040)_{64350} -\dynkin(00120)_{6930} -\dynkin(00200)_{4125} +2\dynkin(00011)_{210} -\dynkin(00000)_{1} \bigr)t^{-4}\nonumber\\&\qquad
     + \bigl(   -\dynkin(10050)_{68640} +\dynkin(01030)_{17280} -2\dynkin(00110)_{1200} +\dynkin(00001)_{16} \bigr)t^{-3}\nonumber\\&\qquad
     + \bigl(   +\dynkin(00060)_{28314} -\dynkin(10040)_{20790} +\dynkin(02020)_{46800} +3\dynkin(01020)_{3696} +2\dynkin(00020)_{126} \bigr)t^{-2}\nonumber\\&\qquad
     + \bigl(   -\dynkin(11030)_{102960} +\dynkin(00050)_{9504} -3\dynkin(10030)_{5280}  +\dynkin(11001)_{3696} \nonumber\\&\qquad\qquad+2\dynkin(02010)_{8064} +2\dynkin(10001)_{144} +6\dynkin(01010)_{560} +8\dynkin(00010)_{16} \bigr)t^{-1}\nonumber\\&\qquad
     + \bigl(   +\dynkin(01040)_{64350} -\dynkin(20011)_{8085} -2\dynkin(11020)_{23040} +3\dynkin(00040)_{2772} \nonumber\\&\qquad\qquad-5\dynkin(10020)_{1050} +\dynkin(03000)_{7644} +4\dynkin(02000)_{770} +10\dynkin(01000)_{45} +9\dynkin(00000)_{1} \bigr)t^{0}\nonumber\\&\qquad
     + \bigl(   -\dynkin(12010)_{43680} +2\dynkin(01030)_{17280} -2\dynkin(20001)_{720} -5\dynkin(11010)_{3696} \nonumber\\&\qquad\qquad+5\dynkin(00030)_{672} -10\dynkin(10010)_{144} \bigr)t^{1}\nonumber\\&\qquad
     + \bigl(   +\dynkin(02020)_{46800} +\dynkin(00031)_{6930} -\dynkin(20100)_{4312} -2\dynkin(12000)_{4410} \nonumber\\&\qquad\qquad+5\dynkin(01020)_{3696} -8\dynkin(11000)_{320} +10\dynkin(00020)_{126} -9\dynkin(10000)_{10} \bigr)t^{2}\nonumber\\&\qquad
     + \bigl(   +\dynkin(10110)_{8800} +2\dynkin(02010)_{8064} +2\dynkin(00021)_{1440} +8\dynkin(01010)_{560} +9\dynkin(00010)_{16} \bigr)t^{3}\nonumber\\&\qquad
     + \bigl(   -\dynkin(10101)_{8800} -2\dynkin(02001)_{8064} -2\dynkin(00012)_{1440} -8\dynkin(01001)_{560} -9\dynkin(00001)_{16} \bigr)t^{5}\nonumber\\&\qquad
     + \bigl(   -\dynkin(02002)_{46800} -\dynkin(00013)_{6930} +\dynkin(20100)_{4312} +2\dynkin(12000)_{4410} \nonumber\\&\qquad\qquad-5\dynkin(01002)_{3696} +8\dynkin(11000)_{320} -10\dynkin(00002)_{126} +9\dynkin(10000)_{10} \bigr)t^{6}\nonumber\\&\qquad
     + \bigl(   +\dynkin(12001)_{43680} -2\dynkin(01003)_{17280} +2\dynkin(20010)_{720} +5\dynkin(11001)_{3696} \nonumber\\&\qquad\qquad-5\dynkin(00003)_{672} +10\dynkin(10001)_{144} \bigr)t^{7}\nonumber\\&\qquad
     + \bigl(   -\dynkin(01004)_{64350} +\dynkin(20011)_{8085} +2\dynkin(11002)_{23040} -3\dynkin(00004)_{2772} \nonumber\\&\qquad\qquad+5\dynkin(10002)_{1050} -\dynkin(03000)_{7644} -4\dynkin(02000)_{770} -10\dynkin(01000)_{45} -9\dynkin(00000)_{1} \bigr)t^{8}\nonumber\\&\qquad
     + \bigl(   +\dynkin(11003)_{102960} -\dynkin(00005)_{9504} +3\dynkin(10003)_{5280} -\dynkin(11010)_{3696} \nonumber\\&\qquad\qquad-2\dynkin(02001)_{8064} -2\dynkin(10010)_{144} -6\dynkin(01001)_{560} -8\dynkin(00001)_{16} \bigr)t^{9}\nonumber\\&\qquad
     + \bigl(   -\dynkin(00006)_{28314} +\dynkin(10004)_{20790} -\dynkin(02002)_{46800} -3\dynkin(01002)_{3696} -2\dynkin(00002)_{126} \bigr)t^{10}\nonumber\\&\qquad
     + \bigl(   +\dynkin(10005)_{68640} -\dynkin(01003)_{17280} +2\dynkin(00101)_{1200} -\dynkin(00010)_{16} \bigr)t^{11}\nonumber\\&\qquad
     + \bigl(   -\dynkin(01004)_{64350} +\dynkin(00102)_{6930} +\dynkin(00200)_{4125} -2\dynkin(00011)_{210} +\dynkin(00000)_{1} \bigr)t^{12}\nonumber\\&\qquad
     + \bigl(   +\dynkin(00103)_{29568} -\dynkin(00012)_{1440} -\dynkin(00110)_{1200} +2\dynkin(00001)_{16} \bigr)t^{13}\nonumber\\&\qquad
     + \bigl(   -\dynkin(00013)_{6930} +\dynkin(00002)_{126} +\dynkin(00100)_{120} \bigr)t^{14}\nonumber\\&\qquad
     + \dynkin(00003)_{672}t^{15}
\end{align}

%%%%%%%%%%%%%%%%%%%%%%%%%%%%%%%%%%%%%%%%%%%%%%%%%%%%%%%%%%%%%%%%
\section{Reducibility conditions for pure spinor constraint}
\label{app:reducibility}

Here, we compute the reducibility coefficients for the pure spinor constraint
up to fourth order.
In section~\ref{sec:reducibility}, we performed the analysis up to
second order and found:
\begin{align}
\text{Original constraint:}&\quad G^{A_{1}} = \lambda\gamma^{\mu}\lambda\,,\quad A_{1}=\dynkin(10000) \\
\begin{split}
\text{First order reducibility:}&\quad
 R_{A_{1}}^{A_{2}} = (\gamma^{\mu}\lambda)_{\alpha}\,,\quad A_{2} = \dynkin(00010)\,, \\
 &\quad \to G^{A_{1}}R_{A_{1}}^{A_{2}} = (\lambda\gamma^{\mu}\lambda)(\gamma_{\mu}\lambda)_{\alpha} = 0\;\text{(strong equality)}\,,
\end{split} \\
\begin{split}
\text{Second order reducibility:}&\quad
 R_{A_{2}}^{A_{3}} = (\gamma^{\mu\nu}\lambda)^{\alpha}\,,\quad A_{3} = \dynkin(01000)\,, \\
&\quad\to
 R_{A_{1}}^{A_{2}}R_{A_{2}}^{A_{3}}
 = (\gamma^{\mu}\lambda)_{\alpha}(\gamma_{\nu\rho}\lambda)^{\alpha} \approx 0\;\text{(weak equality)}\,.
\end{split}
\end{align}
Before proceeding to the third order reducibility,
let us mention a subtle but important technicality
in the BRST construction~\cite{Henneaux:1992ig}.

\subsection{Technicalities and a remark on second order reducibility}
\label{app:redtech}

In order to kill the spurious part of the first generation BRST ghost,
$R_{A_{1}}^{A_{2}}$ must be ``complete'' in the sense that any function $F_{A_1}$ satisfying
\begin{align}
 \label{eq:tech1}
  G^{A_{1}}F_{A_{1}} = 0
\end{align}
can be written as
\begin{align}
  F_{A_{1}} &= R_{A_{1}}^{A_{2}}f_{A_{2}} + G^{B_{1}}f_{A_{1}B_{1}}
\end{align}
for some $f_{A_1}$ and $f_{A_1B_1}$.
In computing $R_{A_{1}}^{A_{2}}$, one can always choose it
so that $f_{A_{1}B_{1}}$ is graded antisymmetric
for arbitrary $F_{A_{1}}$ satisfying~(\ref{eq:tech1}).
Since the pieces of $R_{A_{1}}^{A_{2}}$ that lead to
graded symmetric pieces in $f_{A_{1}B_{1}}$ are irrelevant for
(or decouples from) the BRST construction,
$R_{A_{1}}^{A_{2}}$ should be chosen to meet this condition.

Similarly, the second order reducibility coefficient $R_{A_{2}}^{A_{3}}$ must be chosen so that
the indices $A_{1}B_{1}$ in the relation
\begin{align}
 \label{eq:tech2}
 R_{A_{1}}^{A_{2}}R_{A_{2}}^{A_{3}} = G^{B_{1}}f_{A_{1}B_{1}}{}^{A^{3}} \approx0
\end{align}
are graded antisymmetric.
(There is no analogous symmetric property for the reducibility
coefficients at higher degrees.)

\bigskip
Now, in connection with this, let us explain a subtlety we have not mentioned
when we computed the second order reducibility coefficient $R_{A_{2}}^{A_{3}}$ in the main text.
At first sight, there seems to be another non-trivial relation at this order~\cite{Chesterman:2004xt}
\begin{align}
\begin{split}
 & R_{A_{1}}^{A_{2}}R_{A_{2}}^{\bullet_{3}} \approx 0 \quad\leftrightarrow\quad (\gamma^{\mu}\lambda)_{\alpha}\lambda^{\alpha} = (\lambda\gamma^{\mu}\lambda) \approx0\,,  \\
 & R_{A_{2}}^{\bullet_{3}} = \lambda^{\alpha},\quad \bullet_{3} = \dynkin(00000)\,.
\end{split}
\end{align}
However, the relation $R_{A_{2}}^{\bullet_{3}}$ decouples from the BRST construction
as there is no $\delta$-closed state constructed out of this relation.
That is, although
\begin{align}
 \label{eq:wrongsym}
  \delta(C^{A_{2}}R_{A_{2}}^{\bullet_{3}}) = C^{A_{1}}G^{B_{1}}f_{A_{1}B_{1}}{}^{\bullet_{3}}\approx0
 \quad\leftrightarrow\quad
 \delta(\sigma_{\alpha}\lambda^{\alpha}) = c^{\mu}(\lambda\gamma^{\nu}\lambda)\eta_{\mu\nu}\approx0\,,
\end{align}
there is no way to cancel the weakly zero term on the right hand side
by adding an appropriate term $M^{\bullet_{3}}$.
This means that there is nothing to kill by introducing the next generation ghost,
and so one should not introduce this ghost.
(If one were to introduce the corresponding ghost,
the action of $\delta$ on that ghost would not be nilpotent.)

The reason why there is no $\delta$-closed operator of the form $C^{A_{2}}R_{A_{2}}^{\bullet_{3}}+M^{\bullet_{3}}$
is related to the violation of the assumption~(\ref{eq:tech2}).
As can be seen from the equation~(\ref{eq:wrongsym}),
one cannot construct an appropriate $M^{\bullet_{3}}$
because $f_{A_{1}B_{1}}{}^{\bullet_{3}}=\eta_{\mu\nu}$ is symmetric
in $A_{1}B_{1}$, or in other words because $c^{\mu}c^{\nu}\eta_{\mu\nu}=0$.

This concludes our discussion of the subtlety in the BRST construction,
and let us return to the computation of third and fourth reducibility
coefficients.

\subsection{Third and fourth order reducibilities}

\paragraph{Third order reducibility}

A little computation shows
\begin{align}
\begin{split}
R_{A_{2}}^{A_{3}}R_{A_{3}}^{A_{4}}\approx 0
\quad\to\quad
R_{A_{3}}^{A_{4}}
&=  (\eta_{\underline{\rho}[\mu}\gamma_{\nu]}\lambda)_{\underline{\beta}}
 + {1\over6}(\lambda\gamma_{\underline{\rho}}\gamma_{\mu\nu})_{\underline{\beta}} \\
&= {4\over3} (\eta_{\underline{\rho}[\mu}\gamma_{\nu]}\lambda)_{\underline{\beta}}
 - {1\over6}(\gamma_{\underline{\rho}\mu\nu}\lambda)_{\underline{\beta}}\,,
\\
A_{4} &= \dynkin(10010)=\overline{\mathbf{144}}\,.
\end{split}
\end{align}
The indices newly appeared, which we underline for convenience,
are $\gamma$-traceless and hence in the $\mathbf{144}$ representation.

Indeed, there is a corresponding $\delta$-closed element of the form
$C^{A_{3}}R_{A_{3}}^{A_{4}} + M^{A_{4}}$
where $M^{A_{4}}$ is free of $C^{A_{3}}=c^{\mu\nu}$:
\begin{align}
C^{A_{3}}R_{A_{3}}^{A_{4}} + M^{A_{4}}
&=
{4\over3}c_{\rho \nu}(\gamma^{\nu}\lambda)_{\beta} - {1\over6}c^{\mu\nu}(\gamma_{\rho\mu\nu}\lambda)_{\beta}
  + {3\over2}c_{\rho}\sigma_{\beta} + {1\over6}c^{\nu}(\sigma\gamma_{\rho\nu})_{\beta}\,.
\end{align}
In order to kill this, we introduce the fourth generation ghost
and extend the nilpotent action of $\delta$ as
\begin{align}
\begin{split}
  \delta C^{A_{4}} &=   C^{A_{3}}R_{A_{3}}^{A_{4}} + M^{A_{4}}\,, \\
\leftrightarrow\quad
 \delta\sigma_{\rho \beta}
 &= {4\over3}c_{\rho\nu}(\gamma^{\nu}\lambda)_{\beta} - {1\over6}c^{\mu\nu}(\gamma_{\rho\mu\nu}\lambda)_{\beta}
  + {3\over2}c_{\rho}\sigma_{\beta} + {1\over6}c^{\nu}(\sigma\gamma_{\rho\nu})_{\beta}\,.
\end{split}
\end{align}

\paragraph{Fourth order reducibility}

As we described in section~\ref{sec:spinsofghosts},
the spin contents of the ghosts-for-ghosts are dictated by
the level $0$ partition function $Z_{0}(t,\vec{\sigma})$.
At this level, we expect
\begin{align}
 A_{5}&= \dynkin(11000)+\dynkin(10000)+\dynkin(00010)=\mathbf{320}+\mathbf{10}+\overline{\mathbf{126}} \,.
\end{align}
Indeed, one finds the following reducibility coefficients
\begin{align}
\begin{split}
 R_{A_{4}}^{A_{5a}} &=(\eta^{\kappa\lbra \underline{\mu}}\gamma^{\underline{\nu\rho} \rbra}\lambda)^{\alpha},\quad A_{5a}=\mathbf{320}\,, \\
 R_{A_{4}}^{A_{5a}} &= (\gamma^{\kappa\underline{\rho}}\lambda)^{\alpha},\quad A_{5b} = \mathbf{10}\,, \\
  R_{A_{4}}^{A_{5c}}
 &=  (\gamma^{\kappa}\lambda)_{\lpar\underline{\beta}}{\delta^{\alpha}}_{\underline{\gamma} \rpar},\quad
 A_{5c}=\overline{\mathbf{126}}\,.
\end{split}
\end{align}
Here, indices in $\lpar \beta\gamma \rpar$ are symmetric and (spinorial) $\gamma$-traceless,
and those in $\lbra \mu,\nu\rho\rbra$ are traceless, block-symmetric,
and antisymmetric within each blocks.

Corresponding $\delta$-closed elements are
\begin{align}
\begin{split}
 C^{A_{4}}R_{A_{4}}^{A_{5a}} + M^{A_{5a}}
&= (\sigma^{\lbra \mu}\gamma^{\nu\rho\rbra}\lambda) + {5\over3}c^{\lbra \mu}c^{\nu\rho \rbra} \,,
\\
  C^{A_{4}}R_{A_{4}}^{A_{5b}}+ M^{A_{5b}} &= \sigma_{\nu \alpha}(\gamma^{\nu \rho}\lambda)^{\alpha}  -{4\over3}c_{\nu}c^{\nu\rho} \,,
\\
  C^{A_{4}}R_{A_{4}}^{A_{5b}} + M^{5c}
&=  \sigma_{\mu \lpar \beta}(\gamma^{\mu}\lambda)_{\gamma \rpar}
- {7\over9}\sigma_{\lpar \beta}\sigma_{\gamma \rpar}\,,
\end{split}
\end{align}
and we shall introduce the fifth generation ghosts
\begin{align}
  C^{A_{5}} 
  = (c^{\lbra \mu,\nu\rho \rbra}, c^{\mu}, c_{\lpar \alpha\beta \rpar})
 = (\mathbf{320},\mathbf{10},\overline{\mathbf{126}})\,,
\end{align}
and extend the $\delta$-action as usual.

%%%%%%%%%%%%%%%%%%%%%%%%%%%%%%%%%%%%%%%%%%%%%%%%%%%%%%%%%%%%%%%%

\end{document}